\documentclass{aa}
\usepackage{natbib}
\usepackage{longtable}
\usepackage{lscape} % for 'landscape' environment
\bibpunct{(}{)}{;}{a}{}{,} % to follow the A&A style
\usepackage{graphicx}
\usepackage{txfonts}
\providecommand{\abs}[1]{\lvert#1\rvert}
\newcommand\tab[1][1cm]{\hspace*{#1}}

\begin{document} 

   \title{Properties of central stars of planetary nebulae with distances in Gaia DR2}
   
   \titlerunning{Properties of PNe central stars in GDR2}

   \author{I. Gonz\'alez-Santamar\'{\i}a\inst{1,2} \and M. Manteiga\inst{2,3} \and A. Manchado\inst{4,5} \and A. Ulla\inst{6} \and C. Dafonte\inst{1,2}}
   \authorrunning{I. Gonz\'alez-Santamar\'{\i}a et al.}
   \institute{Universidade da Coru\~na (UDC), Department of Computer Science and Information Technology, Campus Elvi\~na sn, 15071 A Coru\~na, Spain \\
  \email{iker.gongalez@udc.es}
  \and
  CITIC, Centre for Information and Communications Technology Research, Universidade da Coru\~na, Campus de Elvi\~na sn, 15071 A Coru\~na, Spain                    
  \and
   Universidade da Coru\~na (UDC), Department of Nautical Sciences and Marine Engineering, Paseo de Ronda 51, 15011, A Coru\~na, Spain \\
   \email{manteiga@udc.es}
  \and
  Instituto de Astrofísica de Canarias, 38200 La Laguna, Tenerife, Spain
      %\and Consejo Superior de Investigaciones Científicas %(CSIC), 28006, Madrid, Spain
  \and
  Universidad de La Laguna (ULL), Astrophysics Department; CSIC. 38206 La Laguna, Tenerife, Spain
    %  \email{amt@iac.es}
          \and
  Universidade de Vigo (UVIGO), Applied Physics Department, Campus Lagoas-Marcosende, s/n, 36310 Vigo, Spain\\
    %  \email{ulla@uvigo.es}
           }

   \subtitle{}

   \date{Received 24 June 2019/Accepted 6 August 2019}

% \abstract{}{}{}{}{} 
% 5 {} token are mandatory
 
  \abstract
   {We have compiled a catalogue of central stars of planetary nebulae (CSPN) with reliable distances and positions obtained from Gaia Data Release 2 (DR2) astrometry. Distances derived from parallaxes allow us to analyse the galactic distribution and estimate other parameters such as sizes, kinematical ages, bolometric magnitudes, and luminosities.}
   {Our objective is to analyse the information regarding distances together with other available literature data about photometric properties, nebular kinematics, and stellar effective temperatures  to throw new light on this rapid and rather unknown evolutionary phase. We seek to understand how Gaia distances compare with other indirect methods commonly used and, in particular, with those derived from non-local thermodynamic equilibrium (non-LTE) models;  how many planetary nebulae (PNe) populate the Galaxy; and how are they spatially distributed. We also aim to comprehend their intrinsic luminosities, range of physical sizes of the nebulae; how to derive the values for their kinematical ages; and whether those ages are compatible with those derived from evolutionary models.} 
   {We considered all PNe listed in %\citet{kerber03}, \citet{stanghellini10}, \citet{weidmann11} and HASH database \citep{parker16}.\LEt{As per Editor in Chief and journal style, please remove all references from the Abstract.} 
   catalogues from different authors and in Hong Kong/AAO/Strasbourg/$H_{\alpha}$ (HASH) database.
   By X-matching their positions with Gaia DR2 astrometry we were able to identify 1571 objects in Gaia second archive, for which we assumed distances calculated upon a Bayesian statistical approach. 
   %from \citet{bailerJones18}. 
   From those objects, we selected a sample of PNe with good quality parallax measurements and distance derivations, we which refer to as our Golden Astrometry PNe sample (GAPN), and obtained literature values of their apparent sizes, radial and expansion velocities, visual magnitudes, interstellar reddening, and effective temperatures.}
{We found that the distances derived from DR2 parallaxes compare well with previous astrometric derivations of the United States Naval Observatory and Hubble Space Telescope, but that distances inferred from non-LTE model fitting are overestimated and need to be carefully reviewed. From literature apparent sizes, we calculated the physical radii for a subsample of nebulae that we used to derive the so-called kinematical ages, taking into account literature expansion velocities. Luminosities calculated with DR2 distances were combined with literature central stars $T_{eff}$ values in a Hertzsprung-Russell (HR) diagram to infer information on the evolutionary status of the nebulae. We compared their positions with updated evolutionary tracks 
%by  \citet{millerbertolami17} 
finding a rather consistent picture. Stars with the smallest associated nebular radii are located in the flat luminosity region of the HR diagram, while those with the largest radii correspond to objects in a later stage, getting dimmer on their way to become a white dwarf. Finally, we commented on the completeness of our catalogue and calculated an approximate value for the total number of PNe in the Galaxy.}
   {}

%   \keywords{
%       planetary nebulae: general --
%    galactic distances --
%    HR diagram --
%    planetary nebulae population
%               }
\maketitle
%
%-------------------------------------------------------------------
%\end{titlepage}

\section{Introduction}
The planetary nebulae (PNe) phase represents a very short stage in the late evolution of low- and intermediate-mass stars, which occurs while they ionise their envelope to finally enter the white dwarf (WD) cooling track.  This stellar evolutionary phase is interesting for a number of reasons. One reason is that PNe significantly contribute to the chemical enrichment of the interstellar medium by the ejection of processed material in the form of gas and dust. Chemical abundances can be easily derived from PNe spectra to constrain the initial composition of the progenitor star and to provide clues to mixing and nucleosynthesis processes. Emission line spectra of PNe can be used to easily identify them and their luminosity function has been used as an extragalactic distance indicator.  Still, there are some fundamental open issues. In particular, the discrepancy in the distances to the central stars derived by different methods, such as parallaxes, hydrodynamical wind models, or evolutionary models using non-LTE atmospheres. This problem can only be addressed by measuring precise and consistent distances to PNe. Distances can be used to determine intrinsic properties of the nebulae, such as radii and luminosities, and allow the derivation of masses and evolutionary ages by means of evolutionary models.
\\
\\
Gaia satellite Data Release 2 (DR2) contains information on astrometry (parallaxes and proper motions), brightness in three bands, and radial velocities for a limited subsample of red stars for more than a billion galactic sources. These positional and kinematical measurements provide important tools to analyse the composition and evolution of the Milky Way, and they have allowed, for instance, the first complete and accurate census of the Galaxy 
through a Hetzsprung-Russell (HR) diagram \citep{babusiaux18}, and the kinematical mapping of the different populations of stars revealing orbits, substructures, and 
velocity dispersions totally unexpected for an axisymmetric Galaxy in dynamical equilibrium \citep{katz18}. 
Gaia will be scanning the sky for at least four more years, while improving the quality of the obtained astrometric and photometric measurements. 
%Regarding to photometry, sources are observed with a pixel scale of 59 mas/pixel and a point spread function of $\im 180$ mas, up to magnitude $G \sim 20$, this will give different values comparing with the visual magnitude from the common Johnson system.
In the meantime, DR2 parallaxes allow the computation of distances and the derivation of the absolute luminosities of the central stars of PNe (CSPN) and the radii of the nebulae. In this paper we review some properties of the population of PNe in our Galaxy as seen through the eyes of Gaia in DR2.  

Section 2 discusses the errors in the measurement of parallaxes present in DR2, the systematic corrections (zero point), statistical errors, and the way they can be used to derive statistically consistent distances \citep{lindegren18,bailerJones18}. In section 3 we explain how we retrieved the astrometric measurements for available PNe in DR2. From all PNe identified in DR2 with available distances in \citet{bailerJones18}, some statistical properties are presented. In order to study the intrinsic properties of those nebulae with reliable distance derivations, we selected a sample of PNe, which we call the Golden Astrometry PNe (GAPN), with good quality parallax measurements and distance derivations, by imposing quality cuts in the available measurements of parallaxes. Some additional cleaning was carried out to exclude objects from this sample that were misclassified as PNe or post-AGB stars.

In section 4, galactic distribution and distances adopted for GAPN are presented and compared with those obtained with previous determinations of distances using astrometry \citep{harris07}, non-local thermodynamic equilibrium (non-LTE) models \citep{napiwotzki01}, and other methods \citep{stanghellini10,frew16,schonberner18}. Using these distances, nebular radii are also estimated, derived from angular sizes in the Hong Kong/AAO/Strasbourg/$H_{\alpha}$ (HASH) database \citep{parker16}.
Nebular absolute sizes and literature expansion velocities (with a correction) for a suited subsample of nebulae are used to derive kinematical ages, as shown in section 5. 

Section 6 is devoted to analysing the physical properties of some of our GAPN based on their distances and on literature values of their visual magnitudes, interstellar reddening, and effective temperatures. We calculate the luminosities and derive the star temperature versus luminosity positions in the HR diagram, which can be discussed in comparison with updated evolutionary tracks in  \citet{millerbertolami17}.
Finally, an estimation of PNe density, scale height, birth rate, and total number of nebulae in the Galaxy are provided in section 7, using a procedure based on \citet{frew08} and compared with other literature results such as \citet{zijlstrapottasch91} and \citet{Pottasch96}. We also analyse the completeness of our general sample and, finally, summarise our conclusions.

\section{Parallaxes and distances in DR2}

In Gaia DR2, parallaxes and their uncertainties are given with a great accuracy, milliarcseconds (mas), but the formal uncertainties listed in DR2 are estimated from the internal consistency of measurements and they do not represent the total errors. Following \citet{lindegren18}, the 
total error in DR2 parallaxes is the addition of random (internal) and systematic errors, the latter including the parallax zero point, and they are dependent, at least, on position, magnitude, and colour. This is in part due to patterns imprinted by the Gaia scanning law and to the spacecraft attitude errors. 

On the one hand, parallaxes need a bias correction called `zero point', $w_{0}$, and according to the study of Lindegren, as a global average, this parameter takes a value of -0.03 mas. On the other hand, it is necessary to correct the internal error of parallax measurements, $\sigma_{i}$, where values depend on the specific source as given by Gaia DR2. Additionally, DR2 parallax measurements are subject to systematic errors, $\sigma_{s}$,which depend on the brightness of the object (i.e. G magnitude in the Gaia photometric system). The total error, $\sigma_{T}$, can be obtained from

$$\sigma_{T}= \sqrt{k^{2}\cdot\sigma_{i}^{2}+\sigma_{s}^{2}},$$

where $k$ is estimated to be 1.08 \citep{lindegren18}.  Following DR2 documentation, we calculated the total error as follows:

\begin{equation}
\label{eq:aqui-le-mostramos-como-hacerle-la-llave-grande}
\sigma_{T} = \left\{
\begin{array}{ll}
\sqrt{1.08^{2}\cdot\sigma_{i}^{2}+(0.043)^{2}} , & \mathrm{if\ } G > 13 \\
\sqrt{1.08^{2}\cdot\sigma_{i}^{2}+(0.021)^{2}} , & \mathrm{if\ } G \leq 13 \\
\end{array}
\right.
.\end{equation}
\\\\
Once computed, total uncertainties in parallaxes can be used to establish confidence criteria in our selection of stars, which allows us to work with 
reliable distance values derived from them. Additionally, we considered the recommended goodness-of-fit indices for Gaia DR2 astrometry \citep{lindegren18}, the unit weight error (UWE), which is computed from the astrometric chi square test of the measurements, and the renormalised unit weight error (RUWE), which uses an empirical normalisation factor provided in DR2 ESA web page. We chose the limiting values of both quantities that are 
recommended in DR2 documentation, i.e. UWE $< 1.96$ or RUWE $< 1.40$.

The derivation of distances from parallaxes ($\pi_{true}$)  with high uncertainties is not straightforward because distance `r' has a non-linear relationship to the
measured quantity, $r=1/\pi_{true}$, and it is constrained to be positive \citep{luri18}. A useful approach is to consider some assumptions about the 
distribution of the distances in our Galaxy, known as a prior within a statistical Bayesian analysis. As discussed in \citet{bailerJones15} and \citet{Astraatmadja16}, a possibility is to assume that the a priori probability volume density of stars in the Milky Way is exponentially declining with some appropriate distance scale. This exponentially decreasing space density (EDSD) is explicitly endorsed in DR2. We decided to use the 
\citet{bailerJones18} catalogue of estimated distances from DR2 parallaxes, which uses an EDSD with a distance scale L that varies as a function of 
galactic latitude and longitude, according to a model suited for Gaia observations. In the next sections we discuss the distance errors and the distribution of distances obtained for our sample of PNe.

\section{Selection of a sample of PNe central stars with distances and reliable distances in DR2}

\subsection{General selection}

There are several compilations of PNe and CSPN in recent literature, among which we chose the following: Kerber et al. (2003); Stanghellini et al. (2010); Weidmann et al. (2011); and HASH database \citep{parker16}.
Firstly, we considered the objects contained in the first three catalogues (Kerber, Stanghellini, and Weidmann), taking into account both coordinates and names. Next, we complemented our compilation selecting the objects catalogued as 'true PN' in the HASH database, which includes, in addition to the other catalogues, new objects that were detected in several $H_{\alpha}$ surveys.

Thus, in total, we ended up with 2554 sources.
The next step was to verify which of these objects were observed by Gaia in DR2. It is understood that Gaia observations are aimed to detect unresolved sources, in our case PNe central stars; hence, PNe without a visible central star or with a very faint star are not expected to be catalogued in DR2.  For this task we used the ARI's Gaia Services and we did queries by list to the 'gaiadr2.gai\_source' table using the coordinates or the names of the PNes. By obtaining the closest Gaia object for each coordinate/name given in our PNe list, we retrieved Gaia DR2 measurements (parallaxes, G magnitudes…) when available, reaching a number of 1948 sources with measured parallaxes. Among these, we further checked for objects with dubious identifications, keeping only those objects with coordinates not farther than 5 arcsec from those listed in DR2, finding a total number of 1736 objects.

This list of objects was then queried to the Simbad database, doing an X-match with Gaia's objects coordinates to obtain more parameters of the sources (object type, photometry, angular size, radial velocity …) provided by this database. Several sources in our catalogue were identified by Simbad with an object type other than 'PN' and were excluded from our list. So, we ended up with a final list of 1571 objects.

Once the Gaia DR2 ID of the objects were obtained, we proceeded to retrieve their estimated distances from us by querying the 'gaiadr2\_complements.geometric\_distance' table in DR2. This table lists the values of the estimated distances, using the Bayesian approach by \citet{bailerJones18} mentioned in the previous section. Apart from the estimated distances, this approach provides 
lower and upper bounds of the distance values, among other parameters.

\begin{figure}
       \includegraphics[width=0.52\textwidth]{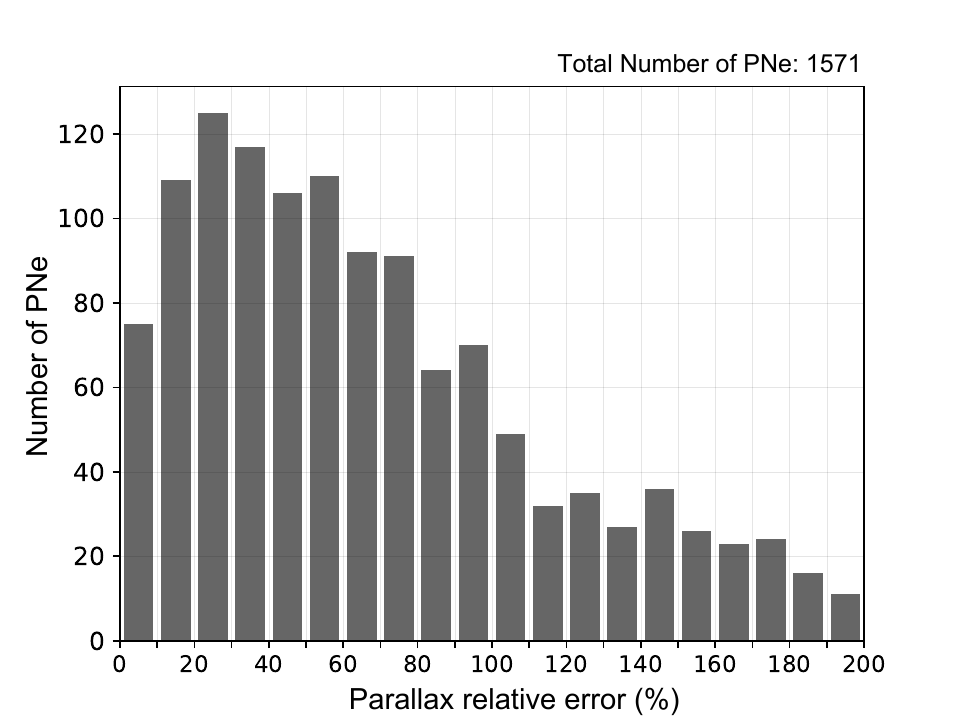}
       \includegraphics[width=0.52\textwidth]{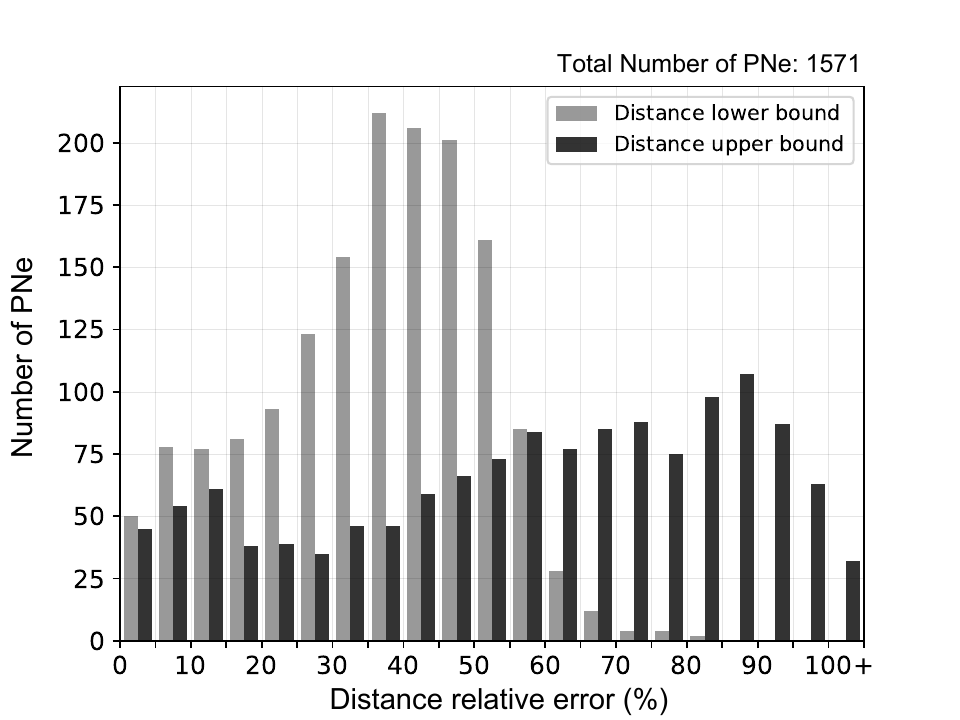}
       \caption{Parallax relative errors (upper) and low and high distances relative errors (lower).}
       \label{fig:parallax_distance_errors}
\end{figure}

\begin{figure}      %\includegraphics[width=0.5\textwidth]{distance/Muestra_total/Histogram_distance_500.pdf}
        \includegraphics[width=0.52\textwidth]{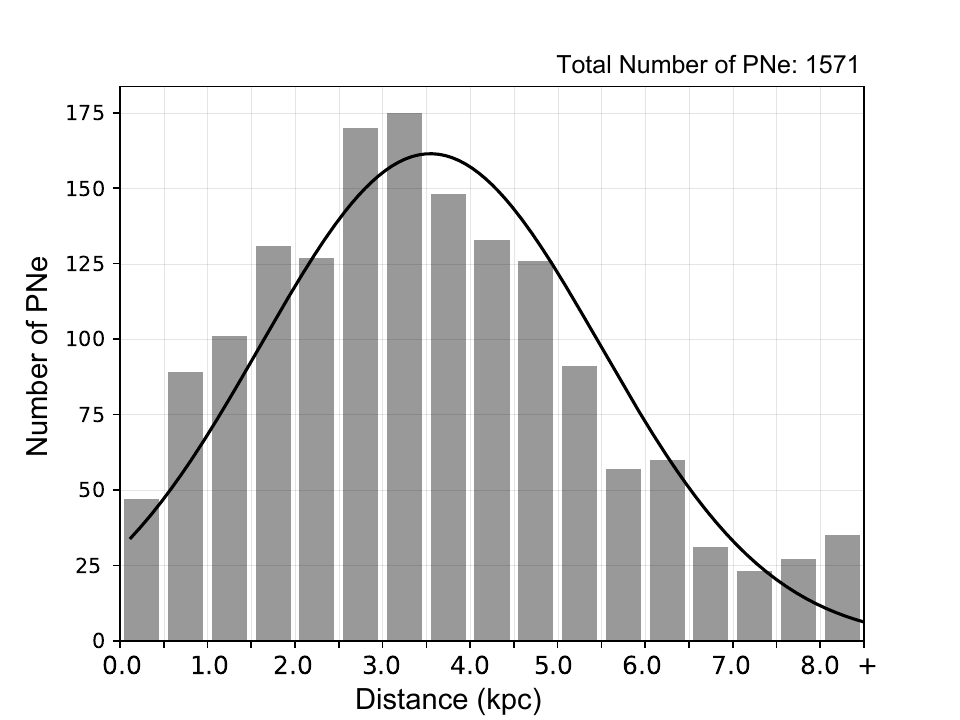}
        \caption{Distances to PNe in DR2 derived with a Bayesian approach (\citet{bailerJones18}).}
        \label{fig:distance_total}
\end{figure}

The distributions of parallaxes relative errors and distance errors (higher and lower bounds) % from the adopted Bayesian approach) 
are shown in  Fig.~\ref{fig:parallax_distance_errors}
%Fig.~\ref{fig:parallax_distance_error}. 
More than 600 PNe have their parallaxes measured with relative errors higher than $100\%$, which translates into errors in the derivation of distances higher than $50\%$ for most of the stars in our sample (for the upper bound of errors). Meanwhile, Fig.~\ref{fig:distance_total} shows the histogram of the derived distances, for the sample of 1571 PNe with distances in DR2.

%In addition, \textit{Figure 3} shows the distribution of stars at several values of estimated distances with errors lower than $10\%$, $30\%$, $50\%$, $70\%$, $90\%$ and for the remaing objects. 
%We would like to comment on the wide range of uncertainties in the distance that the studied sample contains. 
We are aware of the problem that the use of parallaxes with large uncertainties translates into distances and other astrophysical quantities derived from them, such as luminosities and sizes. 
As discussed in \citet{luri18}, truncation of data using a threshold for the parallax relative error, or the exclusion of objects with negative measured parallaxes from a sample, makes the distribution of distances unrepresentative and can lead to wrong conclusions regarding the statistical properties of the sample. Consequently, we decided to keep the complete sample for the discussion regarding general properties of our sample, such as the distance distribution of PNe and the estimation of the number of PNe in our galaxy. Conclusions in such cases can only be formulated among the appropriate errors reported for the measured quantities. 

\subsection{Selection of sample with reliable distances}

A different approach can be followed if we intend to derive individual properties for a subsample with high quality measurements of both parallaxes and the corresponding  distances.%, as derived from a Bayesian approach. 
With this objective in mind, we constructed a sample of PNe with reliable distances in DR2, our GAPN, with constraints in the following properties:

\begin{itemize}
        \item Angular distance: The distance in arc seconds between the PN coordinates and the closest object detected by Gaia. We chose objects with angular distances lower than 5 arcsec from the original coordinates.
        \\
        \item Parallax relative error: Obtained by taking into account all reported internal and systematic errors, as explained in section 2. We chose a threshold for  parallax relative errors of $30\%$.
        \\ 
        \item Low/high distance relative error: Obtained by taking into account both low and high bounds in distance, given by the adopted Bayesian approach. We selected objects with relative errors for both lower and upper distance bounds lower than $30\%$.
        \\ 
        \item Unit weight error (UWE): This is defined as $\sqrt{\frac{\chi^{2}}{N-5}}$, where $\chi^{2}$ is the 'astrometric\_chi2\_al' 
        value and $N$ is the 'astrometric\_n\_good\_obs\_al' value. These parameters are provided by Gaia database. We set a lower threshold value for UWE of 1.96, following  the recommendation in \citet{lindegren18}.
        \\
        \item Renormalised unit weight error (RUWE): This is the UWE value divided by the normalisation function $U_{0} (G,C)$ (or $U_{0} (G)$ for 
        sources without known colour). This function depends on the brightness and colour of the sources, $G$ ('phot\_g\_mean\_mag') and on $C = G_{BP}-G_{RP}$ and has been interpolated 
        from the tables provided in ('ESA Gaia DR2 known issues' page). The $RUWE$ lower threshold value was set to 1.4 \citep{lindegren18}.
\end{itemize}

%The constrains that we applied to select the GAPN sample consider both the quality of the measured parallaxes (goodness of fit parameters and total parallax errors), as well as the errors reported for the distances derived from the Bayesian approach. Specifically, we chose objects with angular distances lower than 5 arcsec from the original coordinates, parallax relative errors lower than $30\%$, a $UWE$ value lower than 1.96 or a $RUWE$ value lower than 1.4, and both lower and upper distance bounds relative errors lower than $30\%$. $UWE$ and $RUWE$ threshold values are recommended in \citet{lindegren18}. After this filtering, we obtained 221 PNe.
%\begin{figure}
%       %\begin{wrapfigure}{r}{0.25\textwidth}
%       \includegraphics[width=0.24\textwidth]{general_data/Muestra_total/hist_UWE.pdf}
%       \includegraphics[width=0.24\textwidth]{general_data/Muestra_total/hist_RUWE.pdf}
%       \caption{\textit{UWE and RUWE parameters.}}
%       \label{fig:muestra_total_uwe_ruwe}
%\end{figure}

%%%%%%%%%%%%%%%%%%%%%%%%%%%%%%%%%%%%%%%%%%%%%%%%%
%Finally, we did a more accurated analysis for objects that: 

In order to check the identification of some dubious sources, we studied in detail those located farther than 2 arcsec from DR2 coordinates, those objects with distances beyond 8 kpc from us, and those whose nebular radius was estimated as larger than 1 pc (using bibliographic angular sizes and the obtained distances, as we explain later in the paper). So, finally, after discarding all dubious cases, we ended up with a catalogue of 201 GAPN.

%Taking into account the few objects we finally selected between 2 and 5 arcescs from the original coordinates, and the decreasing pattern as these distances becames larger (from 2 to 5 arcsec), we considered not to search for distances farher than 5 arcsecs.

\section{Discussion on galactic distribution, distances, physical sizes, and radial velocities}
\begin{figure}
        %\begin{wrapfigure}{r}{0.25\textwidth}
        \includegraphics[width=0.52\textwidth]{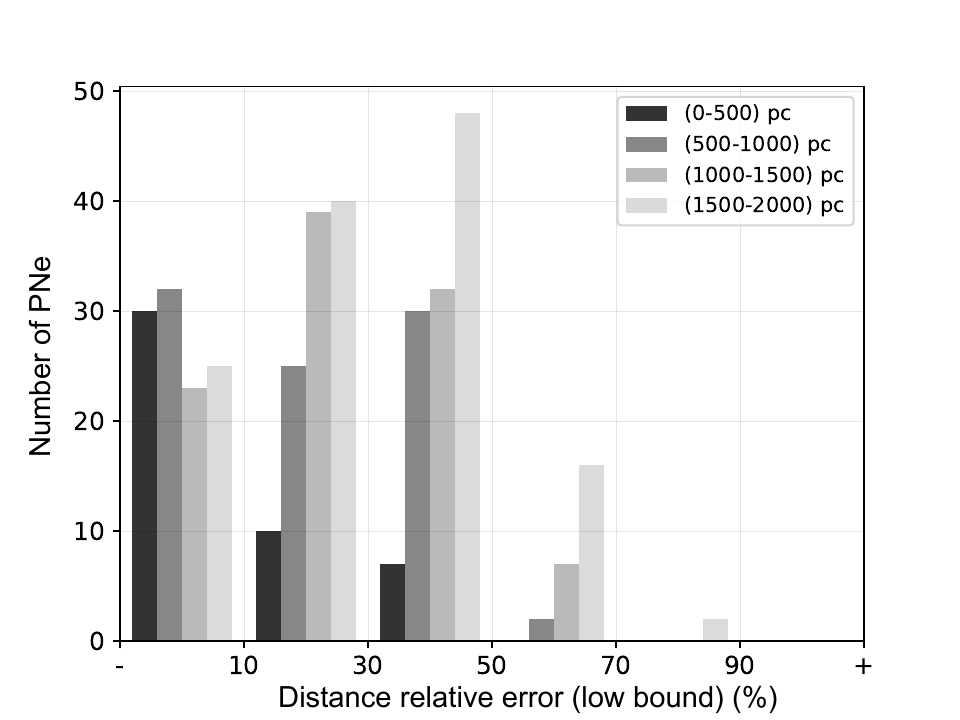}
        \includegraphics[width=0.52\textwidth]{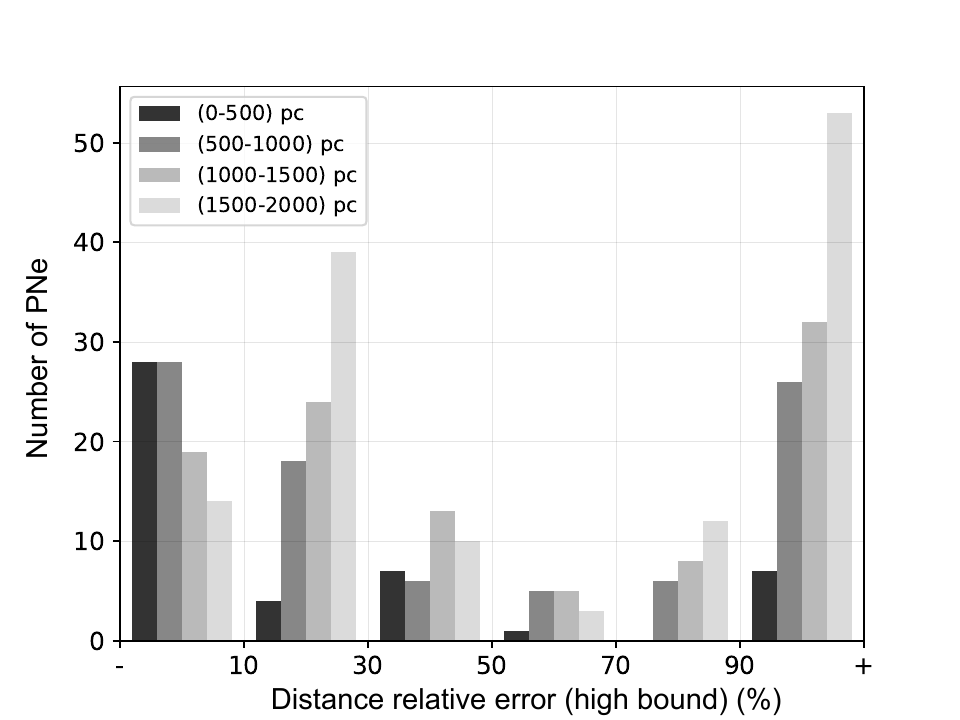}
        \caption{Relative errors in distances for different distance ranges derived from DR2 parallaxes upon a Bayesian approach (Bailer-Jones et al. 2018). Errors are shown for lower (upper panel) and higher (lower panel) bounds.}
        \label{fig:distance_errors}
\end{figure}

As shown in Fig.~\ref{fig:distance_total}, the distribution of distances for the sample of 1571 PNe is rather smooth. This distribution can be fitted with a Gaussian function with maximum value at 3.55 kpc and sigma 1.94 kpc. To visualise how errors are affecting distance derivations, in Fig.~\ref{fig:distance_errors} we show the number of PNe at several distance intervals (0-500 pc, 500-1000 pc, 1000-1500 pc, and 1500-2000 pc); relative errors are binned at $10\%$ or $20\%$ intervals.  This is plotted for both distance error bounds, i.e. low and high. It can be seen that while in the case of the lower bound errors values higher than $50\%$ are found for a very marginal number of PNe, if higher bounds of errors are considered the behaviour is very different; a significant number of nebulae beyond the first 500 pc are affected by relative errors of the order of $50\%$ and higher. In the following sections we focus on the selected sample of 201 PNe with very reliable distances, the GAPN sample. 

\subsection{Galactic distribution, parallaxes, and distances for GAPN}

\begin{figure}
        %\begin{wrapfigure}{r}{0.25\textwidth}
        \includegraphics[width=0.52\textwidth]{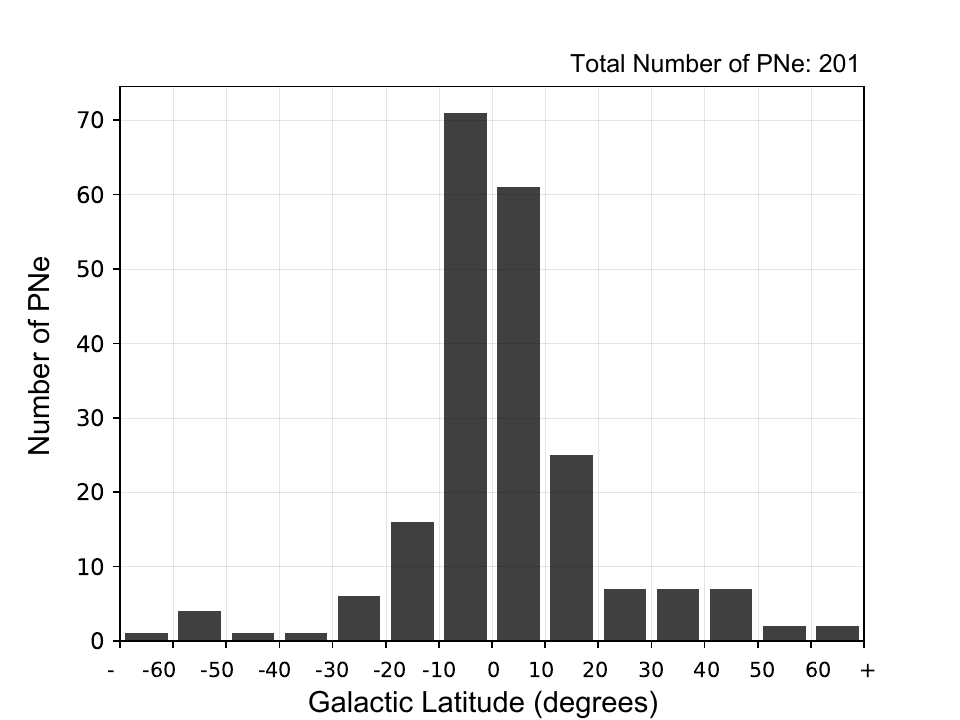}
        \includegraphics[width=0.52\textwidth]{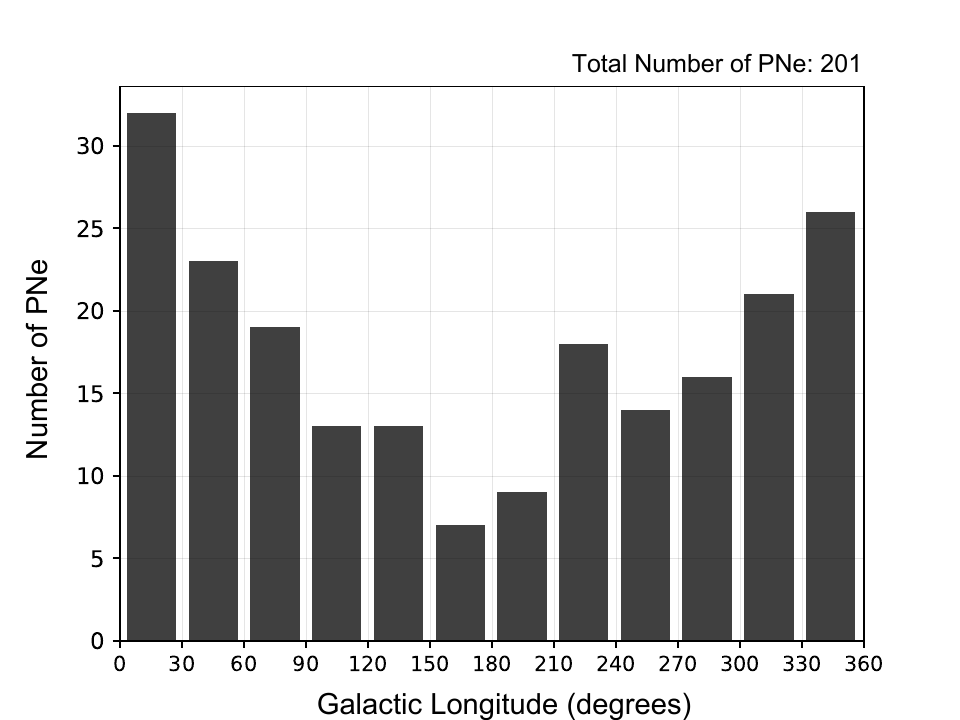}
        \caption{Distribution of PNe in galactic coordinates: latitude (upper panel) and longitude (lower panel).}
        \label{fig:lat}
\end{figure}

Fig.~\ref{fig:lat} shows the spatial distribution in galactic coordinates for GAPN planetaries. As expected, most of the PNe are located close to the galactic plane, and about $60\%$ of the objects are located at latitudes between 10 and -10 degrees. It can also be appreciated that there are more PNe closer to the galactic centre, with more than $25\%$ of GAPN located at galactic longitudes between -30 and 30 degrees. The exact coordinates and Gaia DR2 ID's of this sample objects can be seen in Table \ref{tab:astrometric} \footnote{This table is available in electronic format at the CDS via anonymous ftp to cdsarc.u-strasbg.fr (130.79.128.5)
or via http://cdsweb.u-strasbg.fr/cgi-bin/qcat?J/A+A/.}.
%You can also observe the relation between latitude and longitude positions.
\\
%\end{wrapfigure}

\begin{figure}
        \includegraphics[width=0.52\textwidth]{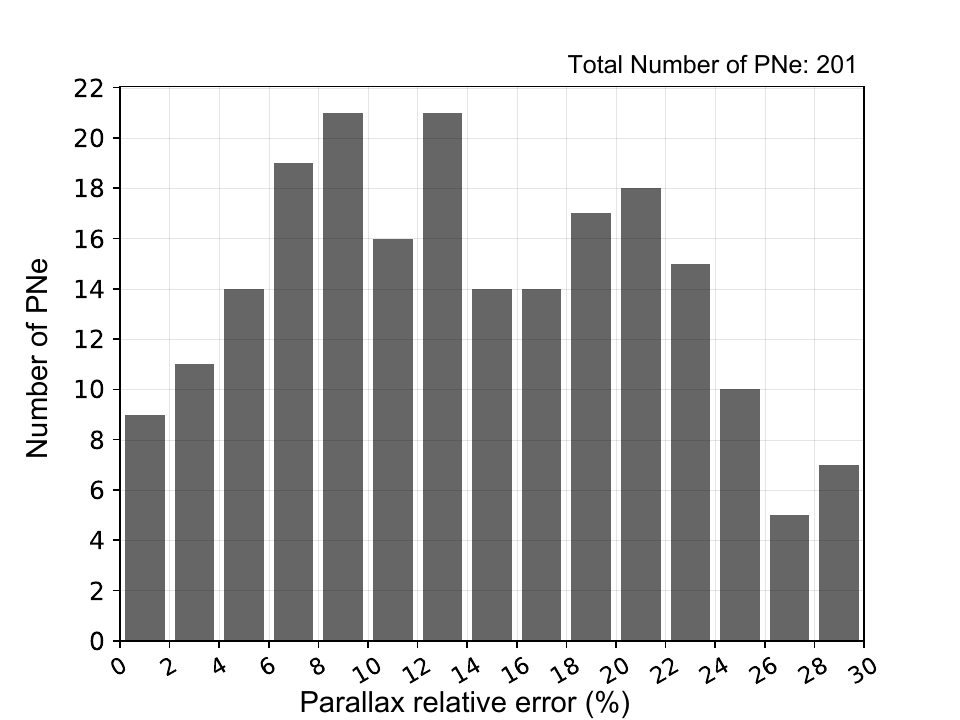}
        \caption{Distribution of parallaxes relative errors for PNe in the GAPN subsample.}
        \label{fig:parallax}
\end{figure}

%The main goal of this work is to obtain and discuss the distances distribution for our catalogue of GAPN. 
Fig.~\ref{fig:parallax} shows the distribution of parallax relative errors for our catalogue of GAPN, which are always below $30\%$ owing to our selection criteria. If we study the relationship between the parallax relative error and the brightness of the star, we do not find a simple trend; but we can conclude that for stars brighter than $G=10$, parallax relative errors are below $5\%$, while for those stars with $G$ values between 10 and 12, parallaxes tend to be bounded below $15\%$.

\begin{figure}
        \includegraphics[width=0.52\textwidth]{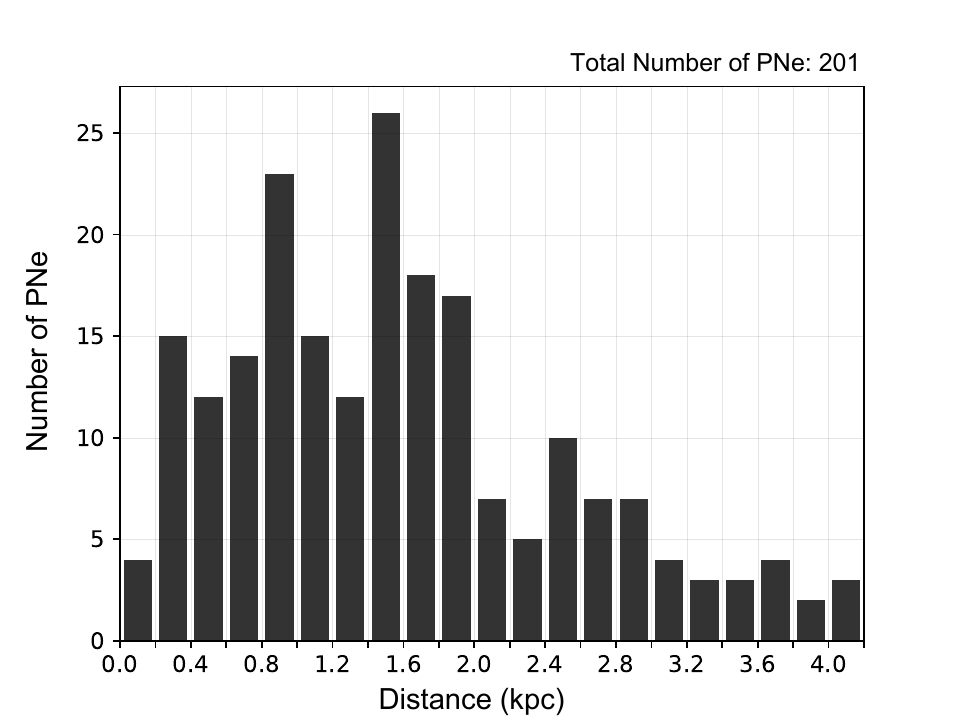}
        \includegraphics[width=0.52\textwidth]{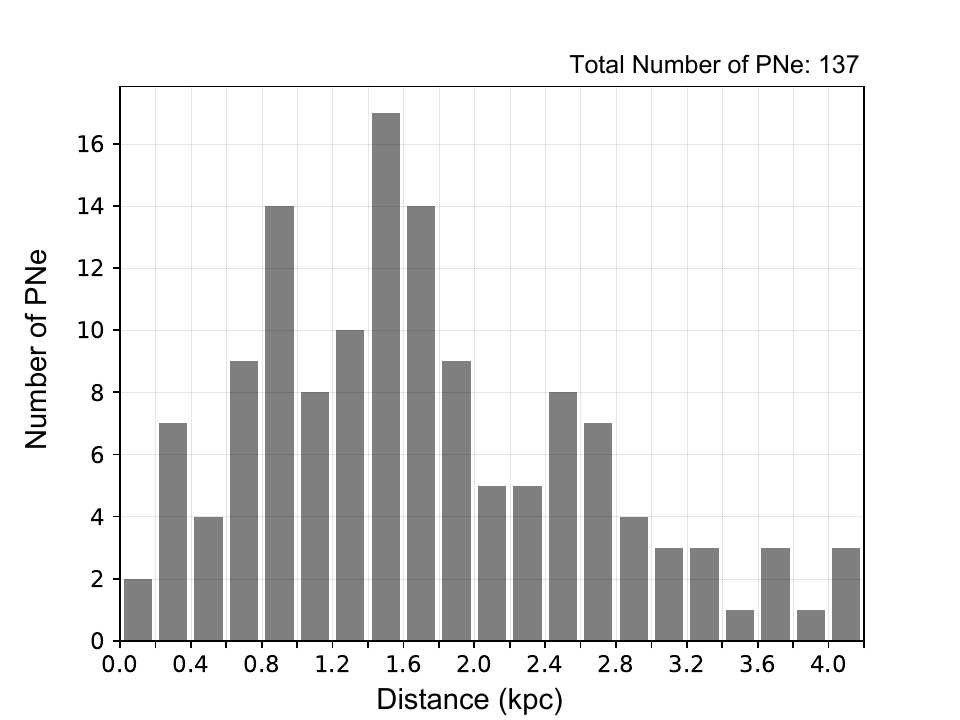}
        \caption{Histogram of distances for GAPN: for the full sample (upper panel) and only for objects in the galactic centre direction with longitudes between -90$^{\circ}$  and +90 $^{\circ}$ (lower panel). }
        \label{fig:distance}
\end{figure}

The upper panel of Fig.~\ref{fig:distance}  presents the distribution of nebulae as a function of distance inferred for our GAPN. %with the Bayesian inference procedure explained in previous sections. 
From a distance close to 2 kpc, the number of nebulae decreases. To analyse if this can be related to the completeness of our sample, % as a function of distance,
we plotted the distribution of sources in the galactic centre direction  in the lower panel of Fig.~\ref{fig:distance}, where the galactic longitudes are -90º<l<90º; there are a total of 137 nebulae. 

%Parallax measurements and their errors allows the estimation of the PN distance using the Bayesian inference probabilistic procedure indicated before. 
%Such method also estimates confidence intervals in the form of low and up boundary limits. In \textit{Figure 5} we represent the estimated distances together with these two boundaries for every object. In the other graphic it is plotted the distribution of distances. As can be seen, for distances longer than 2 Kpc, the density of PN falls down. This could mean that our catalogue is complete approximately until this value.

%One could think that this decrease in the PNe's density could be due to the fact that in the opposite direction to the galactic centre, the PNe density falls down naturally, as it is a very far area from the galactic centre.

%In order to support our idea of completeness till 2 Kpc and to demostrate that is not due to a natural decrease, we did a more accurate study. We selected only the subset of PNe that were located in the galactic centre direction (270º<longitude<90º), a total of 137 PNe. Then we plotted their estimated distances in a histogram (\textit{Figure 5}) and we compared with the results of the whole set (GAPN). 
Both distributions are similar, showing a decrease of the number of sources at distances beyond 2 kpc. From this we can infer that our selection of sources with good astrometric measurements is probably limited in completeness, approximately to such a distance. We return to the %point, regarding the 
completeness of our sample in section 7.4. In Table \ref{tab:astrometric} all numerical values about parallaxes and distances are shown together with their uncertainties.
%So we can say that this decrease is possibly due to a lack of completeness at far distances because of detection difficulties or any similar reasons, and not due to a possible natural decrease in the PNe's density in different directions of the galaxy.
%We have also analysed if there is any relation between the distance and its low or up bounds. So we have plotted distance againts these relative bounds, as it can be seen at \textit{Figure 6}. It seems that there is not a clear relation between these magnitudes, only that the lowest relative errors corresponds to lowest distances.
%\begin{figure}
%       \includegraphics[width=0.24\textwidth]{distance/Muestra_222/Distance_low_error.pdf}
%       \includegraphics[width=0.24\textwidth]{distance/Muestra_222/Distance_high_error.pdf}
%       \caption{\textit{Distances againts low and up relative bounds.}}
%       \label{fig:distance_bounds}
%\end{figure}

\subsection{Comparison with other distance determinations}

\begin{figure*}
        %\centering
        \includegraphics[width=0.52\textwidth]{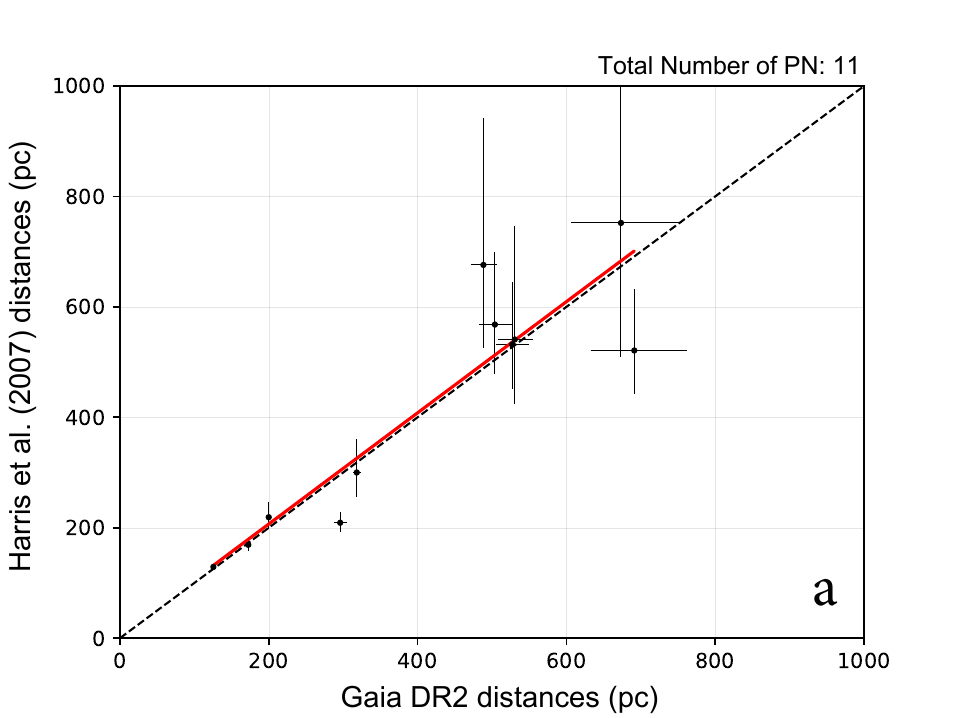}
        \includegraphics[width=0.52\textwidth]{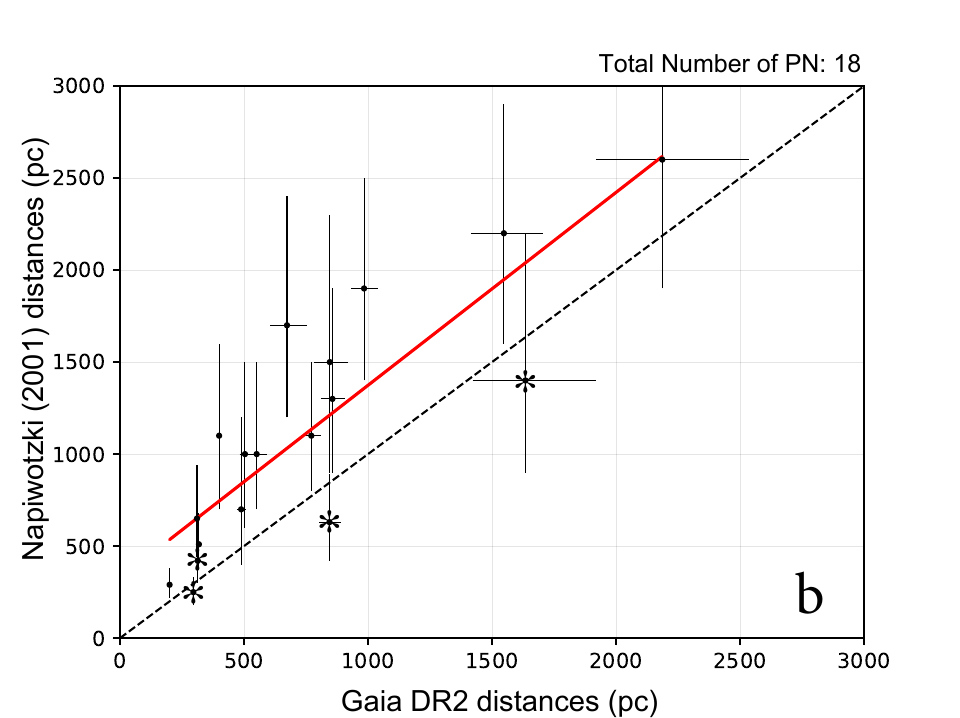}
        \includegraphics[width=0.52\textwidth]{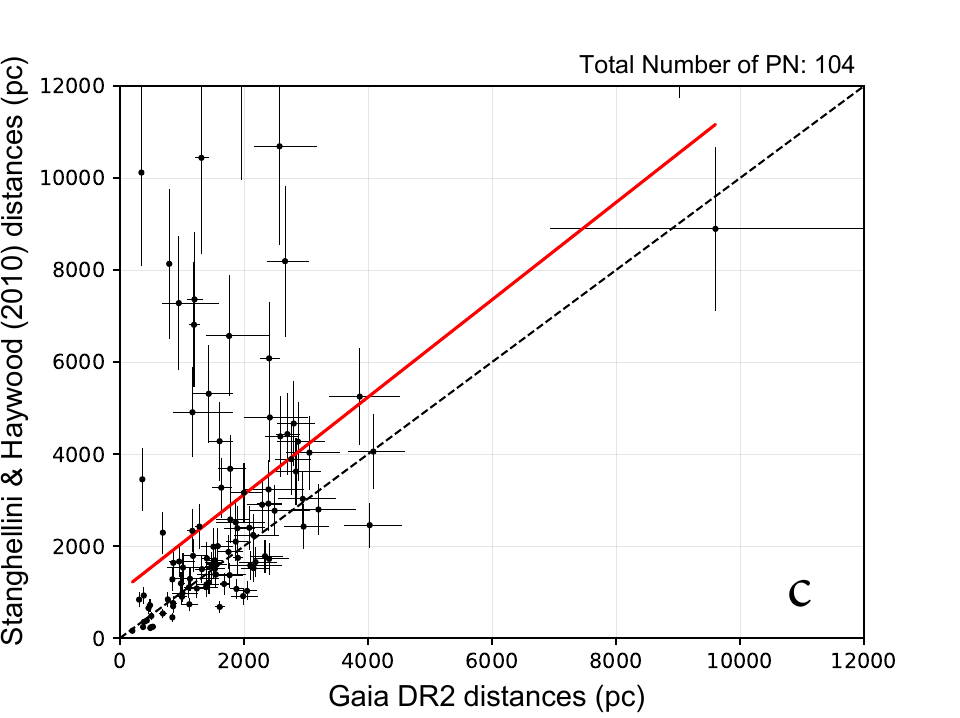}
        \includegraphics[width=0.52\textwidth]{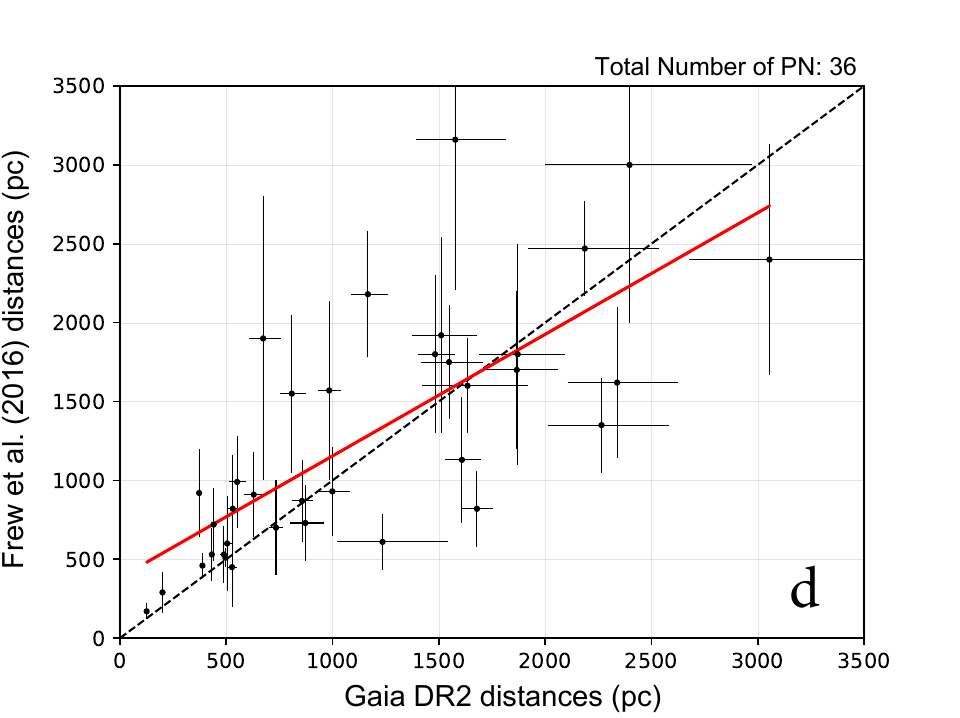}
        \includegraphics[width=0.52\textwidth]{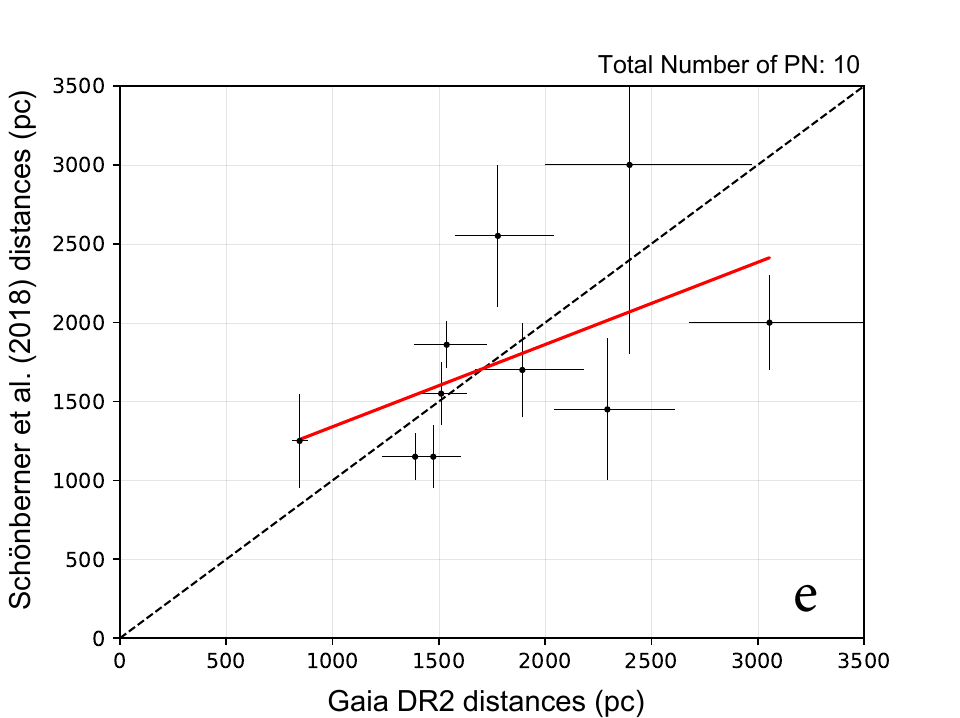}
        \caption{Comparison between DR2 and other distance derivations.}
        \label{fig:distances_comparison}
\end{figure*}

We can compare our DR2 distances %, obtained from Gaia DR2 astrometry, 
with other literature values obtained from astrometric measurements or from other (indirect) methods. In particular, we compared our estimations with those of \citet{harris07}, for astrometry; \citet{napiwotzki01}, for non-LTE model stellar atmosphere fitting; \citet{stanghellini10}, for statistical distances;   \citet{frew16}; for surface brightness versus radius; and \citet{schonberner18}, for hydrodynamical model fitting. This comparison is illustrated in the various panels of Fig.~\ref{fig:distances_comparison}, where the dotted line is the 1:1 relation and the solid line represents the linear regression between the two determinations, which help us to visualise how far the results are from each other. All the points are represented within their error bars.

A comparison with other astrometric distances, such as those in \citet{harris07}, shows a good agreement between uncertainties (upper panel of Fig.~\ref{fig:distances_comparison}). A similar result was found in \citet{Kimeswenger18} synoptic study of PNe distances in DR2. Statistical distances \citep{stanghellini10} do not agree with Gaia distances, showing  overestimated values in many cases. A linear fit to these distances leads to a bias of 1 kpc. However, panel c in Fig.~\ref{fig:distances_comparison} shows that such bias is affected  by the presence of a marginal group of objects displaying wide discrepancies with DR2.  %are two clearly separated groups of PNe, one with a reasonably good fit with Gaia distances, and  whether discrepancies can be more than an order of 3 for the other group.
A possible explanation is that those objects are bipolar or butterfly-like PNe, and such a statistical method cannot be applied to those classes of nebulae.

We now comment on non-LTE model stellar atmosphere fitting to derive distances. From the pioneering work of \citet{mendez88}, followed by \citet{kudritzki06}, and \citet{pauldrach04}, several authors have used non-LTE model stellar atmospheres to derive distances. The analysis of the stellar spectra delivers $T_{eff}$ and surface gravity $g$ values, which are then used to estimate the mass from a $T_{eff}$ vs. $log(g)$ diagram with calculated post-AGB evolutionary tracks.  So far, this method has been applied to 27 CSPN \citep{napiwotzki01}. We find that \citet{napiwotzki01} distances tend to be larger than DR2 distances (see panel b in Fig.~\ref{fig:distances_comparison}). 
%Schoenberner et al (2018i), explained this difference, because the CSPN mass is not a measured quantity.
%This mass depends on the chossen post-AGB evolutionary track. Inclusion of overshooting leads to lower the masses
%for a given luminosity. However, evolutionary track in the literature do not include overshooting.
A linear fit provides a bias around 400 pc with respect to Gaia distances. 
In view of these results we think that it is worthwhile to review current non-LTE model stellar atmospheres, applied to very hot CSPN, as described below.

It is interesting to note that \citet{napiwotzki01} discussed the existence of positive bias between his spectroscopic distances and those obtained from United States Naval Observatory (USNO) and Hubble Space Telescope (HST) parallaxes and concluded that sample truncation in distance or in parallax values can explain such discrepancies between statistical uncertainties of the order of $20\%$. His simulations imply that astrometric distances should be corrected for an undetermined quantity due to a positive bias. Positive bias is a well-known effect that appears when sample truncation is done based, for instance, on the quality of the parallaxes. Our GAPN sample distances were derived using a Bayesian procedure that allows very small or negative parallaxes to have their corresponding distances calculated between confidence intervals. Once distances were derived, we selected useful values by, among others, constraining the goodness of fit indices of Gaia astrometric measurements. Even considering some possible positive selection effect in the distances derived for our GAPN sample, astrometric distances would be overestimated and not underestimated as they are in this case.

We would like to point out that the discrepancies between astrometric and spectroscopic non-LTE distances that we are discussing are evident only for central stars (CS) with $T_{eff}$ larger than 90000 K. In Figure 7b there is a small sample of CS with distances in \citet{napiwotzki01} (star symbols) that are compatible with DR2 derivations, and they all fall in this low $T_{eff}$ regime. A plausible explanation has been indicated by D. Lennon (private communication) in the sense that non-LTE models are not using line-blanketing for metals.

 Finally, when considering the case of \citet{frew16} and \citet{schonberner18} determinations, we found no clear bias between their results and our derivations (see panels d and e of Fig.~\ref{fig:distances_comparison}). The \citet{frew16} distance scale was based on a statistically derived relation of the $H_\alpha$ surface brightness evolution with nebular radius. \citet{schonberner18} calculated the distances to 15 round-shaped PNe by measuring the expansion velocity of the nebular rim and shell edges, and by correcting the velocities of the respective shock fronts with 1D radiation-hydrodynamics simulations of nebular evolution. %These latter authors 
 The latter authors found a reasonable agreement with literature values except with those obtained with non-LTE model atmosphere fitting (spectroscopic gravity distance). 
 %spectroscopic methods. 
 They explained such differences as due to the fact that CSPN mass is not a measured quantity. The mass value depends on the chosen post-AGB evolutionary track and, for instance, the inclusion of overshooting leads to lower masses for a given luminosity. Evolutionary tracks in the literature do not include overshooting. Interestingly, \citet{schonberner18} discussed the use of updated evolutionary models for the calculation of luminosities and masses and concluded that 
 spectroscopic gravity distances are, in general, higher than those derived by other methods and can produce unreasonably high luminosities. This conflicts with the predictions from stellar evolutionary theory because central star masses are even beyond the Chandrasekhar mass limit in some cases.

\subsection{Physical radii}

The knowledge of distances allowed us to obtain the physical size of the PNe from the observed nebular angular sizes. The HASH database lists major and minor axis angular diameters for most of known PNe derived from $H_\alpha$ photometry. A typical radius for the nebulae (R) can then be obtained without taking into account the projection effects or the complexity of some of the nebular shapes, simply by considering the average angular radius in arc seconds from the $H_\alpha$ $10\%$ isophote and DR2 distances.
%$$\tan(\Phi)=\frac{R}{D}, $$ 
%where
%$\Phi$ is the angular radius (Rad),
%$R$ the radius (pc) and
%$D$ the distance (pc). 
%\\\\
%As the angle $\Phi$ is very small we can aproximate the equation to:

%$$\Phi=\frac{R}{D}.$$
%{\bf amt no estoy seguro que haga falta poner esta formula, ya que es trivial}
%$$R=\frac{2\pi}{360}\cdot\frac{1}{3600}\cdot\phi \cdot r ,$$
%where
%$\phi$ is the angular radius (in arcsec). 

\begin{figure}
        \includegraphics[width=0.52\textwidth]{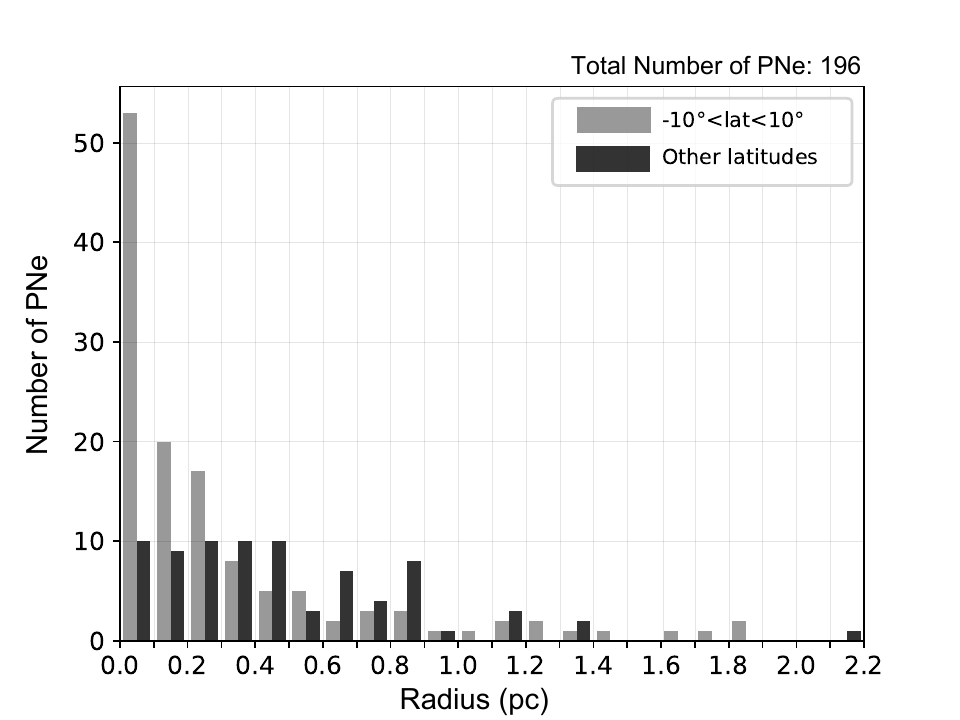}
        \caption{Planetary nebulae physical radii for objects near the galactic plane (-10$^{\circ}$ <lat<10$^{\circ}$ ) and for the rest of galactic latitudes.}
        \label{fig:radius}
\end{figure}

Fig.~\ref{fig:radius} shows the distribution of such typical radii for the case of low latitude nebulae, with latitude values between 10 and -10 degrees, as compared with the remaining objects. Without going into the details about the morphology of the nebulae, which is beyond the scope of this work, it can be noted that $70\%$ of the PNe radii are larger than the typical PNe value of 0.1 pc \citep{osterbrockFerland06}. We found that PNe close to the galactic plane represent a wide range of sizes and do not show any trend to be larger, as would be expected if they mostly evolved from high-mass progenitors with higher expansion velocities \citep{corradiSchwarz95}. Table \ref{tab:astrometric} lists the typical radius for 196 GAPN present in the HASH database. %comprobar y re-escribir y decidir qué tiene de utilidad incluir este dato de la tabla

\subsection{Radial velocities}

\begin{figure}
        \includegraphics[width=0.52\textwidth]{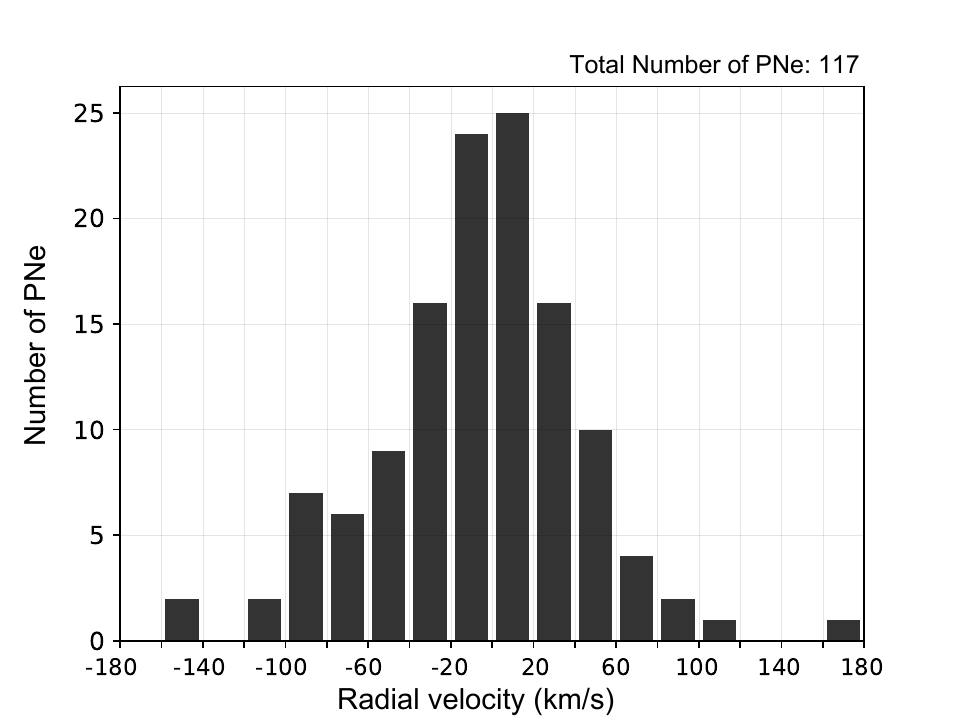}
        \includegraphics[width=0.52\textwidth]{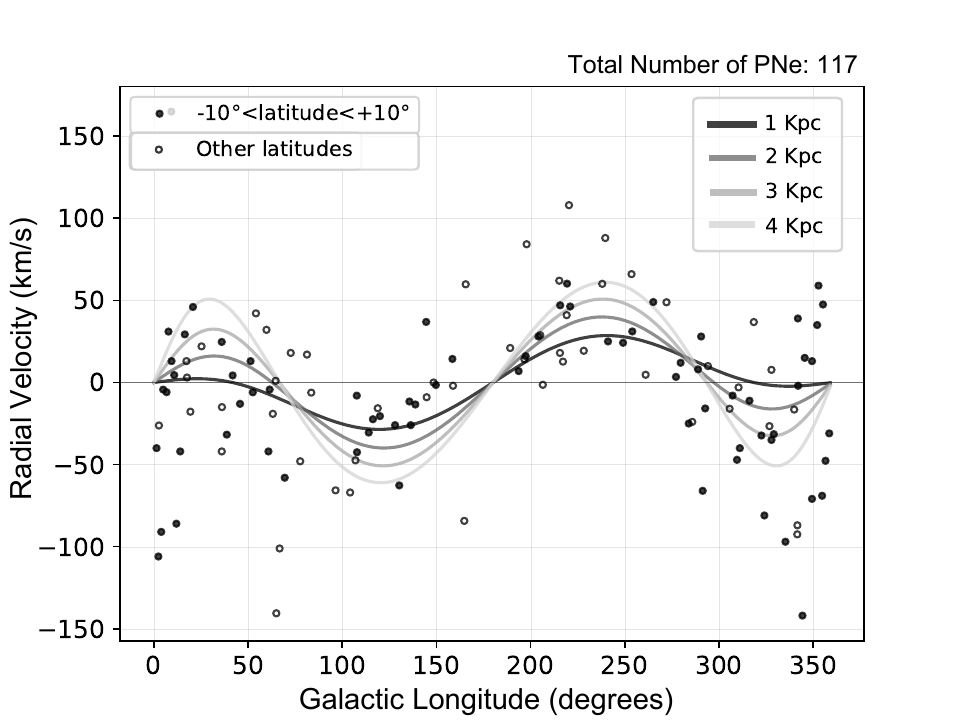}
        \caption{Distribution of systemic radial velocities (upper panel) and radial velocities as a function of galactic longitude (lower panel) for a selection of objects in the GAPN sample.}
        \label{fig:vel_radial}
\end{figure}

It is also interesting to analyse the radial velocities of our GAPN sample and to compare their values with those expected for a pure circular galactic rotation. We retrieved radial velocities from the literature for a total of 117 PNe (see Table \ref{tab:astrometric}). The upper panel of Fig.~\ref{fig:vel_radial} shows the distribution of systemic radial velocities corrected to the local standard of rest (LSR). The lower panel of the figure indicates the radial velocities as a function of the galactic longitude. Filled circles correspond to objects located near the galactic plane (latitudes lower than $\pm10$ $^{\circ}$). We found that, in general, our low latitude GAPN sample follows {\it grosso modo} the radial velocities sinusoidal curves expected for their range of distances, considering the case of pure circular rotation for a flat rotation disc at 230 $km \cdot s^{-1}$. Some nebulae display a wide velocity departure from pure rotation. These are mostly nebulae with galactic longitudes close to zero, i.e. those in the direction of the galactic centre, which are expected to show a high velocity dispersion and a departure from the rotational general trend. A detailed study is also beyond the scope of this work.

%Another point can be to visualise these radial velocities in function of the galactic longitude, as is presented in the same figure. We can apreciate like an oscillation in this distribution.

%\begin{figure}
%       \includegraphics[width=0.5\textwidth]{vel_radial/vel_radial_gal_long.pdf}
%       \caption{\textit{PN radial velocities againts galactic longitude.}}
%       \label{fig:vel_radial_gal}
%\end{figure}
\section{Radii, expansion velocities, and kinematical ages}

The fact of having, for the first time, precise parallaxes that allow us a consistent estimation of distances and physical sizes of a meaningful sample of nebulae, offers us the opportunity to calculate their ages based on nebular sizes and expansion velocities. We also briefly discuss the limitations and hypothesis under which these determinations have been carried out in the literature. %{\bf amt no me gusta esta frase, demasiado larga}

It has been common practice to derive the so-called kinematical ages as the ratio of the nebular size and expansion velocity, this latter calculated from the broadness or splitting of the most brilliant nebular lines, mainly from [OIII], [NII], and $H_\alpha$. As has been reviewed by several authors (\citet{Schonberner14} and references therein), the velocity field of a PN is a complex structure that depends on the CS mass; this velocity field does not vary not linearly with time, apart from other considerations regarding, for instance, the density structure of the nebulae. Furthermore, precise values of expansion velocities also depend on the excitation level of the emission lines used to derive them (the so-called Wilson effect). Usually, expansion velocities are measured in the inner bright rim of the nebular structure, while nebular sizes refer to the outer shell of the objects. There are both observational and theoretical evidences that velocities in the shell and in the rim are different during most of the evolution of the nebulae \citep{Villaver02, Corradi07, Jacob13}. The post-shock velocity, i.e. the flow velocity immediately behind the leading shock of the shell (or the outer edge of the shell), has also been proposed as a simpler proxy of the true nebular expansion speed \citep{schonberner05, Corradi07,Jacob13}, but this value is only available in the literature for a small sample of nebulae because it is difficult to measure.

\begin{figure}
        \includegraphics[width=0.52\textwidth]{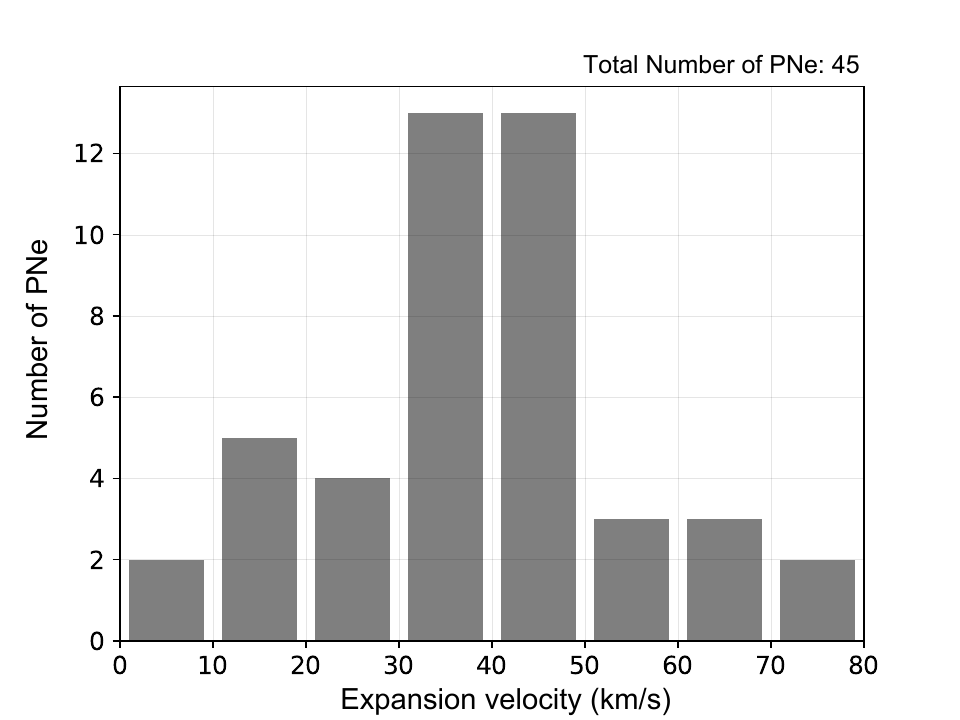}
        \includegraphics[width=0.52\textwidth]{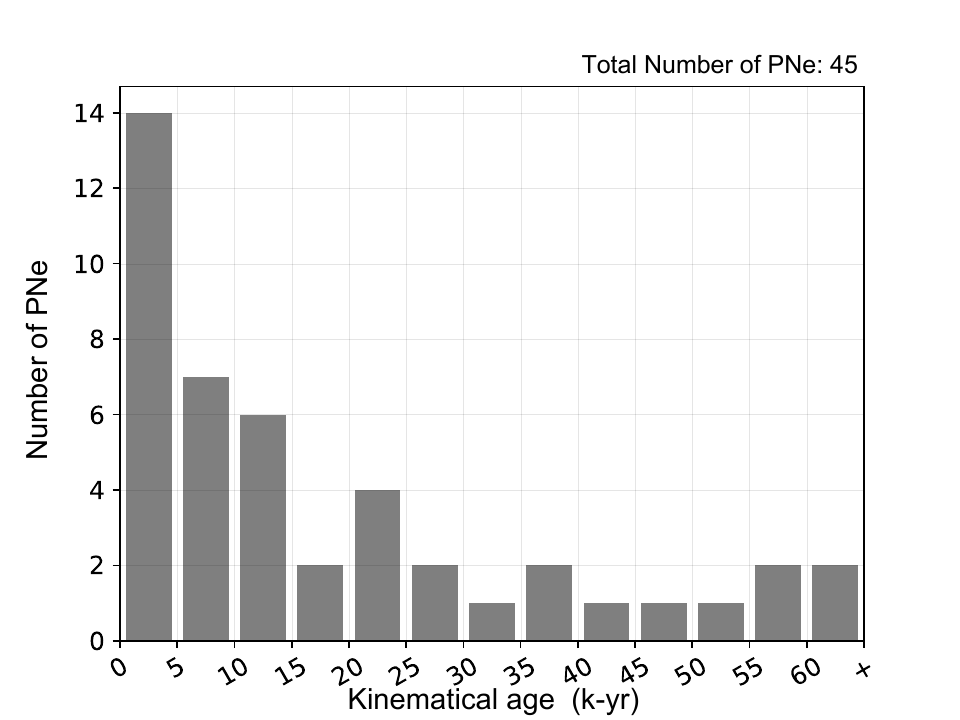}
        \caption{Distribution of expansion velocities (corrected) from the literature (upper panel) and kinematical ages derived from them (lower panel). See text for details.}
        \label{fig:vel_exp}
\end{figure}

\citet{Villaver02} simulations of the dynamical evolution of the circumstellar gas around PNe demonstrated, for the first time, that nebular shells are subject to acceleration during their evolution and that the kinematical ages, as derived from sizes and expansion velocities, can significantly depart from the CS evolutionary time. These authors found that the kinematical ages are always higher than the CS ages when the nebula is younger than 5000 yr while, for intermediate ages (between 5000 and 10000 yr), the ages derived from a dynamical analysis tend to overestimate the age of the CS for high-mass progenitors (in their simulations, 3.5 and 5 $M_\odot$) and to underestimate it for the low-mass progenitors.

\citet{schonberner05}, using 1D hydrodynamic simulations of nebulae envelopes, pointed out that rim and shell velocities are usually very different from each other, unless the nebula is rather old (ages greater than 8000 yr) and the star is approaching its maximum temperature. \citet{Jacob13} showed how hydrodynamic models of the evolution of the envelopes can be used to derive correction factors for the measured expansion velocities (both for rim and post shock velocities) that can, then, be used to derive more realistic values for the true expansion speed of the outer nebular shell.

From the information above, it is evident that while the use of hydrodynamic models and their interpretation with data is being discussed, a consistent set of nebulae data is mandatory to be able to compare observations of nebulae expansion velocities and kinematical ages derived from them with other model-dependent quantities, such as evolutionary ages or the total number of PNe in a stellar population, as derived from population synthesis models.

In Table \ref{tab:velocity} and  Table \ref{tab:photometry}, we show a compilation of 67 PNe with good DR2 distances (i.e. belonging to our GAPN sample) with consistent literature values for their $T_{eff}$, interstellar extinction values, visible magnitudes, expansion velocities, and the corresponding kinematical ages. After examining several literature PNe data sources, we decided to use data on PNe properties in the compilation by \citet{frew08}. Temperatures, in particular, are taken from \citet{frew08} or from \citet{frew16}. Absolute visible magnitudes are based on the \citet{frew08} reported magnitudes, corrected with DR2 distances, and expansion velocities correspond to the values also listed in \citet{frew08} (Table 9.4), while [NII] and post-shock velocities are from \citet{Jacob13}. Additionally, we imposed that the objects are neither known binaries nor H-deficient PNe, and that they have a nearly spherical shape ($R_{min} \geq 0.8 \cdot R_{max}$). Expansion velocities reported in \citet{frew08} were measured as a weighted average of available literature values, and no information about the specific ion or method (line broadness or line splitting) is provided by the author. For those cases where [NII] velocities are available from \citet{Jacob13}, we were able to compare them with \citet{frew08} velocities and we found a general good agreement among them (no significant bias and a mean dispersion around 5 $km \cdot s^{-1}$). We also found that post-shock velocities are always higher than [NII] velocities with a positive bias of about 15 $km \cdot s^{-1}$.

In view of the discussion above, it seems that a reasonable option is to correct the rim expansion velocities using the correction factors in \citet{Jacob13} to account for the fact that the rim velocities are lower than the overall nebular expansion velocities for most of the evolutionary time. In \citet{Jacob13}, correction factors of the order of 1.3 to 1.6 are proposed for the different kinematical scenarios for most of the lifetime of nebulae. The exact value of the correction factor depends on the mass of the CS and on its evolutionary stage and, also, on hydrodynamical modelling. Taking into account the rather high uncertainties in the velocity data, we decided to consider an overall value of 1.5 for the correction factor for rim velocities to derive the corresponding kinematical ages. This correction value is similar to that adopted by \citet{gesicki14},  who used the \citet{perinotto04} hydrodynamical models to calibrate the relation between the average expansion velocity, radius, and age. %Both values are presented in Table XX.

Considering all this information, we were able to select a sample of 45 nebulae with reliable expansion velocities, whose distribution is shown in Fig.~\ref{fig:vel_exp} (upper panel). It can be observed that most of these have expansion velocities between 30 and 50 $km \cdot s^{-1}$.

In addition, we can estimate an average expansion velocity and its typical deviation, so that we obtain

$$<V_{exp}> = (38 \pm 16) \thinspace km \cdot s^{-1}.$$

This estimation is close to the value of $42 \pm 10$ $km \cdot s^{-1}$ given by \citep{Jacob13}.
Once we know the nebular expansion velocity, it is possible to estimate the so-called kinematical age for each PN by a simple relation, such as 

$$T_{age}=\frac{R}{V_{exp}}.$$

%which, in $years$ units, can be expressed as: 

%$$T_{age}=\frac{3.086\cdot10^{13}}{60\cdot60\cdot24\cdot365}\cdot\frac{R}{V_{exp}}.$$

In Fig.~\ref{fig:vel_exp}, we show the distribution of the kinematical ages that we found for our sample of PNe. Although most of the PNe are rather young, with ages under 15000 yrs, we also found nebulae spanning ages well beyond those values.

%Here comes a discussion on kinematical ages, mean values and dispersion, comparison with the mean value of about 22.000 yrs derived by \citet{Jacob13}.

Now, we can also estimate an average value for the kinematical age of the sample, which is known as visibility time ($<T_{VT}>$) of a PNe population \citep{Jacob13}. This can be derived from the average expansion velocity and average radius as follows: 

$$<R> = 0.633 pc.$$

Then, the visibility time can be calculated as %average kinematical age can be expressed as:

$$<T_{VT}> = \frac{<R>}{<V_{exp}>}= 23400 \pm 6800  \thinspace yrs.$$% = 23439\thinspace years,$$

%where $K=\frac{3.086\cdot10^{13}}{60\cdot60\cdot24\cdot365}$.% Its corresponding error can be calculated as:
%$$\Delta T_{kin} = K \cdot <R> \cdot \abs{\frac{-1}{<V_{exp}>^{2}}} \cdot \Delta V_{exp} = 6756 \thinspace years.$$

%We obtained a value of: $$<T_{kin}> = (23400 \pm 6800) \thinspace years.$$ 
This visibility time is very similar to the value of $(21000 \pm 5000)$ yrs given by \citet{Jacob13}.
We should stress the limitations of this derivation. 
Firstly, simply because our statistics is rather poor, and secondly, because we probably have a bias with age, with more young PNe than old because PNe tend to dim as they get older. To study such trend % {bf amt in section 7} 
we analysed separately the ages of those PNe that are located closer than 1 Kpc versus those located farther away. As expected, we found that the distribution of ages for the nearby sample is rather homogeneous while, for the second sample, nebulae tended to be younger. 

\section{Temperatures and luminosities of the central stars}

Reliable distance determinations obtained from Gaia astrometry allow us to consider the exercise of placing these PNe in a HR diagram and to analyse if it is possible to obtain some useful information about their evolutionary status, by comparing the distribution of objects with the positions predicted by the most recent evolutionary models \citep{millerbertolami17}. Furthermore, we can compare the evolutionary ages with kinematical ages obtained in the previous section. To carry out this analysis it is necessary to compile
reliable information on the central star effective temperatures ($T_{eff}$) and the necessary information to calculate the luminosities of the objects. In particular, we need visible magnitudes, distances, extinctions, and bolometric corrections. 

\subsection{Effective temperature}

Central stars $T_{eff}$ are often estimated with the Zanstra method \citep{zanstra28} by measuring H I and He II nebular fluxes, or at least one of them, and taking into account the value of the V magnitude of the star and the nebular extinction. %This method can provide only a lower limit for $T_{eff}$ depending on the detection of He I or the availability of measurements.
%When available, we adopted Zanstra temperatures derived from He II (4686A), and otherwise Zanstra temperatures derived from H I. % We have searched the literature for recent compilations of PNe CS temperatures, and we decided to use those listed in the works by \citet{frew16} and \citet{morenoIbanez16}.
%Due to the very rapid post-AGB evolution, to the successive slowing down in the cooling phase and to the different timescales depending on mass, higher mass cores are expected to be found on average at higher temperatures, but luminosities will be rather similar.
%In this chapter we are going to analyse and present different properties of central stars of the PNe, with the aim to locate as much of them at the Hertzsprung-Russel diagram. So firstly it is necessary to collect information about their temperatures and photometry. This is not an easy task, because sometimes the star is quite faint and it is difficult to separate its ligth from the nebular light.
%Concerning to temperature, there are different methods to estimate it: fitting a blackbody to stellar continuum, with Zanstra temperatures method (hydrogen and helium), with Stoy temperatures method, derived from nebular ionization equilibrium...

\begin{figure}
        \includegraphics[width=0.52\textwidth]{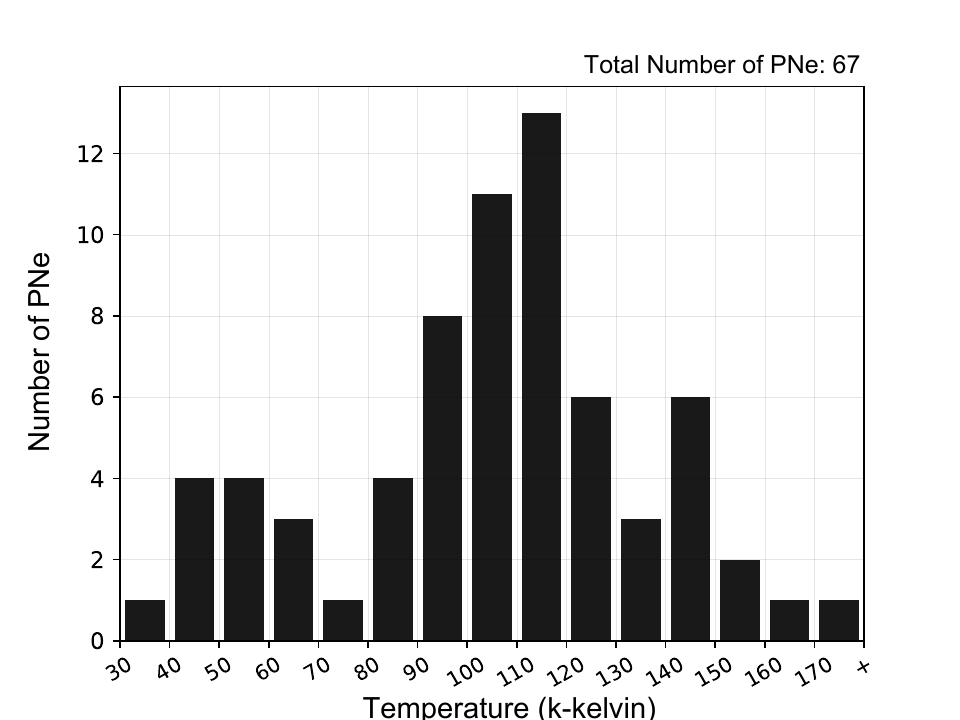}
        \caption{Histogram of CSPN effective temperatures for objects in GAPN with data in \citet{frew08}.}
        \label{fig:temperature}
\end{figure}

\citet{frew08} listed $T_{eff}$ estimations based on different bibliographic sources, focussing on the values obtained from Helium Zanstra method. After examining several literature compilations with PNe data on $T_{eff}$, visual magnitudes, and interstellar reddening derivations, we decided for consistency to 
%also 
centre our analysis on the sample of GAPN in common with the \citet{frew08} compilation. We selected objects with precise values rather than bounded values.  Also, we restricted our selection to those objects that are neither known binaries nor H-deficient PNe. Fig.~\ref{fig:temperature} presents PNe temperatures with their errors for 67 PNe in common with our GAPN sample. Most of these have $T_{eff}$ between 90000 and 120000 K and a bias towards high temperatures because the stars with $T_{eff}$ < 45000 K do not produce He twice ionised \citep{kaler91}.

%\begin{figure}
        %\begin{wrapfigure}{r}{0.25\textwidth}
%       \includegraphics[width=0.45\textwidth]{magnitude/hist_g_mag.pdf}
%       \includegraphics[width=0.45\textwidth]{magnitude/hist_bp_rp.pdf}
%       \caption{Gaia $G$ band magnitudes for objects in the GAPN sample (up) and $G_{BP}-G_{RP}$, Gaia colours.}
%       \label{fig:mag_1}
%\end{figure}

\subsection{Brightness, extinction, and luminosity}

We now focus on the brightness and luminosities of our sample of CSPN. Gaia G-band magnitudes, $G_{BP}-G_{RP}$ colour, and $V$ values (taken from \citet{frew08}) are shown in Table \ref{tab:photometry}.
% and other magnitudes in Johnson filters like $U, B, V, R$ and $I$ (provided by Simbad or another bibliography). 
%As it is shown in %Fig.~\ref{fig:mag_1}, 
Most of the stars have $G$ magnitude values between 12 and 18 , a distribution that peaked at 15.5 magnitudes, and negative $G_{BP}-G_{RP}$ colour values, as corresponds to their high temperatures.
%Concerning $V$ apparent magnitude, we have information about 157 stars, several of them with their error too. 
%On the other hand,
In addition, $V$ values range between 10 and 20 magnitudes and peak around 16.5 magnitudes. We compared the $G$ and $V$ magnitudes object by object to further check for data consistency in our sample. Table \ref{tab:photometry} also lists \citet{frew08} extinction values for the sample of 67 PNe selected from our GAPN. The great majority of stars have very low extinction in the range  0-0.2 magnitudes. 

\begin{figure}
        \includegraphics[width=0.52\textwidth]{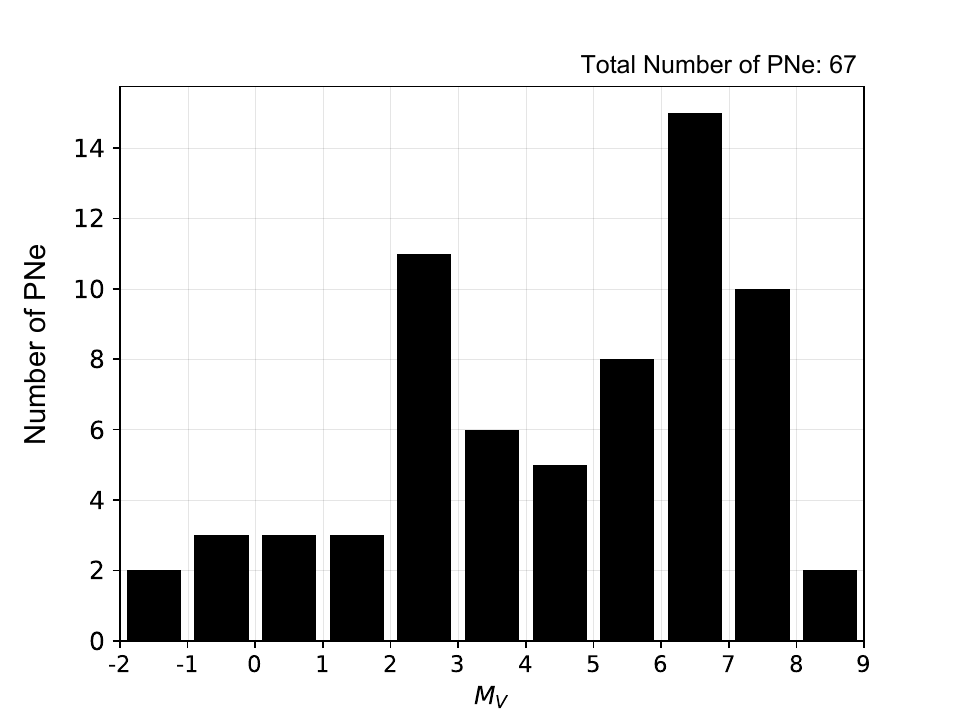}
        \caption{Absolute visible magnitude for stars in GAPN in common with \citet{frew08}.}
        \label{fig:V_Abs}
\end{figure}

Absolute visible magnitudes can then be derived taking into account our DR2 distances and extinctions from \citet{frew08} (Fig.~\ref{fig:V_Abs}). These magnitudes span values between 2 and 8 for most of the stars, which is the expected range during the evolutionary stage of  PNe.

Absolute bolometric magnitudes, $M_{B}$, were derived using the calibration procedure published by \citet{vacca96}, i.e.

$$M_{Bol} = M_{V} + BC,$$ where

$$BC = 27.66 -6.84 \cdot \log (T_{eff}).$$

\begin{figure}
        \includegraphics[width=0.52\textwidth]{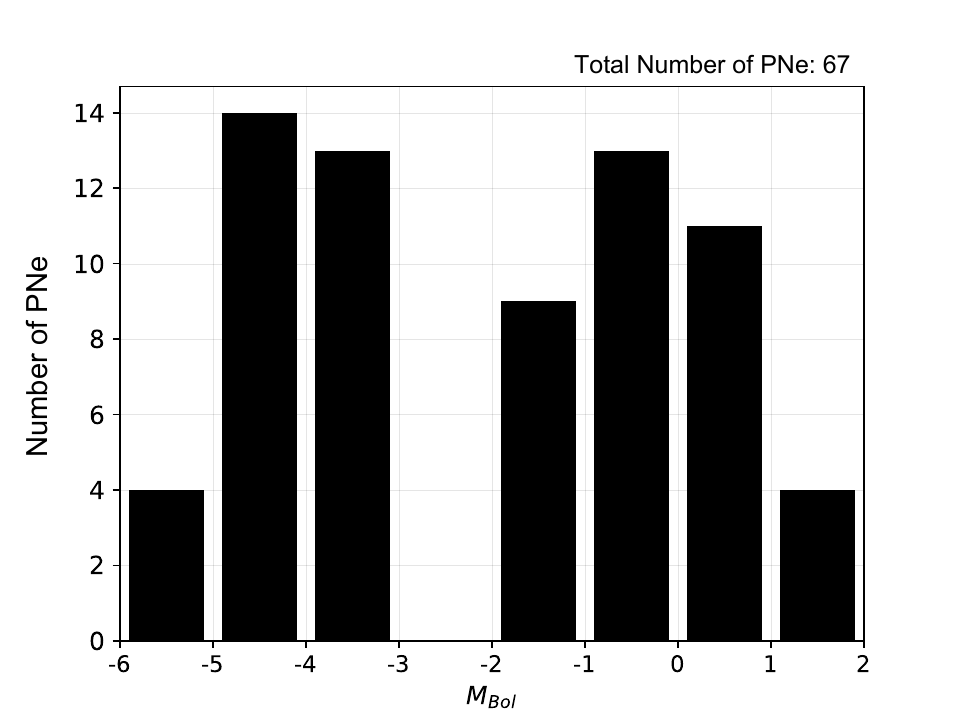}
        \caption{Absolute bolometric magnitude for stars in GAPN in common with \citet{frew08}.}
        \label{fig:Bol_Abs}
\end{figure}

\begin{figure}
        \includegraphics[width=0.52\textwidth]{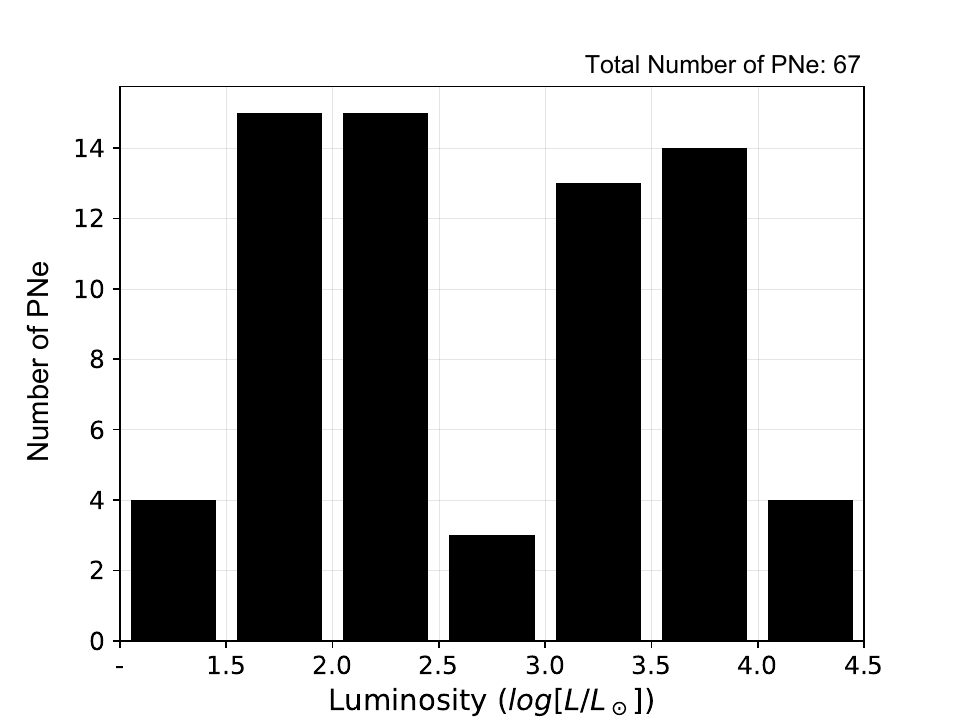}
        \caption{Luminosity for stars in GAPN in common with \citet{frew08}.}
        \label{fig:luminosity}
\end{figure}

These bolometric corrections were calculated for O and early B spectral types assuming a maximum $T_{eff}$ = 50000 K. However, this relation depends only weakly on the surface gravity of the star, so we assumed that it is correct for higher temperatures. From bolometric magnitudes we derived the luminosities. Both quantities are shown in Fig.~\ref{fig:Bol_Abs} and Fig.~\ref{fig:luminosity}.

%where $T_{eff}$ is the effective temperature of the star, whose values were presented before.
%\\\\
%With all this data, we can estimate the $M_{Bol}$ and its uncertainty for an important subset of stars from our catalogue (78). As we know:

%$$M_{Bol} = M_{V} + BC,$$
%\\
%so,

%$$M_{Bol} = m_{v} +5 -5\cdot\log(d) - A(v) + 27.66 -6.84 \cdot \log(T_{eff}),$$
%\\
%and the corresponding uncertainty can be expressed by:

%$$·M_{Bol} = ·m_{v} + \dfrac{5}{\ln(10)} \cdot \dfrac{·d}{d} + \dfrac{6.84}{\ln(10)} \cdot \dfrac{·T_{eff}}{T_{eff}}.$$

%both for low and high bounds of the distance. The obtained results are shown in Fig.~\ref{fig:Bol_Abs}. %It can be observed that this magnitude is much brighter than the visual one (as it brings all the wavelengths of the light), for most of these stars this magnitude moves from -8 to 2.
%\\

%Finally, luminosities were calculated as:

%$$ M_{Bol} -  M_{Bol_{\odot}}= -2.5 \cdot \log(\frac{L}{L_{\odot}})$$

%where
%$M_{Bol_{\odot}}$ is the Sun absolute bolometric magnitude $\equiv 4.75$ and
%$L_{\odot}$ is the Sun Luminosity $\equiv 3.828\cdot10^{26}$ W.
%\\

%This parameter is usually expressed in logarithmic form: %of the number of solar luminosities. Then we can express like: 

%$$\log(\frac{L}{L_{\odot}}) = \dfrac{M_{Bol_{\odot}}-M_{Bol}}{2.5},$$

%as it is shown in Fig.~\ref{fig:luminosity} for our stars. The same way as for $M_{Bol}$, we can estimate the uncertainty for $\log(\frac{L}{L_{\odot}})$. In this case, it only depends on $M_{Bol}$ uncertainties (low and high bounds):

%$$·\log(\frac{L}{L_{\odot}}) = \dfrac{1}{2.5} \cdot ·M_{Bol}$$

%As it can be observed, most of the obtained luminosity values ($\log(\frac{L}{L_{\odot}})$) range between 1.5 and 4.

\subsection{Location in the HR diagram}

Once the luminosities are derived, the stars can be plotted on a HR diagram to compare their distribution with the prediction of evolutionary models for post-AGB stars. We decided to make a comparison with the new evolution tracks by \citet{millerbertolami17} because they include updated opacity values, both for the low and high temperature regimes and for both the C- and O-rich AGB stars. These models also included conductive opacities and nuclear reaction rates that have been updated, and a consistent treatment of the stellar winds for the C- and O-rich regimes. The author explains that the new models reproduce several AGB and post-AGB observables that were not reproduced by the older grids (see \citet{millerbertolami17} for details). From these models, post-AGB timescales are approximately three to ten times shorter than those of old post-AGB stellar evolution models, and luminosities are about $0.1-0.3$ dex brighter than from previous models with similar remnant masses.

\begin{figure*}
\centering
   \includegraphics[width=17cm]{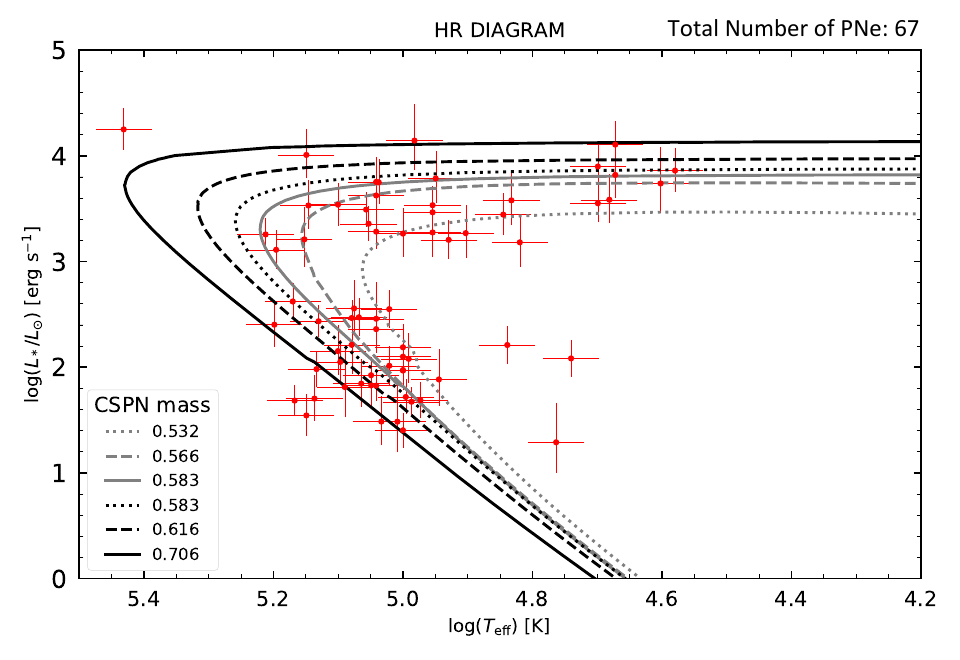}
     \caption{Hertzsprung-Russell diagram for a selection of GAPN stars, together with \citet{millerbertolami17} evolutionary tracks.}
     \label{fig:HR_diagram}
\end{figure*}

\begin{figure}
        \includegraphics[width=0.52\textwidth]{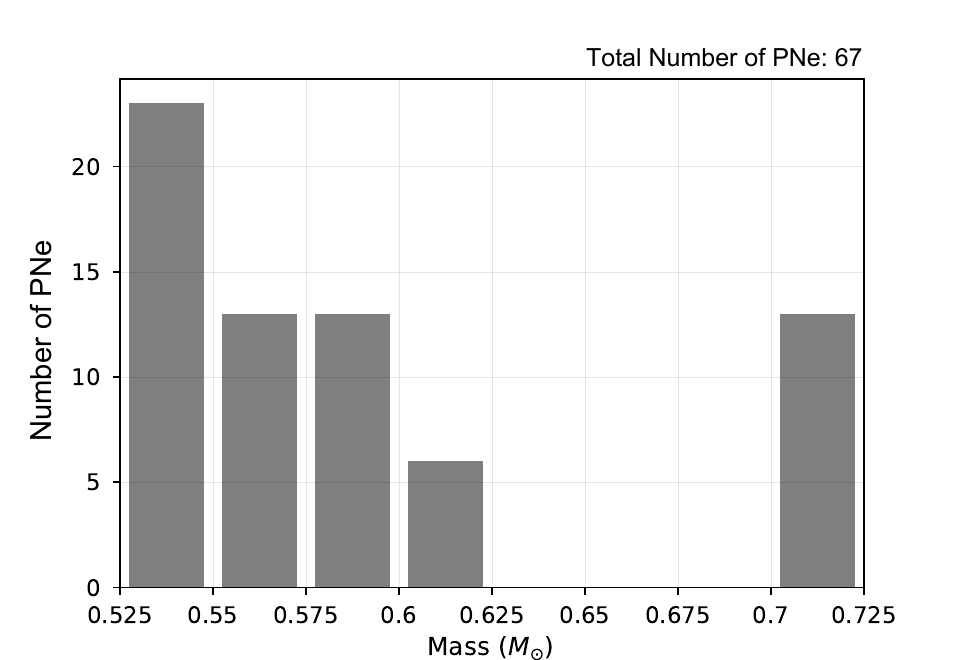}
        \includegraphics[width=0.52\textwidth]{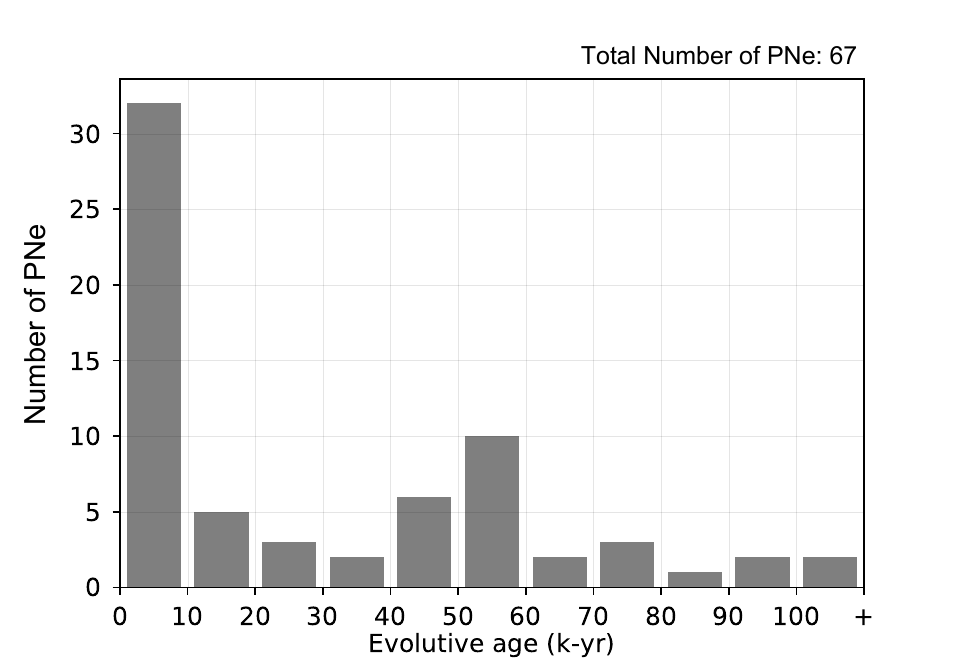}
        \caption{Estimated masses and evolutionary ages derived from \citet{millerbertolami17} evolutionary tracks.}
        \label{fig:mass_age}
\end{figure}

Fig.~\ref{fig:HR_diagram} shows such HR diagram together with the evolutionary tracks for a wide range of masses.
Considering the location on the HR diagram of the stars, we interpolated masses and evolutionary ages, as shown in Fig.~\ref{fig:mass_age}. Numerical values can be consulted in Table \ref{tab:photometry}. Most of the stars have masses between 0.525 and 0.625 $M_{\odot}$.% and evolutionary ages below $10^{4}$ yrs, confirming that, most of them, are in an early stage of their evolution as PNe.

%\begin{figure}
        %\includegraphics[width=0.5\textwidth]{HR_diagram/trazas_evolutivas_frew.pdf}
        %\caption{Hertzsprung-Russel diagram for a selection of GAPN stars, together with \citet{millerBertolami17} evolutionary tracks.}
        %\label{fig:HR_diagram}
%\end{figure}

For a more detailed study, we divided the HR diagram into three regions. The first region corresponds to the stars in a very early stage, while they are increasing their temperature (till $\log({T_{eff}})=4.8$) at a rather constant luminosity value and fulfilling  $\log(\frac{L}{L_{\odot}}) > 3.0$. The second region corresponds to the same flat luminosity part of the HR diagram, but for higher temperatures, from $\log({T_{eff}})=4.8$ until the maximum $T_{eff}$ value. Finally, the third region covers the evolution of objects that have reached their maximum temperatures as PNe and are decreasing in luminosity ($\log(\frac{L}{L_{\odot}}) < 3.0$) on their way to becoming a WD star. We now analyse our Gaia DR2 derived quantities that, in the case of the CSPN, are updated luminosities and, for the nebulae, are the physical radii; this physical quantity increases its value with evolution as the nebulae expand. We aim to see if the new models allow us to draw a consistent picture of the evolutionary stage of the objects. We limit this analysis to a subsample of PNe with expansion velocities obtained from the literature (55 PNe out of 67 PNe, see Table \ref{tab:velocity}). 

In Table \ref{tab:HRD} we present the mean values we obtained for masses, physical radii, and evolutionary ages in each of the three HR diagram regions. The masses mean values are similar in the three regions. Because evolutionary times in the early stages are very short, high-mass objects tend to pile up in the later stage.  Regarding radii and evolutionary ages, as expected, there is a clear increase per region of the mean values in both parameters. We find values for the mean radius of 0.093 pc, 0.298 pc, and 0.804 pc, respectively. The mean values of evolutionary ages in each of the regions are 14.2 kyr, 20.5 kyr, and 33.8 kyr, respectively. Such mean ages show a high dispersion because in each of the regions there are objects with all possible mass values and evolutionary times depend strongly on mass.  Following \citet{millerbertolami17}, to compile such ages, we added a transition time from an early post-AGB stage until  $T_{eff}=7000 K$ (or $\log(T_{eff})$=3.85). Such transition times are about 1 kyr for the highest mass CSPN and 2 kyr for the remaining masses.

%For the high luminosity early phase we found a mean value of 16.7 K-years for the evolutionary age, and a mean radius of 0.24 pc. Such values are clearly lower than the ones obtained in the second region, 33.4 K-years and 0.81 pc, but we found a generally quite high dispersion for those mean values, because we are dealing with stars with different masses that evolve at significantly different velocities .

% It is a logical result as they are located in an earlier stage. While the corresponding expansion velocities mean value (obtained from the radii mean value and evolutionary ages mean value) is quite similar to the expansion velotities given by \citet{frew08} mean value  in the second region, but seems not to have similarity in the first region.

%Focusing on the second stage stars, we have further analysed luminosities and radii by considering massive ($M > 0.55 M_{\odot}$) and less massive ($M < 0.55 M_{\odot}$) stars. As expected, massive stars present a larger mean value for the radii (0.91 pc) and a lower mean evolutionary age (13.0 K-years) than those of the second group (0.72 pc and 49.2 K-years), with a very high dispersion for the evolutionary ages. 

%This can be explained by the fact that although the most massive objects evolve faster than the other ones, they use to present higher expansion velocities than PNe with small mass CS.

Secondly, considering evolutionary ages and radii, we can estimate the mean expansion velocities spanned by the nebulae and compare those values with those reported in the literature and discussed in section 5. The only region where we found consistent values (in mean) is the latest evolutionary phase, when the stars are very evolved objects already cooling towards the white dwarf stage. Noticeably, expansion velocities for our sample of 31 objects in common with \citet{frew08} in this region have a mean value of around 24 $km \cdot s^{-1}$ (without correction factor), which coincides with the mean expansion velocity that we can derive from evolutionary ages and radii. In any case it has to be pointed out that the dispersion of the values of the last quantity is very high. In the other two regions, corresponding to an early evolutionary phase, expansion velocities mean values are very low in comparison with \citet{frew08} observational velocities. It should be taken into account that evolutionary ages depend strongly on the value of $T_{eff}$, and they are very different depending on the mass value adopted for the CS. Estimations of mean expansion velocities in the early evolutionary stages are subjected to a larger uncertainty than those in a more advanced stage. 

It is important to stress again that, in general, we find quite a high dispersion for the mean values of radii, ages and, consequently, evolutionary expansion velocities, because we are dealing with stars with different masses, i.e. stars that evolve at significantly different velocities. Despite this, we found that the position in the HR diagram of the stars provides valuable information about the PN evolutionary state, and that the expansion and size of the envelopes agree in general terms with the evolutionary state of the CS. Mean values of all these parameters and their typical deviations are presented in Table \ref{tab:HRD}. In this study, we do not discard objects according to their geometric shape to obtain their expansion velocity, and therefore we have a selection of 54 PNe (and not 45 as in section 5).
%Note that the sample is composed by 54 and not 55 planetaries because there is one object which falls out of the three regions considered in the HR diagram.

\begin{table}
\caption{Mean values of different parameters in three regions of the HR diagram, together with their dispersion values, in brackets.}  
\label{table:HRD}      
\centering  

\begin{tabular}{l| l l l}          
\hline\hline                       
\textbf{Parameter} & \textbf{Region 1} & \textbf{Region 2} & \textbf{Region 3} \\    
\hline 
    Number of CSs  & 7 & 16 & 31 \\ 
    \hline 
    <M> ($M_{\odot}$) & 0.583 \scriptsize{(0.054)} & 0.587  \scriptsize{(0.061)} & 0.605 \scriptsize{(0.064)} \\
    \hline 
    <R> (pc) & 0.093 \scriptsize{(0.049)} & 0.298  \scriptsize{(0.202)} & 0.804 \scriptsize{(0.518)}\\
    \hline 
    <$T_{evo}$> (Kyr) & 14.2 \scriptsize{(18.3)} & 20.5 \scriptsize{(23.6)}       & 33.8 \scriptsize{(33.3)}\\
    \hline 
    <$V_{exp}^{mod}$> ($km \cdot s^{-1}$) & 6.4 \scriptsize{(3.4)} & 14.3 \scriptsize{(9.6)}     & 23.2 \scriptsize{(15.0)}\\
    \hline 
    <$V_{exp}^{obs}$> ($km \cdot s^{-1}$) & 20.3  \scriptsize{(14.1)} & 27.4  \scriptsize{(5.9)} & 24.6  \scriptsize{(10.0)}\\
\hline 

\label{tab:HRD}

\end{tabular}

\tablefoot{Parameter $<V_{exp}^{mod}>: $ mean expansion velocity from evolutionary age and nebular size;   $<V_{exp}^{obs}>: $ mean expansion velocity from emission lines observations.  
Region 1: $ \log(\frac{L}{L_{\odot}})$ > 3.0 \& $\log({T_{eff}})$ < 4.8; 
Region 2: $\log(\frac{L}{L_{\odot}})$ > 3.0 \& $\log({T_{eff}})$ > 4.8;
Region 3: $\log(\frac{L}{L_{\odot}})$ < 3.0 \& $\log({T_{eff}})$ > 4.9.
}

\end{table}

%As we can see most of these stars falls inside the expected area, while some of them are ubicated a bit far from the expectations. This could happen for different reasons, mainly for a lack of precision in temperature values, in extinction constants or in distance estimations. The corresponding masses for those stars within the evolutionary traces, have values between 0.532 and 0.706 solar masses.

It is interesting to note that three stars in Fig.~\ref{fig:HR_diagram} appear to be located out of the Miller-Bertolami evolutionary tracks, lying to the lower $T_{eff}$ zone. The accompanying PNe are \object{PN We 1-10}, \object{PN K 2-2}, and \object{PN M 2-55}. We searched in detail the available literature about these sources and came to the conclusion that the most plausible hypothesis about their atypical location in the HR diagram is that they are born-again PNe \citep{Herwig99}, two of which (\object{PN We 1-10} and \object{PN K 2-2}) are very large nebulae that have rather low expansion velocities and correspondingly, large kinematical ages. The CS $T_{eff}$ and luminosity values for the three of these fit well with He-burning evolutionary tracks (see for instance \citet{Iben84}). More work, however, is needed to confirm this explanation.

\section{Properties of PNe population in the Galaxy}

The total number of PNe populating our Galaxy is an intrinsically interesting value that can be used to study the underlying population from which they derive. For instance, evolutionary times and progenitor masses can be used to constrain the SFH for the range of ages covered by the PNe.
Using the information regarding 3D positions of the PNe obtained from Gaia DR2, for the complete sample of 1571 PNe, we can estimate the total population of PNe in the Milky Way. 

\subsection{Density}

Firstly, we need to calculate the density of PNe in our neighbourhood and, as a first approximation, we can assume that such value can be extrapolated
to the whole Galaxy. Following the procedure by \citet{frew08}, we calculated the number of stars inside a cylindrical volume around the Sun. We considered a radius of R=2 kpc,  which is close enough to claim for completeness and far enough to have a considerable number of PNe for statistical significance (see Fig.~\ref{fig:distance}). %{\bf amt aqui podriamos hacer referencia a la seccion 4.1 o la Figura 6 (upper), In Figure 6, can be appreciated that the number of PNe further away from 2 kpc drops significantly.} 
We calculated the number of PNe inside a cylinder with radius r fulfilling $r=D\cdot cos(\phi)< 2$ kpc (without height restrictions), where $D$ is the distance and $\phi$ is the latitude in radians, obtaining a total of 374 PNe. 

%On the other hand, to determine the \textit{scale height} ($H_{z}$), firstly, 

\begin{figure}
        \includegraphics[width=0.52\textwidth]{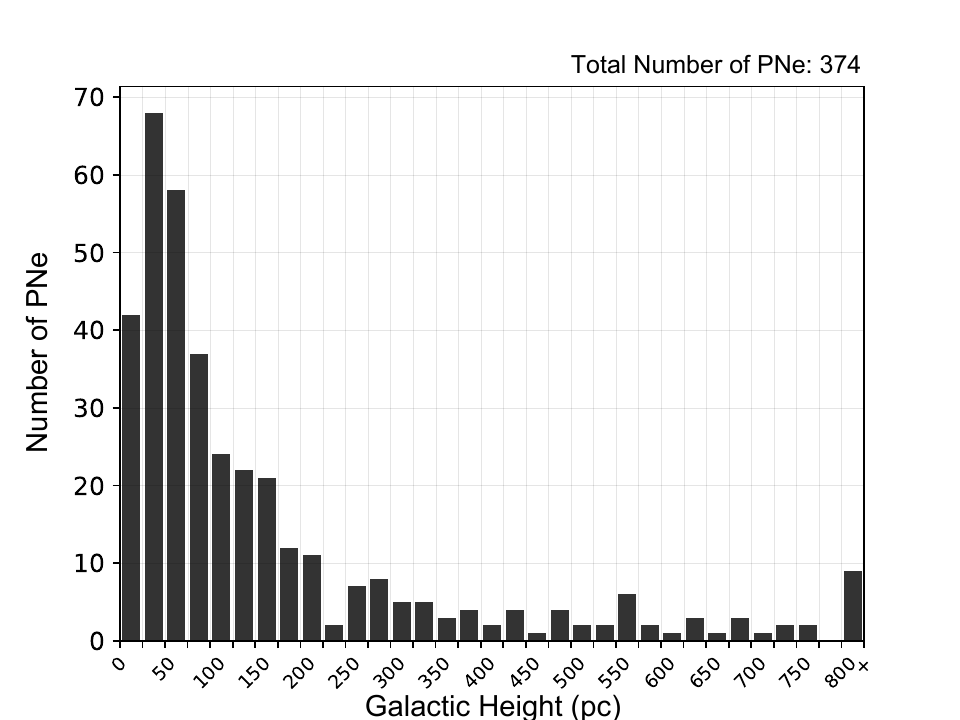}
        \caption{Galactic heights (absolute value) distribution for those PNe inside a cylinder with radius of 2 kpc.}
        \label{fig:galactic_heights}
\end{figure}

\begin{figure}
        \includegraphics[width=0.52\textwidth]{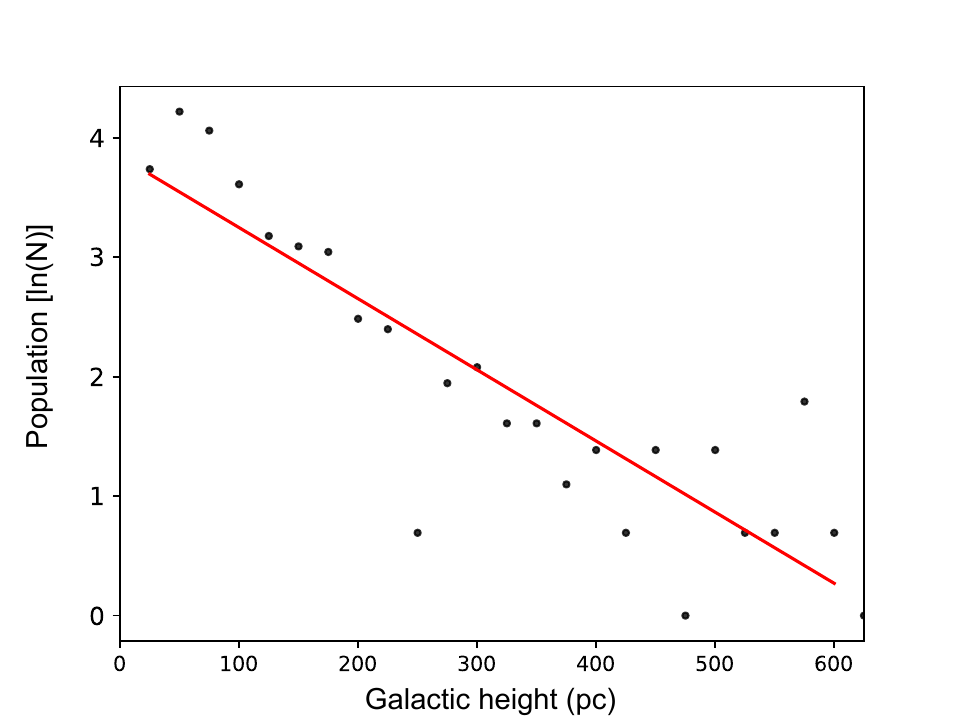}
        \caption{Logarithm of PNe population density as a function of the galactic height (absolute value), together with the linear regression.}
        \label{fig:scale_height}
\end{figure}

Then, we calculated the scale height $H_{z}$, i.e. the galactic height where the PNe population density decreased by a factor $e$ from the galactic plane. We assumed that the Sun is close enough to the galactic plane. The heights from the galactic plane can be calculated as $z = D \cdot sin(\phi)$. The numerical values can be seen in Table \ref{tab:astrometric} and their distribution is shown in  Fig.~\ref{fig:galactic_heights} in bins of $z=25$ pc. This information can be used to derive $H_{z}$ by a linear regression as shown in Fig.~\ref{fig:scale_height}. Only PNe with $\abs{z}<600$ were used for the fit because for higher altitudes the statistic is poor. We obtained the following relationship:

$$ln(N) = -5.96\cdot 10^{-3}\cdot \abs{z}+3.84434,$$

where an average quadratic error of $\sigma^{2}=0.34$ was found. The corresponding value of the scale height is $$H_{z}=168 pc,$$ which has an uncertainty of several tens of parsecs (see section 7.4 for details).
This value is lower than that provided by \citep{frew08} of $217 \pm 20$ pc, but within the range of $180 \pm 20$ pc given by \citet{Pottasch96}.

Once $H_{z}$ has been derived, we can estimate the density of PNe in the Galaxy. If we consider only those PNe with an absolute galactic height below the scale height and inside a cylinder with radius 2 kpc, we obtain a total number of $N_{c}=269$ PNe. The density can then be calculated taking into account the number of stars and the cylindrical volume, ${V_{c}}$, as

$$\rho = \dfrac{N_{c}}{V_{c}} = (6.38 \pm 0.01) \cdot 10^{-8} PNe \cdot pc^{-3}.$$ %\dfrac{N_{c}}{(\pi\cdot R^{2})\cdot 2\cdot H_{z}} = \dfrac{269}{(\pi\cdot 2000^{2})\cdot 2\cdot 167.86},$$ 

%giving a value of:

%$$\rho = 6.376 \cdot 10^{-8} PNe\cdot pc^{-3} ,$$
%
%The corresponding error can be obtained by $\Delta\rho = \abs{\dfrac{\delta\rho(H_{z})}{\delta H_{z}}}\cdot \Delta H_{z} = \dfrac{269}{2\pi \cdot (2000)^{2}} \abs{\dfrac{-1}{167.86^{2}}}\cdot 0.34 = 1.29\cdot 10^{-10} PNe\cdot pc^{-3}.$ So the complete value of the density is:
%
%$$\rho = (6.38 \pm 0.01 )\cdot 10^{-8} PNe \cdot pc^{-3}.$$
%
%$$\rho = 6.376 \cdot 10^{-8} PNe\cdot pc^{-3} ,$$

%$$\rho = (6.38 \pm 0.01 )\cdot 10^{-8} PNe \cdot pc^{-3},$$

%where the corresponding error was obtained by $\Delta\rho = \abs{\dfrac{\delta\rho(H_{z})}{\delta H_{z}}}\cdot \Delta H_{z} = \dfrac{269}{2\pi \cdot (2000)^{2}} \abs{\dfrac{-1}{167.86^{2}}}\cdot 0.34 = 1.29\cdot 10^{-10} PNe\cdot pc^{-3}.$ %So the complete value of the density is:

This value is slightly lower than that given by \citep{zijlstrapottasch91} of $7 \cdot 10^{-8} PN\cdot pc^{-3}$.

\subsection{Total population}
To estimate the total PNe population in our galaxy, we can use the linear regression function calculated in the previous section, i.e.

$$N(z) = e^{-5.96\cdot 10^{-3}\abs{z}} \cdot e^{3.84434} = 46.7278 \cdot e^{-5.96\cdot 10^{-3}\abs{z}}.$$ 

This gives the number of PNe as a  function of absolute galactic height. Thus, a density function can be derived considering this expression per volume ($V=\pi R^{2} \Delta z$), where $R=2$ kpc and $\Delta z = 25$ pc is the height interval used to count PNe, is written as

$$\rho_{z} = \dfrac{N(z)}{V}.$$

%per volume ($V=\pi R^{2} \Delta z$) as a function of the galactic height, where $R=2$ Kpc and $\Delta z = 25$ pc is the height interval used to count stars. We obtain a value for such local density of $\rho_{z} = \dfrac{N(z)}{V}$.% with and error of $e^{\sigma^{2}}=e^{0.34}=1.405.$

If we assume that this density rules for all the Galaxy, $\rho_{G} = \rho_{z}$, and we extrapolate it to the whole galactic volume (approached to a disc of radius $R_{G} = 15$ kpc), we obtain %an approximate value of 
the total PNe population in the Galaxy disc as follows:

$$d(N_{G}) = \dfrac{N(z)}{V} \cdot d(V_{G}),$$

%Substituting the numerical values:

$$d(N_{G}) = \dfrac{46.7278 \cdot e^{-5.96\cdot 10^{-3}\abs{z}}}{\pi\cdot(2000)^{2} \cdot 25} \cdot \pi \cdot (15000)^{2} dz,$$
%\\

$$N_{G} = 105.1376 \cdot \int_{0}^{644} e^{-5.96 \cdot 10^{-3}\abs{z}} dz = 17261 \tab[0.2cm] PNe,$$

where $z = 644 pc$ is the height and the number of PNe becomes zero according to the linear regression fitted to the data.

%$$N_{G} = 17.261 \tab[0.2cm] PNe.$$

To our estimation of 17261 PNe in the disc of the Galaxy, we must add the number of PNe estimated to be populating the bulge, which is about 3500  PNe according to \citet{Peyaud05}; this leads to an estimation of 20761 PNe in the Galaxy, excluding the halo. This number can be considered a lower limit, taking into account that we are certainly loosing, at least, both some compact nebulae and low brightness nebulae. %Despite this number could be subestimated, it is inside the range given by other papers and bibliography.

\subsection{Birth rate}

We can further attempt an estimation of the birth rate of the PNe in our galaxy (within the scale height limits), considering the obtained density of $63.8$ $PNe \cdot kpc^{-3}$. If we estimate which percentage of the PNe have an age up to, for instance, $10^{4}$ yrs, we can calculate how many are born per year and per unit volume. Therefore, analysing data about the evolutionary ages obtained in section 6, 32 PNe, out of a total sample of 68, resulted to be younger than $10^{4}$ yrs. This is an approximately $47\%$ rate of young PNe.

Based on this, we can conclude that the density of PNe younger than $10^{4}$ yrs is approximately $30$  $PNe \cdot kpc^{-3}$. And, dividing by this amount of years, we ended up estimating that the birth rate of the PNe is about $3 \cdot 10^{-3}$  $PNe \cdot kpc^{-3} \cdot yr^{-1}$. This rate is very similar to that reported in the classical work by \citet{Pottasch96}.

\subsection{Completeness of the sample}

When dealing with studies of a large astrophysical sample, a difficult question to tackle is that of its completeness because in general there are several potential sources of incompleteness that cannot be neglected.
%In this section we are aimed to discuss about the completeness of our sample, which is a difficult question as there are several sources of incompleteness. 
The density estimated in section 7.1 was calculated for the region within the scale height limits but in this section we intend to estimate a global galactic density. To accomplish this, we considered a height of 644 pc as the point where the linear regression fitted to the PNe population density, as a function of galactic height, goes to zero (see section 7.2). From that height, we assume that there are an insignificant number of PNe well beyond the galactic disc. 
We found 355 PNe inside the cylinder of 2 kpc radius and a height of $2 \cdot 644$ pc (twice the galactic height).
From these numbers we can estimate an approximately total density of

$$\rho = \dfrac{355}{\pi\cdot 2000^{2}\cdot 2\cdot 644} = 2.19 \cdot 1
0^{-8}  PNe/pc^{3}.$$

\begin{figure}
        \includegraphics[width=0.52\textwidth]{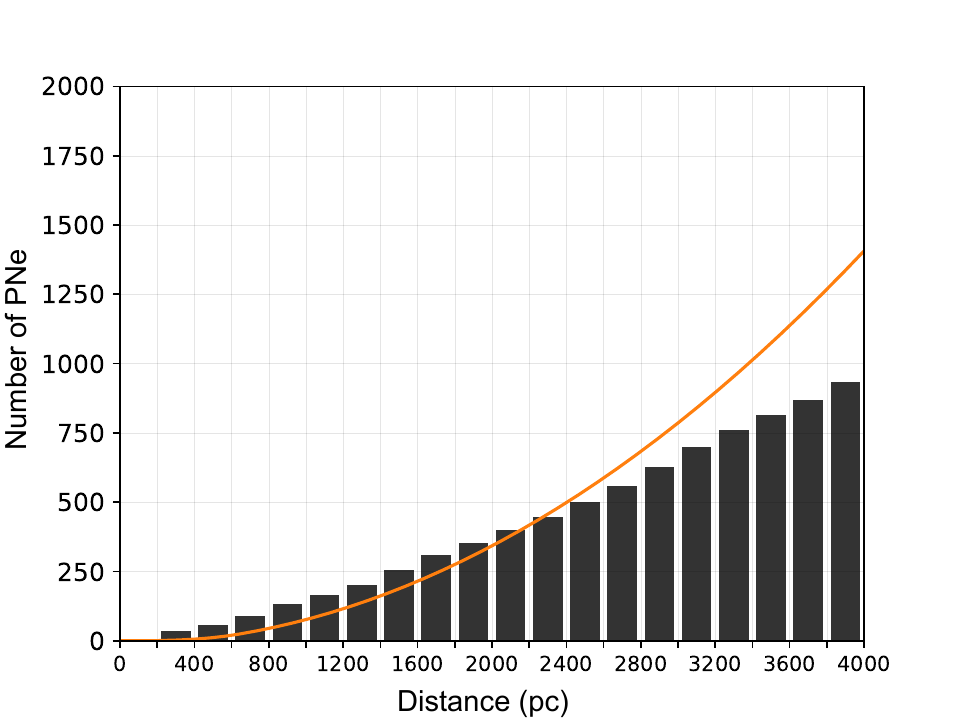}
        \caption{Cumulative population histogram (only objects inside the galactic height) as a function of distance, together with the predictive population function derived using the calculated density. See text for details.}
        \label{fig:completeness}
\end{figure}

%Now, we represent a cumulative distribution of the PNe population as a function of distance. Then, together with this histogram, we represent the increasing function of population according to the density obtained before, and considering spherical volumes of radius equal to the distance. We have to consider that, for spherical radii greater than 644 pc (the height limit for the density), we have to subtract the spherical caps to the volume to fix the population function, as we are assuming no PNe are at higher altitudes than this limit. As it can be seen in Fig.~\ref{fig:completeness}, we can conclude that we have completeness in our sample approximately till 2500 pc.

For this study we used the general sample of 1571 PNe, but discarding the objects with a galactic height beyond 644 pc (the height limit for the density obtained), as we are assuming that an insignificant number of PNe are present at higher altitudes. In Fig.~\ref{fig:completeness}, we represent a cumulative distribution of this PNe population as a function of distance, %{\bf amt No entiendo bien lo que representa el histograma, no son las distancias ni de las 221, ni de las 1571} 
together with the increasing function of population according to the density value obtained before 
and considering spherical volumes of radius equal to the distance. For spherical radii greater than 644 pc, it is necessary to subtract the spherical caps to the volume to fix the population function. As can be seen in Fig.~\ref{fig:completeness}, the prediction is fulfilled up to a distance of about 2300 pc, which becomes the distance for which we can expect that we have completeness. This value is similar to that found in Fig.~\ref{fig:distance} %truncation of the histogram of distances to the GAPN sample (Fig.~\ref{fig:distance}). 
where the number of PNe drops for distances larger than about 2 kpc.

At this point, we reflect on the possible factors that contribute to the incompleteness of our sample. First, some objects are lost from the initial total collection of 2554 possible PNe for a variety of reasons: some nebulae lack a parallax measurement in Gaia DR2, others were detected by Gaia farther than 5 arcsec away from the PN compilation coordinates, and others were not catalogued as PNe in the Simbad database. 

Apart from this, there are other external factors causing incompleteness. The most evident is the difficulty to detect objects that are very far away. %the objects the most difficult it becomes to detect them. 
Also, high extinctions expected close to the galactic plane makes it difficult to detect objects in this region; as can be seen in Fig.~\ref{fig:galactic_heights}, there are fewer PNe in the first 25 pc of height from the galactic plane than in the next 25 pc. Finally, as we already mentioned, nebulae with ages over $10^{4}$ yrs start to loose brightness reducing their detectability.
%causes them to be more difficult to be detected.

%To provide an indication of the order of magnitude of the uncertainty in the parameters that we have been deriving in this section, we have redone the calculations considering the extreme bounds of the distances estimations (low bound and high bound), obtaining: 
To summarise, provided below are the parameters derived in this section together with an estimation of their uncertainties, as calculated considering the extreme bounds of distance estimations (low and high bounds), as follows:
\\
%For the low bound distances we have a total of 823 PNe inside the 2 Kpc radius cylindric volume. After fitting the linear regression (in this case with $abs{z})<700$ pc, as we have more objects) between galactic height and population, we obtain the following results:

%$$H_{z} = 195.2 pc.$$

%$$\rho = 11.07 \cdot 10^{-8} PN\cdot pc^{-3}.$$

%$$N_{G} = 44560 \tab[0.2cm] PNe.$$

%On the other hand, when we consider the high bound distances, we have a total of 196 PNe inside the 2 Kpc radius cylindrical volume. After fitting the linear regression (in this case with $abs{z})<400$ pc, as we have less objects) between galactic height and population, we obtain the following results:

%$$H_{z} = 106.1 pc.$$

%$$\rho = 4.57 \cdot 10^{-8} PN\cdot pc^{-3}.$$

%$$N_{G} = 9743 \tab[0.2cm] PNe.$$

%So, our final results with their uncertainties are:

\tab[0.4cm] \text{Scale height}: $H_{z} = 168^{+27}_{-62} pc$
\\\\
\tab \text{Density}: $\rho = 6.4^{+4.7}_{-1.8} \cdot 10^{-8} PN\cdot pc^{-3}$
\\\\
\tab \text{Galactic population}: $N_{G} = 17261^{+27297}_{-7520} PNe$
\\\\
\tab \text{Birth rate}: $R_{B} = 3.0^{+2.2}_{-0.8} \cdot 10^{3} PNe \cdot kpc^{-3} \cdot yr^{-1}$
\\\\
%As we see, the uncertainties are quite wide, as we are taking the extreme values in estimated distances to calculate them. With the method we have till the moment to estimate distances, we can not be more accurate in determining these quantities. 
We note that density is calculated considering the region within the scale height limits, and that the population is that estimated for the galactic disc. More general values can be found in the corresponding subsections. 

\section{Conclusions}
%We found that the distances derived from DR2 parallaxes compare well with previous astrometric derivations (USNO and HST) but that distances inferred from non-LTE model fitting are overestimated and need to be carefully reviewed. From literature apparent sizes, we calculated the physical radii for a subsample of nebulae that we used to derive the so called kinematical ages taking into account literature expansion velocities. Luminosities calculated with DR2 distances were combined with literature $T_{eff}$ values in a HR diagram to study the evolutionary status of the nebulae. We compare with the new evolutionary tracks by Miller-Bertollami (2017) finding a rather consistent picture. Nebulae with the smallest radii are located in the flat luminosity region of the HR diagram while those with the largest radii correspond to objects in a later stage, getting dimmer on their way to become a WD. Finally we commented on the completeness of our catalogue and calculated an approximate value for the total number of PNe in the Galaxy.

From a total sample of 1571 PNe with parallaxes in Gaia DR2, we obtained reliable distances for 201 PNe to obtain our sample, GAPN. Our reliability criteria arise from a filtering of objects fulfilling different constraints such as, for example, having less than a $30\%$ distance uncertainty, a parallax uncertainty also below $30\%$, and Gaia astrometric goodness-of-fit indexes UWE and RUWE values within the recommended thresholds (\citet{lindegren18}). In addition, a more stringent filtering was adopted for doubtful objects.

%We have obtained reliable distances from Gaia DR2 parallaxes for a total of 201 PNe (our so-called Golden Astrometry Planetary Nebulae, GAPN sample), from a whole sample of 1571 PNe present in DR2. \textbf{ amt no me gusta mucho esta frase, lo intento arreglar This reliability arise from a filtering of objects that fulfill different constrains; {\it e.g.} having an uncertainty in distance below $30\%$, a parallax uncertainty below $30\%$ and Gaia astrometric goodness of fit indexes UWE and RUWE values within the recommended thresholds (\citet{lindegren18}), in addition to a more accurate filtering for doubtful objects}.

Regarding their location, we can conclude that most PNe are located near the galactic plane (small latitudes) and in the galactic centre direction (longitudes close to 0º). Concerning distances, we observe that we can claim completeness up to approximately 2.3 kpc even though we detected some nebulae farther than 4 kpc. When comparing our results with those of other authors, we appreciate a significant similarity with those obtained from astrometric methods (USNO and HST).
%we appreciate that they are quite similar to those estimated from astrometric measurements (USNO and HST).
We found that distances obtained from non-LTE model fitting are overestimated and need to be carefully reviewed. % But we can conclude that the tendency is the same as in most studies, as the linear regression of both distances is parallel and close to the bisector.
 Additionally, we found that in general our low latitude GAPN PNe follow the sinusoidal radial velocity curves expected for their range of distances, considering the case of pure circular rotation for a flat rotation disc at 230 $km \cdot s^{-1}$.  
 
 We calculated the physical radii for a subsample of nebulae and we found that most of them have a radius larger than 0.1 pc and only a few have a radius larger than 1 pc. %According to radial velocity, we can say that for most of the PNe values range around $\pm 40$ $km \cdot s^{-1}$ up to a maximum velocity below  $\pm 200$ $km \cdot s^{-1}$ %(aquí no me queda claro si lo estoy refraseando bien?). 
 Considering physical radii and observational expansion velocities as taken from literature, we derived the so-called kinematical ages of the nebulae and discussed the limitations of such derivations. Although most of the PNe are rather young, with ages under 15000 yrs, we also found nebulae spanning ages well beyond those values. From the average kinematical age value and the mean physical radius of the sample, we obtained a value for the visibility time of the PNe population, $<T_{VT}>$, similar to that derived by \citet{Jacob13}.
%we obtained the visibility time of the PNe population, $<T_{VT}> = 23400$ yrs. {\bf amt esto no se pone en las conclusiones, solo en la discusion a value similar to the one derived by \citet{Jacob13}.}
%On the other hand, the majority of expansion velocities (corrected with a factor proposed by \citet{Jacob13}) are below 50 $km \cdot s^{-1}$, reaching a maximun value of 80 $km \cdot s^{-1}$. While we obtain a mean value of $(38 \pm 16) \hspace{0.1cm} km\cdot s^{1}$.
%The corresponding kinematical ages , derived from these velocities, are below 15000 years for most of PNe. With a mean value of $(23400 \pm 6800) yrs$. Analysing the results on the low histogram of Fig.~\ref{fig:vel_exp}, we can see that the younger the PNe are the larger amount of them we have, so we can conclude that it is more difficult to detect old PNe than young ones, probably because the old ones are fainter and, consecuently, more difficult to detect in general terms. (\textbf{ iker - esta ultima frase quizas sobra aquí})

Luminosities calculated with DR2 distances were combined with literature $T_{eff}$ values in a HR diagram to study the evolutionary status of the stars and their nebulae. We compared the position of the CS in the HR diagram with the new evolutionary tracks by \citet{millerbertolami17} and we found a rather consistent picture. Stars with the smallest nebular radii are located in the flat luminosity region of the HR diagram, while those with the largest radii correspond to objects in a later evolution stage, getting dimmer on their way to become a WD.
For a more detailed analysis, we divided the diagram into three regions and obtained the mean values and dispersion for the mass, radius, and evolutionary age. 
%From these data, we 
We calculated an expansion velocity mean value per region, which can be compared with the mean observational expansion velocity from literature data. The only region in which we found consistent values (in mean) is the latest evolutionary phase, when the stars are very evolved objects already cooling towards the WD stage. In the other two regions, corresponding to early evolutionary phases, the estimations of mean expansion velocities are subjected to larger uncertainty.  %Noticeably, expansion velocities for our sample of 31 objects in common with \citet{frew08} in this region have a mean value of around 24 $km \cdot s^{-1}$ (without correction factor), which coincides with the mean expansion velocity that we can derive from evolutionary ages and radii. In any case it has to be pointed out that the dispersion of the values of the last quantity is very high. In the other two regions, corresponding to an early evolutionary phase, expansion velocities mean values are very low in comparison with \citet{frew08} observational velocities. It should be taken into account that evolutionary ages depend strongly on the value of $T_{eff}$, and they are very different depending on the mass value adopted for the CS. Estimations of expansion velocities in the early evolutionary stages are subjected to larger uncertainty than those in a more advanced stage. 
%{\bf amt no estoy seguro que esto sea para las concluiones, mas bine para la discusion. Quiza resumirlo. It is also important to stress that, in general, we find a quite high dispersion for the mean values of radii, ages and, consequently, evolutionary expansion velocities, because we are dealing with stars with different masses, stars that evolve at significantly different velocities. Despite this, we found that the position in the HR diagram of the nebulae provides valuable information about their evolutionary state, and that the expansion and size of the envelopes agree in general terms with the evolutionary state of the CSPN. Mean values of all these parameters and their typical deviations are presented in Table \ref{tab:HRD}.}
Also, we find it important to stress that in general a rather high dispersion for the mean values of radii, ages, and, consequently, evolutionary expansion velocities is found; this is because we are dealing with CS of different masses, therefore evolving at significantly different velocities. Despite this, the HR diagram positions of the stars provide valuable evolutionary information, and the size of the envelopes and expansion results quite agree with the evolutionary stage of the CSPNe.

%Concerning the CSs properties, after obtaining their temperatures and luminosities, we are able to analyse the location in the HR diagram of 67 of them. We can see that most of these ones are inside, or close to, the evolutionary tracks that determine their life evolution depending on their mass, and so we can have an approximate idea of their mass values. In that diagram, Fig.~\ref{fig:HR_diagram}, masses go between 0.532 and 0.706 solar masses. Using an algorithm to interpolate values for stars between two tracks, we have estimated their masses and evolutionary ages more accurately.  

Finally, we can draw some conclusions about the total PNe population in the Galaxy. Based on the whole sample of 1571 PNe and taking into account only those inside a close volume (2 kpc radius cylinder) around the Sun, we  obtained a density function and extrapolated this value to the whole Galaxy. This procedure has given us a total number of 20761 PNe in the Milky Way. This number is a bit smaller than others in the literature, but the value is inside the uncertainties limits. We also estimated in $3 \cdot 10^{-3}$  $PNe \cdot kpc^{-3} \cdot yr^{-1}$ as the birth rate of PNe in our galaxy. 
%We briefly discussed on the limitations of the quantities that we have derived and on the possible factors that are contributing to the incompleteness of our sample.
We also included a brief discussion on the limitations for all the quantities derived in the present work and on the possible factors contributing to the incompleteness of our chosen sample. 

%In order to be able to show all these results and conclusions, we have used several assumptions, like about the sample completeness that GDR2 allows us to study, about the interestllar extinction correction, about the bolometric correction we used for the calculation of the luminosity values or about the nebular geometry.

%On the other hand, we have to be aware of the possibility that some CSs could be binaries or that we have selected some field stars. Also the quality of our filtering criteria may not be the best possible and that would mean the lose of some stars for the study. Furthermore, the Gaia scan law prints patterns translate in less measurements for some sources, sources that we may be losing. Finally, we have to consider what the influence would be on our statistics of the PNe whose CS has not been detected, and that could include some of the youngest (with a low diluted nebula, hiding the CS) and of the oldest (with a CS that has lost a lot of brightness) PNe in the galaxy.

\begin{acknowledgements}
        This work has made use of data from the European Space Agency (ESA) Gaia mission and processed by the Gaia Data Processing and Analysis Consortium (DPAC). This research has made use of the SIMBAD database, HASH database, and the ALADIN applet. The authors thank D. Lennon for his comments on non-LTE spectroscopic distances derivations, and an anonymous referee for his/her useful comments and suggestions.  Funding from Spanish Ministry projects ESP2016-80079-C2-2-R, RTI2018-095076-B-C22, Xunta de Galicia ED431B 2018/42, and  AYA-2017-88254-P is acknowledged by the authors. MM thanks the Instituto de Astrofísica de Canarias for a visiting stay funded by the Severo Ochoa Excellence programme. IGS acknowledges financial support from the Spanish National Programme for the Promotion of Talent and its Employability grant BES-2017-083126 cofunded by the European Social Fund.
\end{acknowledgements}


\bibliographystyle{aa} % style aa.bst
\bibliography{bibliopn}

\begin{appendix}

%%%%%%%%%% GENERAL DATA
\clearpage
\onecolumn
\begin{landscape}
\section{Tables}
%\scriptsize
%\footnotesize
\small

\begin{longtable}{ l l l l l l l l l l l l l l}         %p{0.85cm} p{0.85cm} l l}

%\hspace{1cm}
\caption{Astrometric data}
\label{tab:astrometric}\\

\hline\hline
 
PNG name & Gaia DR2 ID & Other name & RA & Dec & Parallax & Parallax er. & Distance & Dist. low er. & Dist. high er.& |z| & Radius & $V_{rad}$ \\ 

 &  &  & ($^{\circ}$) & ($^{\circ}$) & (mas) & (mas) & (pc) & (\%) & (\%) & (pc)   & (as) & ($km s^{-1}$) \\ 

\hline
\endfirsthead
\multicolumn{14}{l}
{\tablename\ \thetable\ -- \textit{Continued from previous page}} \\
\hline\hline
PNG name & Gaia DR2 ID & Other name & RA & Dec & Parallax & Parallax er. & Distance & Dist. low er. & Dist. high er.& |Gal. height| & Radius & $V_{rad}$ \\ 

 &  &  & ($^{\circ}$) & ($^{\circ}$) & (mas) & (mas) & (pc) & (\%) & (\%) & (pc)   & (as) & ($km s^{-1}$) \\ 
\hline
\endhead
\hline \multicolumn{14}{r}{\textit{Continued on next page}} \\
\endfoot
\hline
\endlastfoot
PN G000.2+01.7 & 4060845334458792832 & PN K 6-8 & 264.9135 & -27.7902 & 1.259 & 0.136 & 807.3 & 9.21 & 11.26 & 24.2 & 3.39 & \dots				\\
PN G000.5-01.7 & 4056558471795846272 & JaSt 96 & 268.4877 & -29.3377 & 0.614 & 0.06 & 1640.1 & 6.03 & 6.84 & 50.6 & 12.5 & \dots\\
PN G001.1+00.0 & 4057637676787155200 & JaSt 62 & 267.1917 & -27.9606 & 0.614 & 0.058 & 1639.7 & 5.67 & 6.38 & 2.5 & 3 & \dots\\
PN G001.1+00.8 & 4060723421811887104 & JaSt 54 & 266.2963 & -27.5439 & 0.716 & 0.141 & 1460.6 & 17.39 & 26.46 & 20.6 & 7.5 & \dots\\
PN G001.4-03.4 & 4050281914036256768 & PN ShWi 1 & 270.6075 & -29.4181 & 0.845 & 0.1 & 1202.0 & 9.55 & 11.76 & 71.5 & 6.48 & -40\\
PN G002.4+05.8 & 4111368477921050368 & NGC 6369 & 262.3352 & -23.7597 & 0.908 & 0.08 & 1110.9 & 6.64 & 7.64 & 113.2 & 14.75 & -106\\
PN G002.7-52.4 & 6574225217863069056 & IC 5148 & 329.8962 & -39.3857 & 0.812 & 0.103 & 1238.9 & 10.0 & 12.41 & 982.1 & 65.07 & -26\\
PN G002.8-02.2 & 4062786243067043968 & PN Pe 2-12 & 270.2929 & -27.6389 & 0.269 & 0.063 & 3762.8 & 14.56 & 20.19 & 150.5 & 3.88 & \dots\\
PN G003.9-03.1 & 4063052668692883072 & PN KFL 7 & 271.7084 & -27.1046 & 0.525 & 0.1 & 1964.8 & 15.48 & 22.19 & 107.2 & 3.27 & -91\\
%PN G004.9-04.9 & 4052553745525657600 & PN M 1-44 & 274.0723 & -27.0752 & 0.77 & 0.071 & 1313.1 & 7.85 & 9.28 & 113.6 & 2.85 & -4\\
PN G006.3+02.2 & 4070418060753850112 & MPA J1751-2223 & 267.9167 & -22.3885 & 0.533 & 0.102 & 1927.7 & 15.44 & 22.06 & 74.2 & 4 & \dots\\
PN G006.5+03.4 & 4118615354715439872 & PN PBOZ 29 & 266.9519 & -21.5364 & 1.244 & 0.076 & 807.3 & 4.49 & 4.92 & 48.0 & 4.55 & \dots\\
PN G006.7-02.2 & 4066262280359160832 & M 1-41 & 272.376 & -24.2066 & 0.544 & 0.105 & 1892.5 & 15.72 & 22.65 & 74.3 & 40.25 & -6\\
%PN G007.7-05.3 & 4077240461640366336 & SB 14 & 275.9297 & -24.7914 & 1.163 & 0.223 & 905.4 & 17.6 & 27.02 & 84.9 & 9.45 & 31\\
PN G008.7-04.2 & 4089592611425604864 & PTB 29 & 275.2841 & -23.3989 & 0.385 & 0.073 & 2631.3 & 13.22 & 17.76 & 193.2 & 14.5 & \dots\\
PN G009.3-06.5 & 4077540142924415360 & SB 15 & 277.811 & -23.9681 & 0.33 & 0.059 & 3056.1 & 10.89 & 13.81 & 347.4 & 7.05 & 165\\
PN G009.4-05.0 & 4089517157442187008 & NGC 6629 & 276.4269 & -23.2029 & 0.485 & 0.058 & 2087.0 & 9.89 & 12.27 & 183.7 & 8.03 & 13\\
PN G010.8-01.8 & 4094354493205707392 & NGC 6578 & 274.0688 & -20.4508 & 0.571 & 0.086 & 1777.7 & 11.45 & 14.75 & 56.7 & 5.97 & 5\\
PN G011.7+00.0 & 4095521422965004928 & M 1-43 & 272.9538 & -18.7725 & 1.077 & 0.079 & 933.4 & 5.53 & 6.2 & 1.7 & 3.12 & \dots\\
%PN G011.9+04.2 & 4144897913181069184 & M 1-32 & 269.0839 & -16.4847 & 0.423 & 0.093 & 2417.6 & 17.11 & 25.47 & 178.5 & 4.28 & -86\\
%PN G014.0-05.5 & 4092472404177087872 & V-V 3-5 & 279.1344 & -19.3247 & 0.663 & 0.054 & 1515.0 & 4.35 & 4.75 & 145.9 & 5 & -42\\
PN G016.4-01.9 & 4103910524954236928 & M 1-46 & 276.9847 & -15.5485 & 0.354 & 0.05 & 2838.7 & 11.1 & 14.13 & 97.8 & 5.85 & 29\\
PN G016.7-01.8 & 4103933580341644544 & MPA J1828-1516 & 277.0103 & -15.2726 & 0.327 & 0.063 & 3070.5 & 12.09 & 15.75 & 100.1 & 10 & \dots\\
PN G017.3-21.9 & 6864617817991978624 & A 65 & 296.6425 & -23.137 & 0.658 & 0.07 & 1532.7 & 7.38 & 8.63 & 573.4 & 59.52 & 13\\
PN G017.6-10.2 & 4086643583803222400 & A 51 & 285.2558 & -18.2043 & 0.469 & 0.065 & 2155.0 & 9.24 & 11.28 & 383.2 & 29.52 & 3\\
%PN G019.4-19.6 & 6868431267915481088 & K 2-7 & 295.1212 & -20.4517 & 0.601 & 0.045 & 1671.8 & 5.89 & 6.66 & 562.4 & 76 & -18\\
PN G020.7-05.9 & 4102207686291028864 & Sa 1-8 & 282.6848 & -13.5173 & 0.38 & 0.065 & 2662.0 & 11.16 & 14.25 & 276.7 & 3.5 & 46\\
PN G020.7-08.0 & 4101453112078146432 & MPA J1858-1430 & 284.5803 & -14.5072 & 0.803 & 0.141 & 1289.1 & 15.23 & 21.74 & 180.6 & 105 & \dots\\
PN G025.3+40.8 & 4457218245479455744 & IC 4593 & 242.9356 & 12.0714 & 0.41 & 0.088 & 2397.0 & 16.63 & 23.98 & 1567.4 & 7.5 & 22\\
PN G026.5-03.0 & 4252307211263954688 & Pe 1-19 & 282.4361 & -7.0265 & 0.396 & 0.085 & 2573.4 & 16.05 & 23.2 & 126.7 & 2.2 & \dots\\
PN G026.9+04.4 & 4269678120544342144 & FP J1824-0319 & 276.1703 & -3.3333 & 4.918 & 0.09 & 203.5 & 1.48 & 1.52 & 15.7 & 835 & \dots\\
PN G027.6+16.9 & 4376331092036268032 & DeHt 2 & 265.4204 & 3.1159 & 0.612 & 0.087 & 1656.7 & 10.87 & 13.81 & 482.3 & 55 & \dots\\
PN G030.6+06.2 & 4276328581046447104 & Sh 2-68 & 276.2434 & 0.8598 & 2.507 & 0.092 & 399.7 & 2.95 & 3.14 & 43.7 & 230.1 & \dots\\
PN G033.2-01.9 & 4265754371561662976 & Sa 3-151 & 284.7155 & -0.5484 & 0.849 & 0.062 & 1185.1 & 6.11 & 6.94 & 39.3 & 5.5 & \dots\\
PN G035.9-01.1 & 4268419106711771776 & Sh 2-71 & 285.5012 & 2.153 & 0.618 & 0.057 & 1623.8 & 5.44 & 6.09 & 38.7 & 51.83 & 25\\
PN G036.0+17.6 & 4488953930631143168 & A 43 & 268.3845 & 10.6234 & 0.461 & 0.079 & 2186.5 & 12.25 & 16.06 & 661.9 & 40.02 & -42\\
PN G036.1-57.1 & 6628874205642084224 & NGC 7293 & 337.4108 & -20.8372 & 5.006 & 0.093 & 199.9 & 1.51 & 1.56 & 167.9 & 426.25 & -15\\
PN G038.7+01.5 & 4282499452616310912 & PM 1-267 & 284.0941 & 5.8832 & 1.042 & 0.099 & 969.0 & 8.3 & 9.91 & 26.8 & \dots & -32\\
PN G039.0-04.0 & 4292267621344388864 & PN G039.0-04.0 & 289.3183 & 3.5798 & 0.27 & 0.051 & 3696.8 & 8.8 & 10.6 & 264.1 & 16 & \dots\\
PN G041.8-02.9 & 4294123077230164736 & NGC 6781 & 289.617 & 6.5387 & 2.075 & 0.105 & 483.7 & 4.16 & 4.53 & 25.2 & 62.5 & 4\\
%PN G045.6+24.3 & 4556040392088375168 & K 1-14 & 265.6532 & 21.4507 & 2.78 & 0.047 & 359.8 & 0.62 & 0.63 & 148.2 & 26.37 & \dots\\
PN G045.7-04.5 & 4296362443149857920 & NGC 6804 & 292.8964 & 9.2253 & 1.169 & 0.082 & 860.0 & 5.44 & 6.09 & 68.8 & 26.73 & -13\\
PN G046.8+03.8 & 4506484097383382272 & Sh 2-78 & 285.792 & 14.1163 & 1.607 & 0.144 & 629.2 & 7.59 & 8.92 & 42.2 & 297.48 & \dots\\
PN G047.0+42.4 & 1305573511415857536 & A 39 & 246.8905 & 27.9093 & 1.019 & 0.073 & 984.1 & 5.22 & 5.81 & 664.6 & 81 & \dots\\
PN G049.3+02.3 & 4320639728629291776 & VSP 2-30 & 288.3234 & 15.6558 & 3.016 & 0.052 & 331.8 & 0.92 & 0.93 & 13.8 & \dots & \dots\\
PN G051.0+02.8 & 4513629514212361984 & WhMe 1 & 288.7489 & 17.3794 & 0.576 & 0.066 & 1747.6 & 7.62 & 8.95 & 85.7 & \dots & \dots\\
PN G051.3+01.8 & 4514868732516293760 & PM 1-295 & 289.8279 & 17.1969 & 0.369 & 0.054 & 2716.1 & 7.57 & 8.88 & 86.0 & 9.96 & 13\\
PN G051.9+25.8 & 4594460760030429056 & K 1-15 & 266.2357 & 27.3353 & 0.696 & 0.143 & 1455.6 & 16.71 & 24.49 & 634.0 & 21.5 & \dots\\
PN G052.5-02.9 & 4318785810234714752 & Me 1-1 & 294.7909 & 15.9467 & 0.256 & 0.045 & 3866.9 & 12.75 & 16.79 & 199.7 & 2.2 & -6\\
PN G053.8-03.0 & 1820963913284517504 & A 63 & 295.5429 & 17.0873 & 0.391 & 0.057 & 2567.1 & 8.36 & 9.99 & -135.5 & 22.5 & \dots\\
PN G054.1-12.1 & 1803234906762692736 & NGC 6891 & 303.7869 & 12.7043 & 0.437 & 0.059 & 2292.7 & 10.95 & 13.89 & 481.1 & 6.55 & 42\\
PN G055.3+06.6 & 4521168448795699328 & A 54 & 287.1646 & 22.9832 & 0.408 & 0.086 & 2494.3 & 15.95 & 22.98 & 290.7 & 29.4 & \dots\\
PN G055.4+16.0 & 4585381817643702528 & A 46 & 277.8275 & 26.9367 & 0.359 & 0.064 & 2767.5 & 11.22 & 14.27 & 764.3 & 45.25 & \dots\\
PN G058.6+06.1 & 2024098484670541952 & A 57 & 289.2736 & 25.6258 & 0.538 & 0.109 & 1920.3 & 16.76 & 24.86 & 206.6 & 18.3 & \dots\\
PN G059.7-18.7 & 1761341417799128320 & A 72 & 312.5086 & 13.5582 & 0.723 & 0.105 & 1395.2 & 11.48 & 14.76 & 448.0 & 67.8 & 32\\
PN G060.8-03.6 & 1827256624493300096 & NGC 6853 & 299.9016 & 22.7212 & 2.687 & 0.064 & 372.4 & 1.61 & 1.66 & 24.0 & 203.7 & -42\\
PN G061.4-09.5 & 1816547660416810880 & NGC 6905 & 305.5958 & 20.1045 & 0.43 & 0.065 & 2338.6 & 9.95 & 12.33 & 388.9 & 19.73 & -4\\
PN G063.1+13.9 & 2090486618786534784 & NGC 6720 & 283.3962 & 33.0291 & 1.301 & 0.077 & 771.3 & 4.42 & 4.84 & 186.3 & 38.7 & -19\\
PN G064.6+48.2 & 1380199049219990784 & NGC 6058 & 241.1106 & 40.683 & 0.35 & 0.06 & 2763.5 & 9.59 & 11.69 & 2063.3 & 16 & 1\\
%PN G065.0-27.3 & 1745948362385486720 & NGC 7078 648 & 322.4972 & 12.1745 & 2.918 & 0.251 & 346.0 & 7.53 & 8.84 & 158.8 & 1.45 & -141\\
PN G066.7-28.2 & 1770058865674512896 & NGC 7094 & 324.2207 & 12.7886 & 0.645 & 0.077 & 1547.8 & 8.56 & 10.27 & 731.5 & 50.37 & -101\\
PN G069.4-02.6 & 2053683628140774528 & NGC 6894 & 304.0998 & 30.565 & 0.869 & 0.148 & 1178.0 & 14.43 & 20.08 & 53.9 & 27.42 & -58\\
PN G069.7+00.0 & 2055052657577415936 & K 3-55 & 301.7345 & 32.277 & 0.63 & 0.087 & 1606.4 & 10.69 & 13.52 & 0.1 & 4.5 & \dots\\
PN G072.7-17.1 & 1840395547924993152 & A 74 & 319.2181 & 24.1475 & 1.504 & 0.174 & 673.4 & 9.88 & 12.25 & 198.6 & 400.8 & 18\\
PN G075.9+11.6 & 2125895669204184832 & PN AMU 1 & 292.787 & 43.416 & 0.677 & 0.06 & 1482.4 & 5.59 & 6.28 & 299.4 & 99.75 & \dots\\
PN G076.3+14.1 & 2127040806264002944 & Patchick 5 & 289.8771 & 44.7618 & 0.439 & 0.071 & 2265.1 & 11.03 & 13.98 & 552.4 & 77.75 & \dots\\
PN G076.6-05.7 & 1869422453048750336 & LS II +34 26 & 312.0693 & 34.4567 & 0.299 & 0.045 & 3314.0 & 11.1 & 14.08 & 331.8 & 1.8 & \dots\\
PN G077.6+14.7 & 2127684982639844224 & A 61 & 289.7925 & 46.2478 & 0.614 & 0.103 & 1634.9 & 13.03 & 17.37 & 417.0 & 99.6 & -48\\
PN G080.3-10.4 & 1855295171732158080 & MWP 1 & 319.2845 & 34.2077 & 2.021 & 0.069 & 495.4 & 2.43 & 2.55 & 89.5 & 336.25 & \dots\\
PN G081.2-14.9 & 1850685091269441792 & A 78 & 323.8724 & 31.696 & 0.663 & 0.082 & 1511.0 & 9.12 & 11.08 & 388.8 & 59.01 & 17\\
PN G083.5+12.7 & 2135352396915239808 & NGC 6826 & 296.2006 & 50.525 & 0.665 & 0.055 & 1510.8 & 6.85 & 7.91 & 334.5 & 12.75 & -6\\
PN G085.4+52.3 & 1595941441250636672 & Jacoby 1 & 230.444 & 52.3678 & 1.365 & 0.079 & 733.9 & 4.36 & 4.77 & 581.1 & 330 & \dots\\
PN G086.1+05.4 & 2179832585761932032 & We 1-10 & 307.9682 & 48.8805 & 0.661 & 0.136 & 1572.8 & 17.73 & 27.1 & 149.7 & 95.7 & \dots\\
PN G089.3-02.2 & 1971995510535755648 & M 1-77 & 319.7807 & 46.3131 & 0.416 & 0.036 & 2405.7 & 6.31 & 7.2 & 95.1 & 3.87 & \dots\\
PN G093.9-00.1 & 2171652769005709568 & IRAS 21282+5050 & 322.4936 & 51.0667 & 0.279 & 0.047 & 3573.1 & 6.29 & 7.16 & 7.3 & 2.63 & \dots\\
PN G094.0+27.4 & 2160562927224840576 & K 1-16 & 275.4671 & 64.3648 & 0.496 & 0.071 & 1985.9 & 9.62 & 11.78 & 914.8 & 56.5 & \dots\\
PN G096.4+29.9 & 1633325248915154176 & NGC 6543 & 269.6392 & 66.633 & 0.645 & 0.079 & 1536.8 & 10.03 & 12.41 & 767.4 & 12.5 & -66\\
PN G096.8+31.9 & 1634314740658456320 & TK 2 & 264.5108 & 66.8965 & 5.803 & 0.066 & 172.4 & 0.79 & 0.8 & 91.3 & 1050 & \dots\\
PN G102.9-02.3 & 2004936573978252672 & A 79 & 336.5719 & 54.8272 & 0.317 & 0.074 & 3087.8 & 15.39 & 21.51 & 125.1 & 27 & \dots\\
PN G104.2-29.6 & 2871119705335735552 & Jn 1 & 353.9722 & 30.4684 & 1.24 & 0.106 & 808.0 & 6.92 & 8.0 & 399.6 & 163 & -67\\
PN G107.0+21.3 & 2288467186442571008 & K 1-6 & 301.0595 & 74.4267 & 3.722 & 0.035 & 268.8 & 0.71 & 0.72 & 98.0 & 89.5 & -47\\
PN G107.7+07.8 & 2218688261534278912 & IsWe 2 & 333.3438 & 65.8987 & 1.153 & 0.137 & 881.7 & 10.04 & 12.51 & 119.8 & 432.48 & -8\\
PN G107.8+02.3 & 2201080755349789568 & NGC 7354 & 340.0828 & 61.2857 & 0.484 & 0.102 & 2101.3 & 16.45 & 23.99 & 84.9 & 16 & -43\\
PN G114.0-04.6 & 1998212476247082880 & PN G114.0-04.6 & 356.4489 & 57.0662 & 0.527 & 0.049 & 1902.7 & 3.97 & 4.3 & 155.0 & 56.64 & -31\\
PN G114.7-01.2 & 2011797217286839168 & PN G114.7-01.2 & 356.0166 & 60.5449 & 0.638 & 0.053 & 1571.7 & 4.24 & 4.63 & 34.3 & 11.05 & \dots\\
%PN G116.2+08.5 & 2215132685747022464 & M 2-55 & 352.9704 & 70.3693 & 1.453 & 0.082 & 690.9 & 4.28 & 4.68 & 102.5 & 24.5 & -22\\
PN G118.8-74.7 & 2376592910265354368 & NGC 246 & 11.7639 & -11.872 & 1.982 & 0.111 & 505.7 & 4.96 & 5.49 & 487.8 & 121.75 & -16\\
PN G120.0+09.8 & 537481007814722688 & NGC 40 & 3.2541 & 72.5219 & 0.534 & 0.038 & 1878.2 & 5.27 & 5.87 & 321.9 & 22.5 & -21\\
PN G120.3+18.3 & 2286296686066976896 & LBN 120.29+18.39 & 356.2588 & 80.9499 & 3.443 & 0.058 & 290.6 & 1.03 & 1.05 & 91.9 & 442.5 & \dots\\
PN G124.0+10.7 & 535357713421191168 & HDW 1 & 16.7821 & 73.5565 & 3.223 & 0.075 & 310.5 & 1.74 & 1.8 & 57.8 & 137.4 & \dots\\
PN G128.0-04.1 & 509206447837376128 & Sh 2-188 & 22.6383 & 58.414 & 1.157 & 0.116 & 872.1 & 8.32 & 9.95 & 61.9 & 327.9 & -26\\
PN G130.2+01.3 & 511904404556131200 & IC 1747 & 29.3987 & 63.3218 & 0.338 & 0.07 & 2948.8 & 13.57 & 18.28 & 71.9 & 6.5 & -63\\
PN G135.6+01.0 & 465640807845756160 & WeBo 1 & 40.0599 & 61.1546 & 0.631 & 0.051 & 1588.7 & 3.99 & 4.32 & 27.8 & 21.25 & -12\\
PN G136.1+04.9 & 467936205865972352 & A 6 & 44.6745 & 64.5017 & 1.034 & 0.196 & 1007.4 & 16.88 & 25.24 & 86.6 & 93.6 & \dots\\
PN G136.3+05.5 & 468033345145186816 & HFG 1 & 45.946 & 64.9098 & 1.427 & 0.049 & 701.7 & 1.5 & 1.55 & 67.9 & 240 & -26\\
PN G138.8+02.8 & 463228376251556224 & IC 289 & 47.5804 & 61.3169 & 0.658 & 0.078 & 1532.7 & 8.78 & 10.6 & 75.0 & 22.5 & -13\\
PN G144.5+06.5 & 473712872456844544 & NGC 1501 & 61.7475 & 60.9206 & 0.597 & 0.051 & 1679.1 & 4.1 & 4.46 & 191.6 & 26.75 & 37\\
PN G144.8+65.8 & 786919754746647424 & LTNF 1 & 179.4369 & 48.9384 & 0.633 & 0.056 & 1573.6 & 5.05 & 5.6 & 1435.8 & 111.3 & -9\\
PN G148.4+57.0 & 843950873117830528 & NGC 3587 & 168.6988 & 55.019 & 1.167 & 0.083 & 856.6 & 5.47 & 6.12 & 718.8 & 102.3 & 0\\
PN G149.4-09.2 & 241918950690107264 & HDW 3 & 51.8142 & 45.4057 & 1.189 & 0.109 & 846.1 & 7.44 & 8.7 & 136.4 & 274.8 & \dots\\
PN G149.7-03.3 & 250358801943821952 & IsWe 1 & 57.2747 & 50.0041 & 2.276 & 0.099 & 440.5 & 3.55 & 3.81 & 26.1 & 362.5 & -2\\
PN G156.3+12.5 & 268129413812162816 & HDW 4 & 84.4845 & 55.5376 & 3.384 & 0.113 & 296.0 & 2.8 & 2.96 & 64.3 & 83.7 & \dots\\
PN G156.9-13.3 & 223404549266115456 & HaWe 5 & 56.3613 & 37.8145 & 3.195 & 0.136 & 313.7 & 3.64 & 3.92 & 72.3 & 17 & \dots\\
PN G158.5+00.7 & 254092090595748096 & Sh 2-216 & 70.8387 & 46.7016 & 7.97 & 0.074 & 125.5 & 0.83 & 0.84 & 1.0 & 2985 & 14\\
PN G158.8+37.1 & 1040417211407001984 & A 28 & 130.3982 & 58.2301 & 2.61 & 0.11 & 383.7 & 3.5 & 3.76 & 231.9 & 159.9 & -2\\
PN G164.8+31.1 & 936605992140011392 & ARO 121 & 119.4651 & 53.4214 & 1.019 & 0.118 & 979.2 & 9.34 & 11.39 & 507.0 & 184.65 & -84\\
PN G165.5-15.2 & 168937010969340160 & NGC 1514 & 62.3207 & 30.776 & 2.175 & 0.047 & 460.3 & 1.77 & 1.83 & 121.4 & 92.5 & 60\\
PN G189.1+19.8 & 885587110718845568 & NGC 2371 & 111.3945 & 29.4906 & 0.562 & 0.083 & 1749.7 & 10.46 & 13.06 & 593.9 & 19.88 & 21\\
PN G193.6-09.5 & 3340384082588168960 & H 3-75 & 85.1876 & 12.3565 & 0.261 & 0.06 & 3686.8 & 12.52 & 16.27 & 613.3 & 15.25 & 7\\
PN G196.6-10.9 & 3336140414380977664 & NGC 2022 & 85.5258 & 9.0863 & 0.561 & 0.095 & 1778.7 & 12.91 & 17.13 & 337.6 & 13.35 & 14\\
PN G197.4-06.4 & 3341996555048041856 & WeDe 1 & 89.8536 & 10.6946 & 1.829 & 0.144 & 551.2 & 6.71 & 7.73 & 61.8 & 465 & 16\\
PN G197.8+17.3 & 865037169677723904 & NGC 2392 & 112.2949 & 20.9118 & 0.53 & 0.057 & 1866.7 & 8.6 & 10.31 & 558.2 & 22.5 & 84\\
PN G204.1+04.7 & 3158419684195782656 & K 2-2 & 103.0965 & 9.9655 & 1.188 & 0.077 & 844.8 & 4.89 & 5.41 & 69.7 & 217.5 & 28\\
PN G205.1+14.2 & 3163546505053645056 & A 21 & 112.2613 & 13.2468 & 1.89 & 0.097 & 530.6 & 4.13 & 4.5 & 130.5 & 316.2 & 29\\
PN G206.4-40.5 & 3189152962633165056 & NGC 1535 & 63.5657 & -12.7394 & 0.853 & 0.071 & 1166.3 & 6.94 & 8.02 & 758.4 & 16.35 & -1\\
PN G208.5+33.2 & 660071056749861888 & A 30 & 131.7227 & 17.8795 & 0.391 & 0.07 & 2405.7 & 10.66 & 13.25 & 1320.4 & 63.48 & \dots\\
PN G214.9+07.8 & 3135710272253699584 & A 20 & 110.7403 & 1.7594 & 0.636 & 0.097 & 1577.4 & 11.76 & 15.21 & 214.5 & 31.92 & \dots\\
PN G215.2-24.2 & 2985789113026163584 & IC 418 & 81.8675 & -12.6973 & 0.675 & 0.062 & 1473.8 & 7.47 & 8.74 & 606.1 & 6.25 & 62\\
PN G215.5-30.8 & 2986220396462236032 & A 7 & 75.7814 & -15.6063 & 2.05 & 0.089 & 488.5 & 3.44 & 3.69 & 250.5 & 391.2 & 18\\
PN G215.6+03.6 & 3109444657456300288 & NGC 2346 & 107.3438 & -0.8066 & 0.716 & 0.051 & 1401.5 & 5.83 & 6.58 & 88.5 & 45.75 & 47\\
PN G215.7-03.9 & 3103822540968315264 & PN G215.7-03.9 & 100.5767 & -4.2969 & 2.841 & 0.384 & 362.2 & 12.21 & 16.11 & 25.2 & 309.9 & \dots\\
PN G216.0+07.4 & 3110803653827201280 & PHR J0723+0036 & 110.952 & 0.6089 & 0.28 & 0.058 & 3473.5 & 10.97 & 13.81 & 452.0 & 35 & \dots\\
PN G217.1+14.7 & 3088991026757468800 & A 24 & 117.9065 & 3.0059 & 1.457 & 0.146 & 691.2 & 8.5 & 10.19 & 176.0 & 189 & 13\\
PN G218.9-10.7 & 3001563840710096512 & HDW 5 & 95.9048 & -10.2233 & 0.934 & 0.099 & 1078.8 & 8.51 & 10.21 & 201.7 & 46.98 & \dots\\
PN G219.1+31.2 & 597324024095840512 & A 31 & 133.5548 & 8.898 & 1.989 & 0.105 & 503.9 & 4.29 & 4.68 & 261.7 & 465 & 41\\
PN G219.3+01.1 & 3053413132594418688 & K 1-9 & 106.815 & -5.1688 & 0.637 & 0.108 & 1592.6 & 13.45 & 18.19 & 32.0 & 19 & 60\\
PN G220.3-53.9 & 5084896688945791232 & NGC 1360 & 53.311 & -25.8715 & 2.582 & 0.092 & 387.9 & 3.12 & 3.33 & 313.5 & 220 & 108\\
PN G221.0-01.4 & 3052395775097859072 & PHR J0701-0749 & 105.2892 & -7.8233 & 0.339 & 0.054 & 2950.3 & 8.27 & 9.86 & 72.8 & 33.25 & 46\\
PN G221.5+46.3 & 615161091995252864 & EGB 6 & 148.2457 & 13.743 & 0.764 & 0.185 & 1235.0 & 17.28 & 24.91 & 893.8 & 360 & \dots\\
PN G222.1+03.9 & 3058094200264637312 & PFP 1 & 110.5737 & -6.3628 & 1.682 & 0.1 & 597.0 & 4.86 & 5.38 & 40.7 & 562.47 & \dots\\
PN G224.9+01.0 & 3047204259149524480 & We 1-6 & 109.3585 & -10.1769 & 0.616 & 0.074 & 1633.2 & 8.68 & 10.45 & 30.3 & 39.25 & \dots\\
PN G228.2-22.1 & 2917223705359238016 & LoTr 1 & 88.7775 & -22.9007 & 0.554 & 0.035 & 1802.2 & 4.52 & 4.96 & 679.3 & 70.8 & 19\\
PN G231.1+03.9 & 3030005560828868096 & BMP J0739-1418 & 114.9606 & -14.3072 & 0.503 & 0.072 & 1989.7 & 9.93 & 12.29 & 135.5 & 75.75 & \dots\\
PN G238.0+34.8 & 3827045525522912128 & A 33 & 144.788 & -2.8084 & 1.072 & 0.087 & 931.6 & 7.15 & 8.31 & 532.6 & 135.03 & 60\\
PN G239.6+13.9 & 5709928951521751168 & NGC 2610 & 128.3476 & -16.1494 & 0.449 & 0.092 & 2148.6 & 14.19 & 19.21 & 517.9 & 24.33 & 88\\
PN G241.0+02.3 & 5710725616423348352 & M 3-4 & 118.7975 & -23.6368 & 0.252 & 0.055 & 3883.8 & 11.38 & 14.48 & 159.5 & 15.75 & 25\\
PN G244.5+12.5 & 5703415268543295744 & A 29 & 130.0788 & -20.9102 & 0.879 & 0.191 & 1152.6 & 17.96 & 27.17 & 250.7 & 210 & \dots\\
PN G248.7+29.5 & 5690534730341025408 & A 34 & 146.3973 & -13.1711 & 0.891 & 0.116 & 1118.1 & 10.32 & 12.87 & 551.3 & 143.52 & \dots\\
PN G249.0+06.9 & 5646499053433949440 & Hen 3-172 & 127.9287 & -27.7588 & 0.401 & 0.036 & 2490.1 & 6.39 & 7.3 & 302.1 & 0.6 & 24\\
PN G253.5+10.7 & 5647809392112960000 & K 1-2 & 134.4415 & -28.9602 & 0.503 & 0.131 & 1938.6 & 19.19 & 29.33 & 362.6 & 40 & 66\\
PN G253.9+05.7 & 5639472001599302528 & M 3-6 & 130.1676 & -32.376 & 0.344 & 0.065 & 2871.5 & 11.83 & 15.26 & 289.1 & 4.8 & 31\\
PN G255.3-59.6 & 4755520010701141504 & Lo 1 & 44.2436 & -44.1716 & 1.267 & 0.063 & 789.6 & 3.27 & 3.49 & 681.2 & 209 & \dots\\
PN G255.8+10.9 & 5635240088724202496 & FP J0905-3033 & 136.2722 & -30.5534 & 1.371 & 0.09 & 731.9 & 5.22 & 5.82 & 139.2 & 385.5 & \dots\\
PN G258.0-15.7 & 5509004952576699904 & WRAY 17-1 & 108.7059 & -46.9609 & 0.548 & 0.125 & 1806.6 & 17.33 & 25.51 & 490.3 & 47.01 & \dots\\
PN G258.5-01.3 & 5527934587141415168 & RCW 24 & 126.4481 & -40.2194 & 1.227 & 0.156 & 830.3 & 10.93 & 13.93 & 18.7 & 271.25 & \dots\\
PN G261.0+32.0 & 5668874905325843456 & NGC 3242 & 156.1922 & -18.6423 & 0.712 & 0.098 & 1388.3 & 11.27 & 14.34 & 736.7 & 15.55 & 5\\
PN G263.1+04.3 & 5620675957697739648 & FP J0904-4023 & 136.0096 & -40.3722 & 0.946 & 0.121 & 1070.2 & 10.62 & 13.4 & 80.7 & 270 & \dots\\
PN G265.1-04.2 & 5521499734013833984 & ESO 259-10 & 128.5294 & -47.277 & 0.907 & 0.055 & 1105.4 & 3.41 & 3.66 & 81.2 & 13.98 & 49\\
PN G272.1+12.3 & 5420219732228461184 & NGC 3132 & 151.7573 & -40.4364 & 1.187 & 0.058 & 844.8 & 4.11 & 4.48 & 181.4 & 36.5 & 49\\
PN G273.6+06.1 & 5411328565822822528 & HbDs 1 & 148.1855 & -46.2798 & 1.458 & 0.056 & 687.1 & 3.21 & 3.43 & 74.1 & 64 & \dots\\
PN G277.1-03.8 & 5307241785737968256 & NGC 2899 & 141.7622 & -56.106 & 0.492 & 0.058 & 2049.5 & 6.99 & 8.1 & 136.9 & 32.07 & 3\\
PN G277.7-03.5 & 5307054800062198400 & WRAY 17-31 & 142.8355 & -56.2943 & 0.688 & 0.127 & 1505.7 & 15.67 & 22.63 & 93.4 & 73.26 & \dots\\
PN G279.6-03.1 & 5306043043221261568 & Hen 2-36 & 145.8568 & -57.2821 & 0.249 & 0.037 & 4027.2 & 10.46 & 13.1 & 223.9 & 10.04 & 12\\
PN G283.9+09.7 & 5362804330246457344 & DS 1 & 163.6689 & -48.7841 & 1.324 & 0.049 & 756.6 & 3.02 & 3.21 & 127.8 & 167.25 & -25\\
PN G285.4-01.1 & 5255353117640172672 & Pe 2-5 & 157.1441 & -59.0565 & 0.919 & 0.054 & 1090.5 & 3.21 & 3.43 & 21.3 & \dots & \dots\\
PN G285.7-14.9 & 5222772389050179840 & IC 2448 & 136.7763 & -69.9418 & 0.318 & 0.065 & 3055.3 & 12.34 & 16.03 & 788.3 & 11 & -24\\
PN G286.1-02.0 & 5255143175335512320 & MPA J1029-6014 & 157.2828 & -60.2345 & 0.684 & 0.075 & 1476.8 & 7.99 & 9.48 & 53.8 & 34.75 & \dots\\
PN G286.5+11.6 & 5374348102823679744 & Lo 5 & 168.4757 & -47.9501 & 0.659 & 0.117 & 1535.4 & 14.34 & 19.81 & 313.7 & 75.5 & \dots\\
PN G288.8-05.2 & 5251802141752703360 & Hen 2-51 & 158.9404 & -64.3199 & 0.373 & 0.051 & 2698.8 & 6.49 & 7.44 & 245.3 & 4.5 & 8\\
PN G290.5+07.9 & 5348195948185156736 & Fg 1 & 172.1508 & -52.9344 & 0.505 & 0.079 & 1999.9 & 11.43 & 14.69 & 275.8 & 23.75 & 28\\
PN G291.3+08.4 & 5345597458614614400 & PHR J1134-5243 & 173.6606 & -52.7256 & 0.243 & 0.044 & 4039.5 & 13.02 & 17.21 & 590.5 & 19.5 & -66\\
PN G292.6+01.2 & 5336133687170599040 & NGC 3699 & 171.991 & -59.9579 & 0.678 & 0.114 & 1506.4 & 13.62 & 18.54 & 32.5 & 21 & -16\\
PN G292.9+01.0 & 5335989479389919104 & PHR J1129-6012 & 172.4594 & -60.2022 & 0.47 & 0.091 & 2159.2 & 14.82 & 20.73 & 40.7 & 9.5 & \dots\\
PN G294.1+43.6 & 3519614068578061568 & NGC 4361 & 186.1281 & -18.7849 & 1.004 & 0.092 & 999.0 & 7.15 & 8.31 & 689.2 & 58.5 & 10\\
PN G302.2-03.1 & 5861458902488176256 & PHR J1244-6601 & 191.2481 & -66.0178 & 0.377 & 0.05 & 2665.2 & 5.97 & 6.76 & 146.6 & 10.75 & \dots\\
PN G305.6-13.1 & 5790663302917359872 & ESO 40-11 & 203.5595 & -75.7754 & 0.438 & 0.077 & 2279.8 & 12.39 & 16.23 & 517.8 & 32.5 & -16\\
PN G306.7-01.5 & 5864899205704369280 & MPA J1326-6407 & 201.6366 & -64.1192 & 0.884 & 0.103 & 1146.6 & 9.42 & 11.56 & 30.2 & 4.5 & \dots\\
PN G307.2-03.4 & 5863702868275424384 & NGC 5189 & 203.3869 & -65.9742 & 0.624 & 0.055 & 1608.6 & 4.91 & 5.44 & 96.9 & 67.75 & -8\\
PN G308.6-12.2 & 5796437220731788288 & Hen 2-105 & 213.8531 & -74.2128 & 0.307 & 0.069 & 3201.1 & 14.09 & 19.12 & 680.9 & 20.55 & \dots\\
PN G309.5-02.9 & 5852194829848899840 & MaC 1-2 & 208.6128 & -64.9936 & 0.776 & 0.055 & 1293.3 & 4.06 & 4.42 & 66.2 & 5.6 & -47\\
PN G310.3+24.7 & 6162542191540458112 & Lo 8 & 201.4062 & -37.6041 & 0.984 & 0.131 & 1030.7 & 11.55 & 14.92 & 431.9 & 60.5 & -3\\
PN G311.0+02.4 & 5870592987893097984 & SuWt 2 & 208.9301 & -59.3777 & 0.408 & 0.044 & 2459.4 & 8.46 & 10.13 & 106.3 & 32.48 & -40\\
PN G312.6-01.8 & 5854138766383247232 & Hen 2-107 & 214.6805 & -63.1195 & 0.243 & 0.052 & 4087.3 & 10.05 & 12.45 & 135.5 & 4.75 & \dots\\
PN G315.7+05.5 & 5895881038094691200 & LoTr 8 & 215.4997 & -55.038 & 1.4 & 0.26 & 748.4 & 17.09 & 25.81 & 72.6 & 13.38 & \dots\\
PN G316.1+08.4 & 5897352631316651264 & Hen 2-108 & 214.537 & -52.1776 & 0.389 & 0.047 & 2586.3 & 9.37 & 11.45 & 380.1 & 6.47 & -11\\
PN G316.3+08.8 & 5897438771177986432 & PHR J1418-5144 & 214.6074 & -51.7441 & 0.8 & 0.122 & 1278.0 & 12.74 & 16.99 & 196.5 & 194.75 & \dots\\
PN G318.4+41.4 & 6292074655679874688 & A 36 & 205.1724 & -19.882 & 2.32 & 0.092 & 432.0 & 3.5 & 3.75 & 286.2 & 191.25 & 37\\
%PN G319.0-02.7 & 5876133461344048768 & MPA J1511-6108 & 227.8386 & -61.1479 & 2.715 & 0.215 & 371.9 & 6.94 & 8.05 & 17.7 & 3 & \dots\\
PN G319.5-04.1 & 5875163902566884096 & BMP J1521-6203 & 230.3192 & -62.0596 & 1.16 & 0.223 & 905.1 & 17.54 & 26.86 & 65.0 & 4 & \dots\\
PN G322.4-02.6 & 5881838006914886784 & Mz 1 & 233.5693 & -59.1523 & 0.733 & 0.156 & 1426.7 & 18.78 & 29.63 & 65.0 & 21.15 & -32\\
PN G324.0+03.5 & 5888049732191403904 & IRAS 15154-5258 & 229.7865 & -53.1638 & 0.3 & 0.057 & 3331.7 & 10.94 & 13.85 & 205.2 & 13.5 & -81\\
PN G326.4+07.0 & 5902564934924665984 & NeVe 3-2 & 229.9328 & -48.9986 & 0.288 & 0.063 & 3475.8 & 13.6 & 18.37 & 423.2 & 16.5 & \dots\\
PN G326.6+42.2 & 6291509197468092672 & IC 972 & 211.108 & -17.2279 & 0.533 & 0.12 & 1867.1 & 17.15 & 25.18 & 1255.2 & 23.5 & -27\\
PN G327.8+10.0 & 6000019147299399936 & NGC 5882 & 229.2081 & -45.6496 & 0.537 & 0.084 & 1892.7 & 11.78 & 15.3 & 331.3 & 7.13 & 8\\
PN G327.8-01.6 & 5836606709883180672 & Hen 2-143 & 240.2477 & -55.0943 & 0.726 & 0.149 & 1431.2 & 17.82 & 27.3 & 41.7 & 1.86 & -35\\
PN G329.0+01.9 & 5982072132545824128 & Sp 1 & 237.9206 & -51.5246 & 0.698 & 0.05 & 1435.5 & 3.25 & 3.47 & 49.0 & 36 & -32\\
PN G331.0+01.2 & 5981613872402237312 & IRAS 16005-5041 & 241.0745 & -50.8341 & 0.566 & 0.103 & 1795.7 & 14.5 & 20.15 & 37.7 & 1.35 & \dots\\
PN G332.1-05.8 & 5930256749426536064 & PHR J1642-5501 & 250.7327 & -55.0311 & 1.224 & 0.193 & 845.4 & 13.97 & 19.3 & 86.6 & 23 & \dots\\
PN G332.5-16.9 & 5911656865276078080 & HaTr 7 & 268.5393 & -60.8327 & 0.54 & 0.074 & 1870.9 & 9.72 & 12.0 & 544.2 & 92 & \dots\\
PN G335.2-03.6 & 5937103069115240192 & HaTr 4 & 251.2508 & -51.2059 & 0.393 & 0.095 & 2597.6 & 18.71 & 28.98 & 164.0 & 12.25 & -97\\
PN G335.5+12.4 & 6008325614048284416 & DS 2 & 235.7709 & -39.3041 & 1.238 & 0.07 & 811.3 & 4.88 & 5.4 & 174.9 & 93 & \dots\\
PN G339.9+88.4 & 3958428334589607552 & LoTr 5 & 193.8905 & 25.8918 & 2.007 & 0.054 & 498.8 & 2.27 & 2.38 & 498.6 & 258.75 & -17\\
PN G341.5+12.1 & 6010805807350513920 & WR 72 & 241.6186 & -35.7535 & 0.735 & 0.06 & 1366.7 & 5.21 & 5.8 & 286.6 & \dots & -93\\
PN G341.6+13.7 & 6011169161583903488 & NGC 6026 & 240.338 & -34.5433 & 0.34 & 0.06 & 2961.3 & 10.85 & 13.74 & 701.6 & 24.63 & -87\\
PN G341.8+05.4 & 6017034570775817984 & NGC 6153 & 247.8774 & -40.2535 & 0.764 & 0.082 & 1321.7 & 8.08 & 9.6 & 125.3 & 12.8 & 39\\
PN G341.9+08.8 & 6021420630046381440 & SB 27 & 244.8084 & -37.7912 & 6.598 & 0.063 & 151.6 & 0.65 & 0.65 & 23.4 & 5.7 & -2\\
PN G343.3-00.6 & 5966150998003866368 & HaTr 5 & 255.367 & -43.0987 & 0.9 & 0.079 & 1118.8 & 6.61 & 7.6 & 12.9 & 52 & \dots\\
%PN G344.2+04.7 & 5971001081288743808 & Hen 2-178 & 250.6395 & -38.9112 & 0.352 & 0.067 & 2856.0 & 12.4 & 16.29 & 236.5 & 2.5 & -142\\
PN G345.4+00.1 & 5966769881320062208 & IC 4637 & 256.2936 & -40.8856 & 0.784 & 0.051 & 1280.9 & 5.34 & 5.96 & 3.1 & 8.1 & 15\\
PN G349.3-01.1 & 5972577055062637056 & NGC 6337 & 260.5653 & -38.4838 & 0.572 & 0.079 & 1768.0 & 10.09 & 12.57 & 34.4 & 23.52 & -71\\
PN G349.3-04.2 & 5960852520245103744 & Lo 16 & 263.9245 & -40.1907 & 0.577 & 0.079 & 1758.1 & 10.12 & 12.63 & 129.5 & 209 & 13\\
PN G352.0-06.7 & 4036463835977400064 & SB 36 & 268.5868 & -39.1772 & 0.63 & 0.081 & 1610.2 & 9.68 & 11.95 & 189.9 & 3.75 & 35\\
PN G352.7-08.4 & 4035884045392107392 & SB 38 & 270.871 & -39.3575 & 0.266 & 0.055 & 3786.5 & 11.06 & 14.08 & 553.7 & 7.5 & 59\\
PN G354.7-07.2 & 4037160548405771776 & SB 40 & 270.732 & -37.1369 & 0.935 & 0.108 & 1087.4 & 9.48 & 11.65 & 137.3 & 9.75 & -69\\
PN G355.1-02.9 & 4041720699831756160 & H 1-31 & 266.3832 & -34.5652 & 1.263 & 0.062 & 794.0 & 3.17 & 3.38 & 40.4 & 0.88 & 48\\
PN G355.3+03.8 & 4058948024160146176 & MPA J1719-3043 & 259.8336 & -30.7279 & 0.692 & 0.069 & 1458.4 & 6.97 & 8.07 & 96.6 & 32.52 & \dots\\
PN G355.5+04.7 & 4059241211414866304 & PHR J1716-3002 & 259.0174 & -30.0374 & 0.562 & 0.099 & 1830.8 & 14.22 & 19.72 & 152.3 & 7.38 & \dots\\
PN G356.5-02.3 & 4053955824662571648 & M 1-27 & 266.6895 & -33.1431 & 0.361 & 0.059 & 2802.4 & 9.78 & 12.09 & 117.1 & 9 & -48\\
PN G357.7-04.8 & 4042513447762858880 & BMP J1759-3321 & 269.9384 & -33.3535 & 1.223 & 0.218 & 856.8 & 16.24 & 23.97 & 72.4 & 287.49 & \dots\\
PN G358.5-07.3 & 4039600536544395392 & NGC 6563 & 273.0108 & -33.8685 & 1.103 & 0.209 & 955.0 & 17.37 & 26.52 & 122.0 & 25.5 & -31\\
PN G358.8+03.4 & 4059708851903583488 & MPA J1729-2804 & 262.3615 & -28.0769 & 1.641 & 0.292 & 640.7 & 16.43 & 24.42 & 38.7 & 7 & \dots\\

\end{longtable}

%\tablefoot{
%This table is available in electronic form
%at the CDS via anonymous ftp to cdsarc.u-strasbg.fr (130.79.128.5)
%or via http://cdsweb.u-strasbg.fr/cgi-bin/qcat?J/A+A/.
%}

\tablebib{
Radii are calculated from minor and major axis angular sizes in the HASH (\citet{parker16}) database.
Radial velocities are from the Simbad database: 
(1) \citet{Beaulieu_1999}; (2) \citet{1953GCRV..C......0W}; (3) \citet{2011yCat.2306....0A}; (4) \citet{1995A&AS..114..269D}; (5) \citet{1992ApJS...83...29S}; (6) \citet{Bianchi_2001}; (7) \textit{Gaia DR2 (2018)}; (8) \citet{Feibelman_1999}; (9) \citet{Gontcharov2006}; (10) \citet{Ahn_2012}; (11) \citet{Zhang_2013}; (12) \citet{1994A&AS..108..603B}; (13) \citet{refId0}; (14) \citet{2016MNRAS.462.1393A}; (15) \citet{2015MNRAS.448.1789M}; (16) \citet{2004MNRAS.349.1069K}; (17) \citet{2014MNRAS.442.1379K}.
}

\end{landscape}

\clearpage
\onecolumn

% EXPANSION VELOCITIES - KINEMATICAL AGES
\hspace{3cm}
%\small
%\begin{center}
\begin{longtable}{ l r r r}
\caption{ Expansion velocity - Radius - Kinematical age.}\label{tab:velocity}\\

\hline\hline
 
PNG name & $V_{exp}$ ($km  s^{-1}$) & Radius (pc) & Age$_{kin}$ (yr)        \\ 

\hline
\endfirsthead
\multicolumn{4}{l}
{\tablename\ \thetable\ -- \textit{Continued from previous page}} \\
\hline\hline
PNG Name & $V_{exp}$ ($km s^{-1}$) & Radius (pc) & Age$_{kin}$ (yr) \\ 

\hline
\endhead
\hline \multicolumn{4}{r}{\textit{Continued on next page}} \\
\endfoot
\hline
\endlastfoot
        
\object{PN G002.7-52.4} & 79.5 & 0.391 & 4811                        \\
\object{PN G009.4-05.0} & 9 & 0.081 & 8828                           \\
\object{PN G025.3+40.8} & 18 & 0.087 & 4738                          \\
\object{PN G035.9-01.1} & 31.5 & 0.408 & 12674                       \\
\object{PN G036.1-57.1} & 31.5 & 0.413 & 12832                       \\
\object{PN G041.8-02.9} & 18 & 0.147 & 7968                          \\
\object{PN G045.7-04.5} & 37.5 & 0.111 & 2908                        \\
\object{PN G046.8+03.8} & 30 & 0.907 & 29599                         \\
\object{PN G047.0+42.4} & 43.5 & 0.386 & 8694                        \\
\object{PN G060.8-03.6} & 48 & 0.368 & 7498                          \\
\object{PN G061.4-09.5} & 60 & 0.224 & 3647                          \\
\object{PN G063.1+13.9} & 33 & 0.145 & 4291                          \\
\object{PN G069.4-02.6} & 64.5 & 0.157 & 2376                        \\
\object{PN G072.7-17.1} & 39 & 1.308 & 32830                         \\
\object{PN G077.6+14.7} & 45 & 0.789 & 17167                         \\
\object{PN G080.3-10.4} & 45 & 0.808 & 17563                         \\
\object{PN G081.2-14.9} & 40.5 & 0.432 & 10444                       \\
\object{PN G083.5+12.7} & 24 & 0.093 & 3808                          \\
\object{PN G096.4+29.9} & 30 & 0.093 & 3038                          \\
\object{PN G107.7+07.8} & 18 & 1.849 & 100506                        \\
\object{PN G107.8+02.3} & 37.5 & 0.163 & 4254                        \\
\object{PN G128.0-04.1} & 27 & 1.386 & 50245                         \\
\object{PN G148.4+57.0} & 51 & 0.425 & 8152                          \\
\object{PN G149.4-09.2} & 16.5 & 1.127 & 66854                       \\
\object{PN G149.7-03.3} & 18 & 0.774 & 42088                         \\
\object{PN G158.5+00.7} & 7.5 & 1.816 & 236961                       \\
\object{PN G164.8+31.1} & 36 & 0.877 & 23827                         \\
\object{PN G197.4-06.4} & 25.5 & 1.243 & 47681                       \\
\object{PN G197.8+17.3} & 79.5 & 0.204 & 2506                        \\
\object{PN G204.1+04.7} & 15 & 0.891 & 58112                         \\
\object{PN G205.1+14.2} & 48 & 0.813 & 16582                         \\
\object{PN G206.4-40.5} & 31.5 & 0.092 & 2872                        \\
\object{PN G215.2-24.2} & 21 & 0.045 & 2081                          \\
\object{PN G215.5-30.8} & 43.5 & 0.926 & 20841                       \\
\object{PN G217.1+14.7} & 30 & 0.633 & 20659                         \\
\object{PN G219.1+31.2} & 43.5 & 1.136 & 25554                       \\
\object{PN G220.3-53.9} & 51 & 0.414 & 7938                          \\
\object{PN G221.5+46.3} & 37.5 & 2.155 & 56247                       \\
\object{PN G238.0+34.8} & 48 & 0.516 & 10513                        \\
\object{PN G239.6+13.9} & 36 & 0.253 & 6888                          \\
\object{PN G244.5+12.5} & 30 & 1.173 & 38277                         \\
\object{PN G248.7+29.5} & 52.5 & 0.778 & 14501                       \\
\object{PN G255.3-59.6} & 45 & 0.800 & 17398                         \\
\object{PN G261.0+32.0} & 42 & 0.105 & 2439                          \\
\object{PN G272.1+12.3} & 31.5 & 0.149 & 4644                        \\
\object{PN G277.1-03.8} & 37.5 & 0.319 & 8317                        \\
\object{PN G277.7-03.5} & 42 & 0.535 & 12460                         \\
\object{PN G283.9+09.7} & 45 & 0.614 & 13342                         \\
\object{PN G294.1+43.6} & 48 & 0.283 & 5776                          \\
\object{PN G318.4+41.4} & 54 & 0.401 & 7258                          \\
\object{PN G327.8+10.0} & 34.5 & 0.065 & 1854                        \\
\object{PN G339.9+88.4} & 46.5 & 0.626 & 13168                       \\
\object{PN G341.8+05.4} & 25.5 & 0.082 & 3148                        \\
\object{PN G345.4+00.1} & 31.5 & 0.050 & 1563                        \\
\object{PN G358.5-07.3} & 16.5 & 0.118 & 7002                        \\

\end{longtable}

%\end{center}

\tablefoot{
Expansion velocities are corrected with a factor of 1.5, following the recommendations in \citet{Jacob13}.
}

\tablebib{ 
$V_{exp}$ from \citet{frew08}. %; $V_{NII}$/$V_{post}$ from \citet{Jacob13}.
}

% PHOTOMETRY - TEMPERATURE - MASS - EVOLUTIONARY AGE
\clearpage

\begin{center}
%\small

%\begin{longtable}{ l l l l l l l l l l l} 
\begin{longtable}{ l r r r r r r r r r r}
\caption{ Photometry - Temperature - Mass - Evolutionary age.}\label{tab:photometry}\\

\hline\hline
PNG name & $G$  & $G_{BP}-G_{RP}$ & $m_{V}$ & $A(V)$ & $M_{V}$ & $M_{bol}$ & $log(L/L_{\odot})$ & $T_{eff}$ (kK) & Mass ($M_{\odot}$) & $Age_{evo}$ (yr) \\
\hline
\endfirsthead
\multicolumn{11}{l}
{\tablename\ \thetable\ -- \textit{Continued from previous page}} \\
\hline\hline
PNG name & $G$  & $G_{BP}-G_{RP}$ & $m_{V}$ & $A(V)$ & $M_{V}$ & $M_{bol}$ & $log(L/L_{\odot})$ & $T_{eff}$ (kK) & Mass ($M_{\odot}$) & $Age_{evo}$ (yr) \\
\hline
\endhead
\hline \multicolumn{11}{r}{\textit{Continued on next page}} \\
\endfoot
\hline
\endlastfoot 

\object{PN G002.7-52.4} & 16.1 & -0.64 & 16.16 & 0.02 & 5.67 & -1.15 & 2.36 & 110 & 0.5319 & 73618         \\
\object{PN G009.4-05.0} & 12.68 & 0.45 & 12.87 & 1.77 & -0.5 & -4.8 & 3.82 & 47 & 0.5832 & 2829          \\
\object{PN G025.3+40.8} & 11.17 & -0.33 & 11.33 & 0.21 & -0.78 & -4.6 & 3.74 & 40 & 0.5660 & 4608        \\
\object{PN G035.9-01.1} & 13.47 & 1.21 & 19 & 3.1 & 4.85 & -3.03 & 3.11 & 157 & 0.5832 & 5919            \\
\object{PN G036.0+17.6} & 14.68 & -0.28 & 14.74 & 0.53 & 2.51 & -4.31 & 3.63 & 110 & 0.5660 & 7625       \\
\object{PN G036.1-57.1} & 13.48 & -0.63 & 13.53 & 0.02 & 7.01 & 0.19 & 1.83 & 110 & 0.6087 & 7388        \\
\object{PN G041.8-02.9} & 16.75 & 0.35 & 16.88 & 1.64 & 6.82 & -0.06 & 1.92 & 112 & 0.6005 & 4533        \\
\object{PN G045.7-04.5} & 14 & 0.49 & 14.17 & 1.7 & 2.79 & -3.26 & 3.21 & 85 & 0.5319 & 59020            \\
\object{PN G046.8+03.8} & 17.64 & -0.05 & 17.78 & 1.73 & 7.06 & 0.18 & 1.83 & 112 & 0.6160 & 7150        \\
\object{PN G047.0+42.4} & 15.59 & -0.53 & 15.6 & 0.06 & 5.58 & -1.43 & 2.47 & 117 & 0.5660 & 19415       \\
\object{PN G055.4+16.0} & 14.94 & -0.3 & 14.8 & 0.49 & 2.1 & -3.2 & 3.18 & 66 & 0.5319 & 57647           \\
\object{PN G060.8-03.6} & 14.03 & -0.59 & 14.09 & 0.14 & 6.09 & -1.34 & 2.44 & 135 & 0.5828 & 4776       \\
\object{PN G061.4-09.5} & 14.55 & -0.28 & 14.6 & 0.47 & 2.29 & -5.27 & 4.01 & 141 & 0.7061 & 1162        \\
\object{PN G063.1+13.9} & 15.61 & -0.57 & 15.78 & 0.44 & 5.9 & -1.81 & 2.62 & 148 & 0.5826 & 4239        \\
\object{PN G069.4-02.6} & 18.17 & 0.41 & 18.32 & 1.92 & 6.04 & -0.5 & 2.1 & 100 & 0.5319 & 83787         \\
\object{PN G072.7-17.1} & 17.05 & -0.52 & 17.2 & 0.25 & 7.8 & 1.04 & 1.49 & 108 & 0.7061 & 36071         \\
\object{PN G077.6+14.7} & 17.29 & -0.56 & 17.42 & 0.15 & 6.2 & 0.04 & 1.88 & 88 & 0.5319 & 137938        \\
\object{PN G080.3-10.4} & 13.05 & -0.59 & 13.13 & 0.06 & 4.6 & -3.39 & 3.26 & 163 & 0.5832 & 5952        \\
\object{PN G081.2-14.9} & 13.15 & -0.36 & 13.26 & 0.17 & 2.19 & -4.63 & 3.75 & 110 & 0.5832 & 4019       \\
\object{PN G083.5+12.7} & 10.58 & -0.34 & 10.68 & 0.3 & -0.52 & -5 & 3.9 & 50 & 0.5826 & 2420            \\
\object{PN G086.1+05.4} & 18.01 & -0.25 & 17.85 & 0.42 & 6.45 & 1.53 & 1.29 & 58 & 0.5319 & 409367       \\
\object{PN G094.0+27.4} & 14.99 & -0.63 & 15.08 & 0.13 & 3.46 & -4.08 & 3.53 & 140 & 0.5660 & 8057       \\
\object{PN G096.4+29.9} & 11.19 & -0.39 & 11.29 & 0.21 & 0.15 & -4.21 & 3.58 & 48 & 0.5319 & 42630       \\
\object{PN G107.7+07.8} & 18.13 & -0.08 & 17.71 & 1.39 & 6.6 & -0.63 & 2.15 & 126 & 0.6007 & 3573        \\
\object{PN G107.8+02.3} & 17.13 & 0.97 & 16.2 & 3.78 & 0.81 & -5.61 & 4.14 & 96 & 0.7061 & 1115          \\
%\object{PN G116.2+08.5} & 16.94 & 1.77 & 16.15 & 2.65 & 4.31 & -0.46 & 2.08 & 55 & 0.5319 & 92191        \\
\object{PN G124.0+10.7} & 16.42 & -0.19 & 16.39 & 0.71 & 8.22 & 0.54 & 1.68 & 147 & 0.7061 & 11356       \\
\object{PN G128.0-04.1} & 17.4 & -0.03 & 17.44 & 1.09 & 6.64 & -1.26 & 2.4 & 158 & 0.7061 & 1479         \\
\object{PN G144.8+65.8} & 15.19 & 0.42 & 16.15 & 0.1 & 5.06 & -1.62 & 2.55 & 105 & 0.5319 & 68281        \\
\object{PN G148.4+57.0} & 15.73 & -0.63 & 16.1 & 0.04 & 6.4 & -0.28 & 2.01 & 105 & 0.5660 & 47101        \\
\object{PN G149.4-09.2} & 17.16 & -0.19 & 17.08 & 0.61 & 6.83 & -0.37 & 2.05 & 125 & 0.6063 & 3994       \\
\object{PN G149.7-03.3} & 16.5 & -0.26 & 16.55 & 0.55 & 7.78 & 1.24 & 1.4 & 100 & 0.7061 & 50555         \\
\object{PN G158.5+00.7} & 12.63 & -0.56 & 12.64 & 0.13 & 7.02 & 0.57 & 1.67 & 97 & 0.5750 & 31675        \\
\object{PN G164.8+31.1} & 17.09 & -0.63 & 17.14 & 0.07 & 7.12 & 0.14 & 1.84 & 116 & 0.6160 & 6151        \\
\object{PN G197.4-06.4} & 17.2 & -0.48 & 17.35 & 0.19 & 8.45 & 0.89 & 1.54 & 141 & 0.7061 & 25156        \\
\object{PN G197.8+17.3} & 10.67 & -0.3 & 10.63 & 0.49 & -1.22 & -5.52 & 4.11 & 47 & 0.7061 & 1059        \\
\object{PN G204.1+04.7} & 14.22 & -0.53 & 14.3 & 0 & 4.66 & -0.78 & 2.21 & 69 & 0.5319 & 74528           \\
\object{PN G205.1+14.2} & 15.96 & -0.59 & 16 & 0.13 & 7.25 & -0.21 & 1.98 & 136 & 0.7061 & 2715          \\
\object{PN G206.4-40.5} & 12.07 & -0.48 & 12.11 & 0.16 & 1.62 & -3.86 & 3.44 & 70 & 0.5319 & 49749       \\
\object{PN G214.9+07.8} & 16.45 & -0.44 & 16.56 & 0.16 & 5.42 & -1.64 & 2.56 & 119 & 0.5511 & 28652      \\
\object{PN G215.2-24.2} & 10.05 & 0.07 & 10.23 & 0.62 & -1.24 & -4.9 & 3.86 & 38 & 0.5826 & 2356         \\
\object{PN G215.5-30.8} & 15.45 & -0.53 & 15.49 & 0.08 & 6.96 & 0.45 & 1.72 & 99 & 0.5829 & 13754        \\
\object{PN G217.1+14.7} & 17.41 & -0.61 & 17.36 & 0.19 & 7.97 & 0.49 & 1.7 & 137 & 0.7061 & 11356        \\
\object{PN G219.1+31.2} & 15.48 & -0.55 & 15.52 & 0.12 & 6.88 & 0.53 & 1.69 & 94 & 0.5746 & 44631        \\
\object{PN G220.3-53.9} & 11.26 & -0.64 & 11.34 & 0.03 & 3.37 & -3.46 & 3.28 & 110 & 0.5319 & 58235      \\
\object{PN G221.5+46.3} & 15.97 & -0.51 & 16.03 & 0.15 & 5.43 & -1.4 & 2.46 & 110 & 0.5319 & 69576       \\
\object{PN G238.0+34.8} & 15.96 &  & 16.03 & 0 & 6.18 & -0.37 & 2.04 & 100 & 0.5499 & 58763              \\
\object{PN G239.6+13.9} & 15.9 & -0.55 & 15.97 & 0 & 4.31 & -3.27 & 3.21 & 142 & 0.5660 & 11287          \\
\object{PN G244.5+12.5} & 18.27 & -0.58 & 18.33 & 0.38 & 7.64 & 1.04 & 1.48 & 102 & 0.7061 & 36071       \\
\object{PN G248.7+29.5} & 16.42 & -0.6 & 16.4 & 0.13 & 6.03 & -0.45 & 2.08 & 98 & 0.5319 & 92191         \\
\object{PN G255.3-59.6} & 15.16 & -0.65 & 15.16 & 0.01 & 5.67 & -1.41 & 2.47 & 120 & 0.5660 & 21472      \\
\object{PN G261.0+32.0} & 12.2 & -0.56 & 12.32 & 0.13 & 1.48 & -4.72 & 3.79 & 89 & 0.5832 & 3632         \\
\object{PN G272.1+12.3} & 10.03 & 0.16 & 15.76 & 0.31 & 5.82 & -0.72 & 2.19 & 100 & 0.5319 & 74528       \\
\object{PN G273.6+06.1} & 12.46 & -0.36 & 12.53 & 0.39 & 2.95 & -3.98 & 3.49 & 114 & 0.5660 & 7986       \\
\object{PN G277.1-03.8} & 15.74 & 1.02 & 16.5 & 1.33 & 3.62 & -5.88 & 4.25 & 270 & 0.7061 & 1230         \\
\object{PN G277.7-03.5} & 17.94 & -0.35 & 17.94 & 0.75 & 6.3 & -0.78 & 2.21 & 120 & 0.5797 & 11255       \\
\object{PN G283.9+09.7} & 12.18 & -0.32 & 12.16 & 0.45 & 2.31 & -3.91 & 3.47 & 90 & 0.5319 & 52185       \\
\object{PN G290.5+07.9} & 14.5 & -0.31 & 14.5 & 0.54 & 2.46 & -3.42 & 3.27 & 80 & 0.5319 & 55982         \\
\object{PN G294.1+43.6} & 13.1 & -0.6 & 13.26 & 0.13 & 3.13 & -4.1 & 3.54 & 126 & 0.5660 & 8018          \\
\object{PN G310.3+24.7} & 12.93 & -0.54 & 12.95 & 0.1 & 2.79 & -3.44 & 3.28 & 90 & 0.5319 & 57033        \\
\object{PN G318.4+41.4} & 11.51 & -0.5 & 11.53 & 0.09 & 3.26 & -3.64 & 3.36 & 113 & 0.5319 & 56386       \\
\object{PN G327.8+10.0} & 13.08 & -0.17 & 13.42 & 0.84 & 1.2 & -4.2 & 3.58 & 68 & 0.5660 & 5830          \\
\object{PN G332.5-16.9} & 14.86 & -0.4 & 14.66 & 0.17 & 3.13 & -3.41 & 3.26 & 100 & 0.5319 & 58235       \\
\object{PN G335.5+12.4} & 12.35 & -0.29 & 12.37 & 0.68 & 2.14 & -4.08 & 3.53 & 90 & 0.5660 & 7285        \\
\object{PN G339.9+88.4} & 8.63 & 1.04 & 14.88 & 0.04 & 6.36 & -0.18 & 1.97 & 100 & 0.5496 & 58370        \\
\object{PN G341.8+05.4} & 15.18 & 0.64 & 15.55 & 2.77 & 2.17 & -4.62 & 3.75 & 109 & 0.5832 & 4046        \\
\object{PN G345.4+00.1} & 12.51 & 0.61 & 12.7 & 1.81 & 0.35 & -4.13 & 3.55 & 50 & 0.5319 & 43511         \\
\object{PN G358.5-07.3} & 17.17 & -0.4 & 17.49 & 0.22 & 7.37 & 0.22 & 1.81 & 123 & 0.7061 & 5920         \\

%\end{minipage}
%\end{threeparttable}
\end{longtable}

\end{center}

\tablefoot{
The value $G$ is the integrated magnitude in Gaia photometric instrument band and $G_{BP}-G_{RP}$ is the colour in the two Gaia photometric bands. Mass and evolutionary ages are estimated from \citet{millerbertolami17} evolutionary tracks.
}

\tablebib{ 
$m_{V}$, $A(V)$ and $T_{eff}$ from \citet{frew08}.
}

\end{appendix}

\end{document}